\documentclass[useAMS,usenatbib]{mnras}
\usepackage{caption}
\usepackage{tabularx}
\usepackage{lmodern}
\usepackage{color}
\usepackage{graphicx}
\usepackage{txfonts}
\usepackage{multirow}
\usepackage{supertabular}
\usepackage{longtable}
\usepackage{rotating}
\usepackage{lscape}
\usepackage{amssymb}
\usepackage{dcolumn}
\usepackage{dcolumn}  
\usepackage{changepage}
\usepackage{arydshln}
\usepackage{natbib}

\DeclareMathAlphabet{\mathsc}{OT1}{cmr}{m}{sc}
\def\testbx{bx}%
\DeclareRobustCommand{\ion}[2]{%
	\relax\ifmmode
	\ifx\testbx\f@series
	{\mathbf{#1\,\mathsc{#2}}}\else
	{\mathrm{#1\,\mathsc{#2}}}\fi
	\else\textup{#1\,{\mdseries\textsc{#2}}}%
	\fi}

\newcommand{\HI}{\ion{H}{i}}
\newcommand{\HII}{\ion{H}{ii}}

\newcommand{\FeII}{\ion{Fe}{ii}}
\newcommand{\MgI}{\ion{Mg}{i}}
\newcommand{\MgII}{\ion{Mg}{ii}}
\newcommand{\MnII}{\ion{Mn}{ii}}
\newcommand{\ZnII}{\ion{Zn}{ii}}
\newcommand{\TiII}{\ion{Ti}{ii}}
\newcommand{\CaII}{\ion{Ca}{ii}}

\newcommand{\CrII}{\ion{Cr}{ii}}
\newcommand{\NII}{\ion{N}{ii}}

\newcommand{\OII}{\ion{O}{ii}}
\newcommand{\OIII}{\ion{O}{iii}}

\newcommand{\kms}{km s$^{-1}$}
\newcommand{\zabs}{$z_{\rm abs}$}

\title[CGM study using DLA-galaxies at $z\sim0.6$]{A study of the circum-galactic medium at z $\sim$ 0.6 using DLA-galaxies\thanks{Based on data obtained under the ESO programme 092.A-0690(A) at the European Southern Observatories with X-Shooter at the 8.2 m telescopes operated at the Paranal Observatory, Chile.}}
\author[Rahmani, H. et al.]{Hadi Rahmani$^{1}$\thanks{E-mail:
hadi.rahmani@lam.fr}, C\'{e}line P\'{e}roux $^{1}$, 
David A. Turnshek$^{2}$, Sandhya M. Rao$^{2}$, 
\newauthor Samuel Quiret$^{1}$, Timothy S. Hamilton$^{3}$, 
Varsha P. Kulkarni$^{4}$, 
Eric M. Monier$^{5}$ 
\newauthor  ~~~~~~~~~~~~~~~~~~~~~~~~~~~~~~~~~~~~~~~~~~~~ and Tayyaba Zafar$^{6}$\\
$^{1}$Aix Marseille Universit\'{e}, CNRS, LAM (Laboratoire d'Astrophysique de Marseille) UMR 7326, 13388, Marseille, France\\
$^{2}$Department of Physics and Astronomy and PITTsburgh Particle physics, Astrophysics, and Cosmology Center (PITT PACC),\\ University of Pittsburgh, Pittsburgh, PA 15260, USA\\
$^3$Department of Natural Sciences, Shawnee State University, Portsmouth, Ohio 45662, USA\\
$^4$University of South Carolina, Dept. of Physics \& Astronomy, Columbia, USA\\
$^5$Department of Physics, The College at Brockport, State University of New York, Brockport, NY 14420, USA\\
$^6$Australian Astronomical Observatory, PO Box 915, North Ryde, NSW 1670, Australia
}
\begin{document}

\date{}

\pagerange{\pageref{firstpage}--\pageref{lastpage}} \pubyear{2002}

\maketitle

\label{firstpage}

\begin{abstract}
We present the study of a sample of nine QSO fields, with damped-Ly$\alpha$ (DLA) or sub-DLA systems at $z\sim0.6$, observed with the X-Shooter spectrograph at the Very Large Telescope. By suitably positioning the X-Shooter slit based on high spatial resolution images of HST/ACS we are able to detect absorbing galaxies in 7 out of 9 fields ($\sim$ 78\% success rate) at impact parameters from 10 to 30 kpc. In 5 out of 7 fields the absorbing galaxies are confirmed via detection of multiple emission lines at the redshift of DLAs where only 1 out of 5 also emits a faint continuum. In 2 out of these 5 fields we detect a second galaxy at the DLA redshift. Extinction corrected star formation rates (SFR) of these DLA-galaxies, estimated using their H$\alpha$ fluxes, are in the range 0.3--6.7 M$_\odot$ yr$^{-1}$. The emission metallicities of these five DLA-galaxies are estimated to be from 0.2 to 0.9 Z$_\odot$. Based on the Voigt profile fits to absorption lines we find the metallicity of the absorbing neutral gas to be in a range of 0.05--0.6 Z$_\odot$. 
The two remaining DLA-galaxies are quiescent galaxies with SFR $<$ 0.4 M$_\odot$ yr$^{-1}$ (3$\sigma$) presenting continuum emission but weak or no emission lines. 
Using X-Shooter spectrum we estimate i-band absolute magnitude of $-19.5\pm0.2$ for both these DLA-galaxies that indicates they are sub-L$^\star$ galaxies. Comparing our results with that of other surveys in the literature we find a possible redshift evolution of the SFR of DLA-galaxies.

\end{abstract}


\section{Introduction}
Inter-galactic Medium (IGM) gas accretion and galactic outflows are processes invoked in large scale structure simulations to regulate the galaxy growth \citep{Springel05,Sijacki07,Booth09,Oppenheimer10,Haas13,Vogelsberger14_1,Schaye15}. Active galactic nuclei (AGN) and supernovae winds are the processes which control the galaxy growth in respectively, bright and faint end of the luminosity function (LF)  \citep{van-de-Voort11,Puchwein13}. While outflows are observationally ubiquitous in star forming galaxies over cosmic time \citep{Lehnert96,Heckman00,Martin05,Weiner09,Nestor11,Martin12} there are only few examples of galactic inflow observations \citep{Rubin12,Bouche13}. Currently there have been successful progresses in implementing the feedback processes in models of galaxy formation and evolution \citep{Ceverino09,Piontek11,Dalla-Vecchia12,Hopkins-p12,Simpson-c15}. However the true nature of such processes remain poorly understood which is mainly due to the lack of stringent observational constraint on the distribution of gas and metals around galaxies.  

The circumgalactic medium (CGM) at the interface between the IGM and galaxies extends from approximately  20 -- 300 kpc around galaxies. The CGM is a complicated site of entwined gas from the IGM accretion and galactic outflows through which the baryon exchange between the IGM and galaxies occurs \citep{Suresh15}. Hence, it provides a suitable site to study modes of gas accretion \citep{Keres05,Keres09,Stewart11}, galactic outflows and chemical enrichment of the IGM \citep{Oppenheimer06,Simcoe06,Ryan-Weber09,Wiersma10,Steidel10,Tumlinson11,Dodorico13,Shull14}.

Absorption systems in the spectra of background QSOs are among the best tools to study CGM of galaxies over a wide range of redshifts and physical conditions. In particular a subclass of QSO absorbers with Hydrogen column densities of $\log$ (N (\HI) [cm$^{-2}$]) $\gtrsim$ 17.0 called the Lyman limit systems (LLS) likely traces the cool ($T \sim10^4$ K) phase of CGM \citep{Tytler82,Sargent89,Prochaska99,Songaila10}. There exist signatures of bimodality in the metallicity distribution of such systems that may represent galactic outflows in the metal rich branch and cold accretion in the metal poor branch \citep[submitted]{Lehner13,Quiret16}. However, accurate metallicity determinations are difficult for the lower N(\HI) LLS due to uncertainties in N(\HI) measurements while they fall in the flat part of the curve of growth and ionization correction.

Two subclasses of higher column density LLS with $\log$ (N (\HI) [cm$^{-2}$]) $\geq$ 20.3 and 19.0 $\leq$ $\log$ (N (\HI) [cm$^{-2}$]) $<$ 20.3 are known as Damped Lyman-$\alpha$ absorbers (DLA) and sub-DLAs, respectively \citep{Peroux01,Wolfe05}. Owing to such high column density of \HI\ they produce Lorentzian wings in their Lyman-$\alpha$ absorption profiles and are dominantly neutral due to self shielding \citep{Wolfe86,Petitjean92,Dessauges-Zavadsky03,Meiring07}. Optical spectroscopic surveys of QSOs have demonstrated that these absorbers dominate the mass density of neutral gas in the universe \citep{Wolfe05,Prochaska05,Noterdaeme09dla,Noterdaeme12dla,Zafar13,Crighton15,Sanchez-Ramirez16}. Therefore, they form the main neutral gas reservoir for the formation of stars at high redshift. 

Better constraint on the physical state of the CGM or ISM can be obtained where the absorbing galaxies are detected in emission \citep[e.g.,][]{Moller98a,Moller02,Chen05,Bouche07_simple,Fynbo10,Rao11,Peroux11a,Krogager12,Bouche12,Bouche13,Schroetter15,Kacprzak15}. \citet{Rao11} studied a large sample of absorbing galaxies, primarily identified based on photometric redshifts, to demonstrate that there exists a statistically significant anti-correlation between the N(\HI) and the impact parameter \citep[see also][for similar studies]{,Moller98a,Christensen07,Monier09,Krogager12}. \citet{Peroux11a} used VLT/SINFONI to search for host galaxies of DLAs and sub-DLAs at $z\sim1$ and found them to be faint galaxies with low SFR of the order of a few M$_\odot$ yr$^{-1}$. Benefiting from 3D observations they extracted morpho-kinematic properties of such galaxies that allowed unraveling the nature of absorbing gas to be outflows or extensions of a rotating disk \citep[also see][]{Schroetter15}. In a similar study \citet{Bouche13} could trace the low metallicity inflowing gas seen in absorption at $\sim26$ kpc from the host galaxy at $z\sim$2.3. Higher rates of star formation ($\gtrsim10$ M$_\odot$ yr$^{-1}$) but with smaller detection rates have been measured for DLA host galaxies at $z\sim2$ \citep{Noterdaeme12,Peroux12,Fynbo13}. There exists further evidences that indicate SFR of DLA-galaxies decreases from $z\sim2$ to $z\sim1$ \citep{Bouche12}. However, such results are limited by the low success rates in detection of DLA-galaxies which is usually attributed to the faint nature of DLA-galaxies near the position of bright background QSOs.  
%

The identification of a substantially large sample of DLAs at low redshifts is difficult because the Lyman-$\alpha$ still lies in the UV wavelengths and the incidence of DLAs is low. As shown by \citet{Rao06}, strong \MgII\ absorbers are unbiased tracers of DLAs, and can be used to obtain higher detection rates of DLAs in the UV. Their HST-UV surveys of QSOs with strong \MgII\ absorbers led to the identification of 41 DLAs and 81 sub-DLAs at redshifts $0.1<z<1.65$. Subsequently, the absorbing galaxies associated with a sample of 80 of them with redshifts $z<1$ were identified using variety of techniques \citep{Rao11}. More recently, \citet{Turnshek15} used the Advanced Camera for Surveys (ACS) HRC-P200L prism aboard the Hubble Space Telescope (HST) to obtain UV spectra of QSOs with strong \MgII\ absorbers. They found $\approx35$ high-probability DLAs. The dispersion of this prism is non-linear and extremely low at red wavelengths. Therefore, in addition to the dispersed UV light of the QSO which enables detection of DLAs, the spectra of galaxies in the field show little dispersion because they are predominantly red, thus producing deep images of galaxies in the field. Subtracting QSOs from these HST/ACS images produces residual frames in which faint objects appear at impact parameters as small as $b = 2$ kpc ($\sim0.5''$). These data provide direct information on galaxies including sky positions, impact parameters, sizes and morphologies when possible. In this paper we describe the results of a VLT/X-shooter study aimed at identifying the galaxies responsible for DLAs or sub-DLAs in a subset of these QSO fields at $z \sim0.6$. 

This paper is organized as following. In Section (2) we describe our sample, the strategy of the observations and X-Shooter data reduction. In Section (3) we demonstrate  the identification of the absorbing galaxies in the X-Shooter spectra and provide the flux measurements for individual objects. In Section (4) we present the X-Shooter QSO absorption line analysis of our DLAs and sub DLAs. Results are presented in Section (5) and we conclude in section (6).
\section[sample]{observations and data reduction}
\subsection[]{QSO subtraction of HST/ACS prism images} 
 To identify the absorbing galaxy candidates from HST/ACS high resolution images we carried out an accurate Point Spread Function (PSF) subtraction of QSOs. This procedure is applied on the reduced and sky background subtracted prism images of the QSOs where the fluxes are normalized with respect to the head of the PSF. Details of HST/ACS observations and data reduction of this dataset were described in \citet{Turnshek15}. The quasar light is removed by subtracting a model of the telescope's PSF. We generate such a model PSF by making a stacked image out of QSOs in our sample. A median combined image is made out of all QSOs after they are co-aligned based on a cross-correlation technique. That is, for a given pixel in the PSF model, its flux is the median of all the fluxes in the input images for that pixel. In subtracting the quasar light, the PSF model is shifted to align with the quasar and scaled in flux so that the residual is flat just outside the quasar core. As a result the QSO core is well removed. The  dispersed part of the spectrum, forming the PSF's tail, is not well matched by the median model and shows a larger variation from one quasar to another. 
The differences in the QSOs’ UV spectra, both continuum and lines, cause residuals in the tail that could obscure a compact galaxy lying on the dispersion axis.

Fig. \ref{fig_FC_1} presents the results of PSF subtraction on HST/ACS observations. North and East are respectively towards up and left for all panels. Name of QSO fields are provided at the top left corner of each panel. Absorbing galaxy candidates falling in the slit and the QSO are marked with respectively circles and dashed squares. Slit width and length is 1$''$ and 11$''$, respectively and slit position angles (PA) are drawn to match our X-Shooter observations. Detected galaxies in our X-Shooter observations are marked with arrows where longer red ones correspond to galaxies associated with absorbers and shorter black ones are other galaxies. (see section \ref{dla_emission}).

In summary we have targeted 9 QSO fields having DLAs at $z\sim0.6$ with X-Shooter to cover 25 absorbing galaxy candidates. The position angle of one of the slits in the field of 0958+0549 has been mis-aligned (see Fig. \ref{fig_FC_1}) and hence we have missed G3. As a result number of targeted galaxies is 24 that means 2.7 galaxy candidates per field. 
\subsection[]{X-Shooter observation and data reduction}\label{xsh_obs}
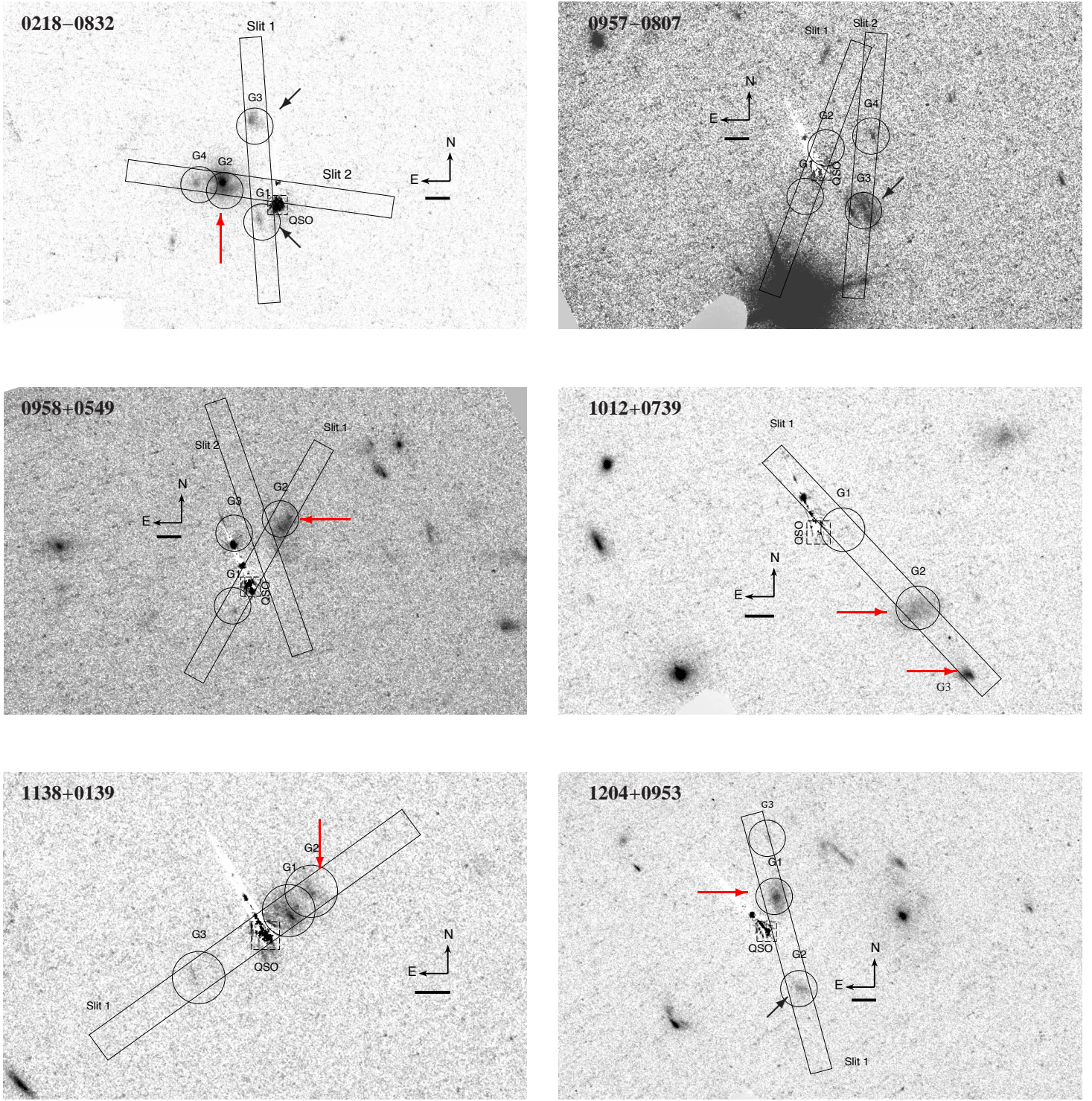
\begin{figure*}
	\centering	
	\hspace*{-1cm}
	\vbox{
		\hbox{
		\includegraphics[width=0.55\hsize,bb=0 0 541 379,clip=,angle=0]{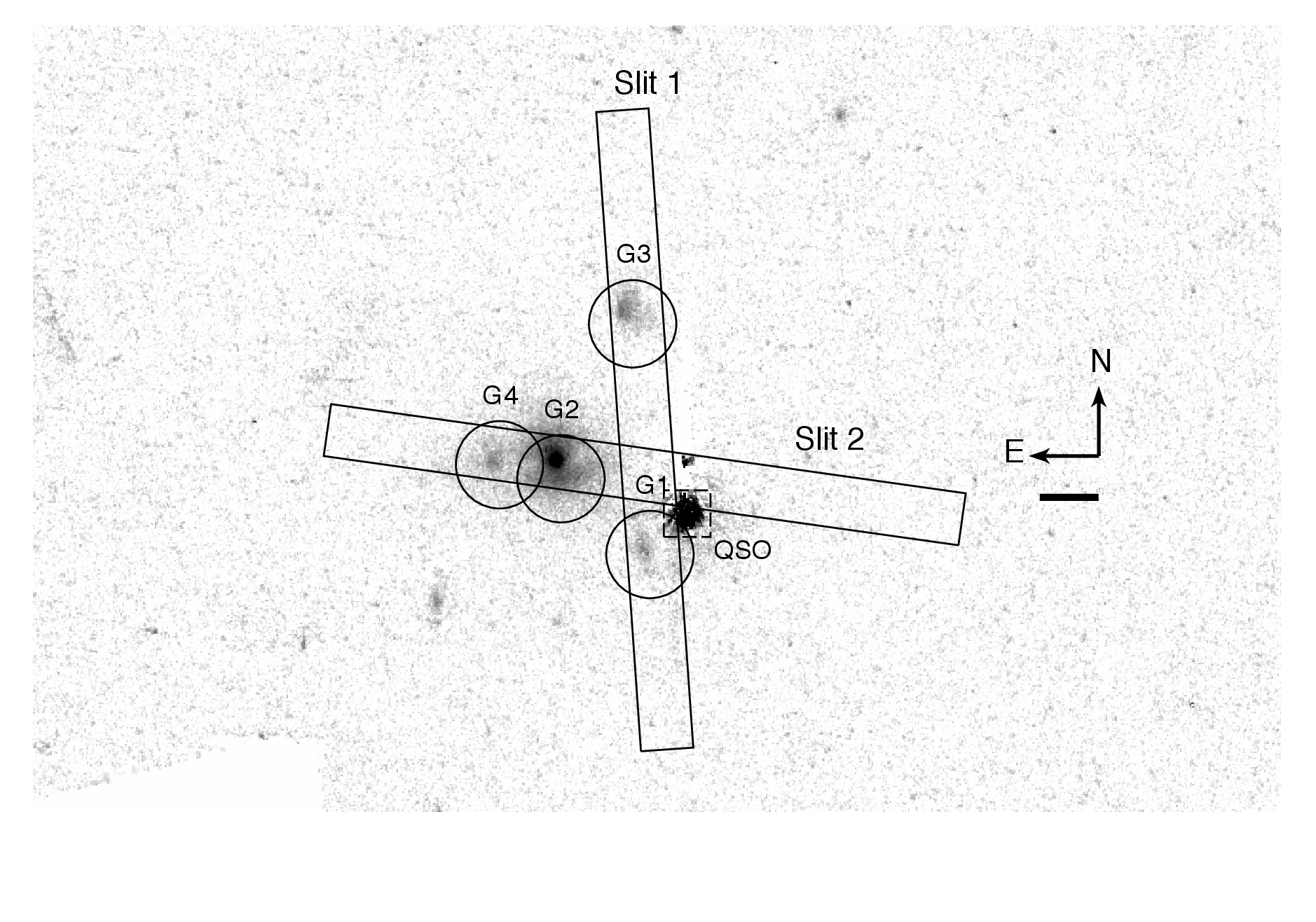}
		\includegraphics[width=0.55\hsize,bb=0 0 541 379,clip=,angle=0]{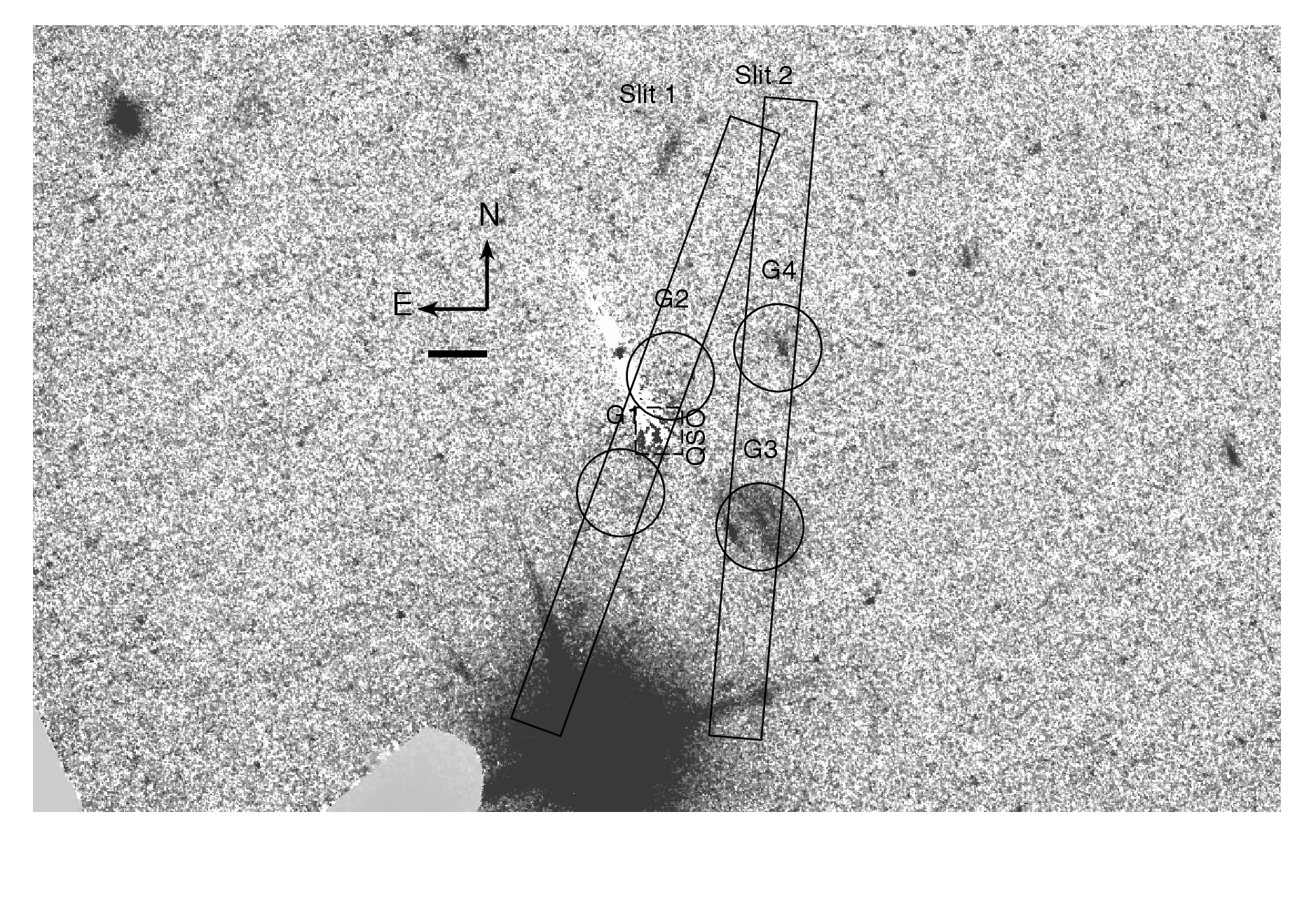}
	}
	\hbox{
		\includegraphics[width=0.55\hsize,bb=0 0 541 379,clip=,angle=0]{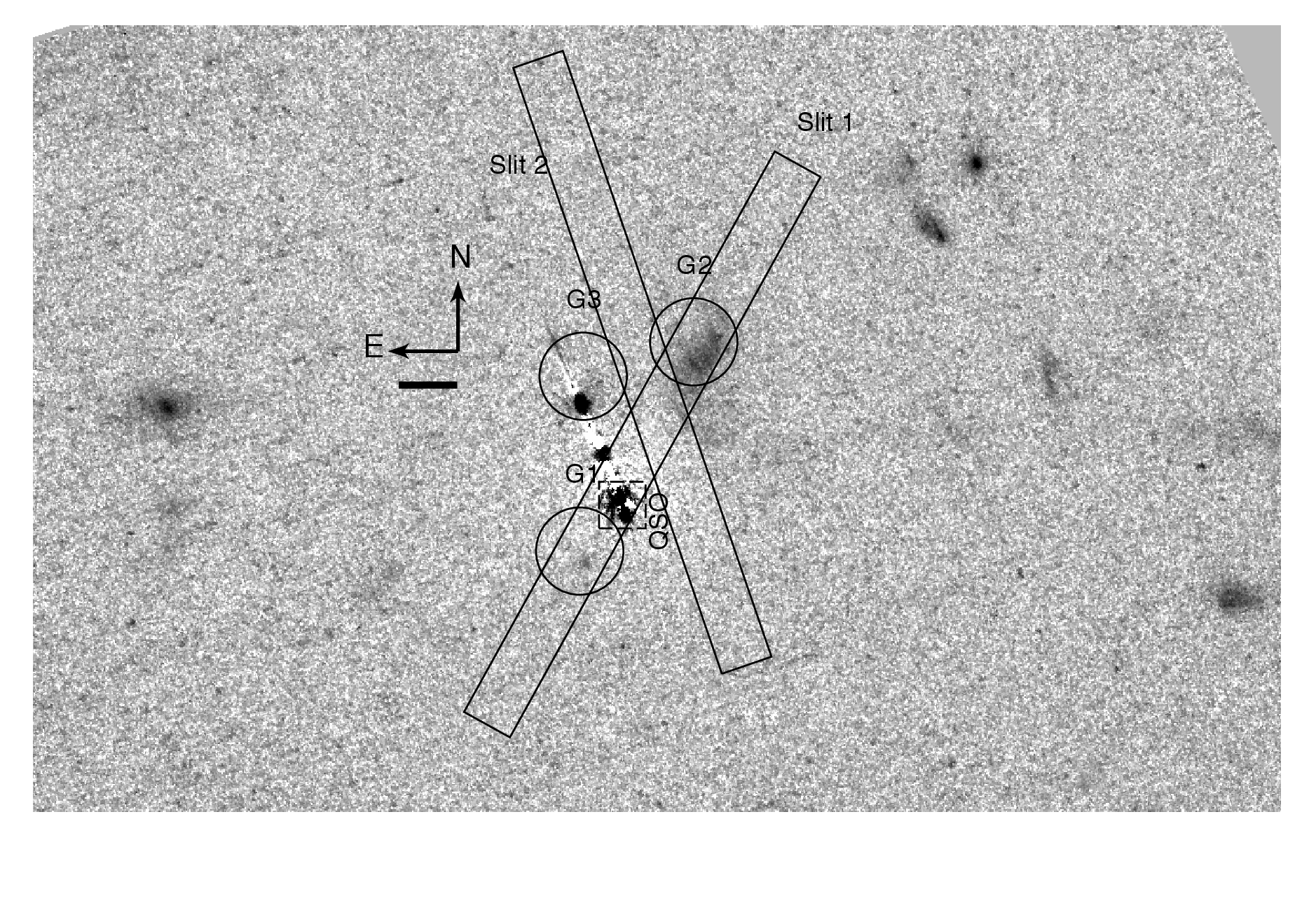}
		\includegraphics[width=0.55\hsize,bb=0 0 541 379,clip=,angle=0]{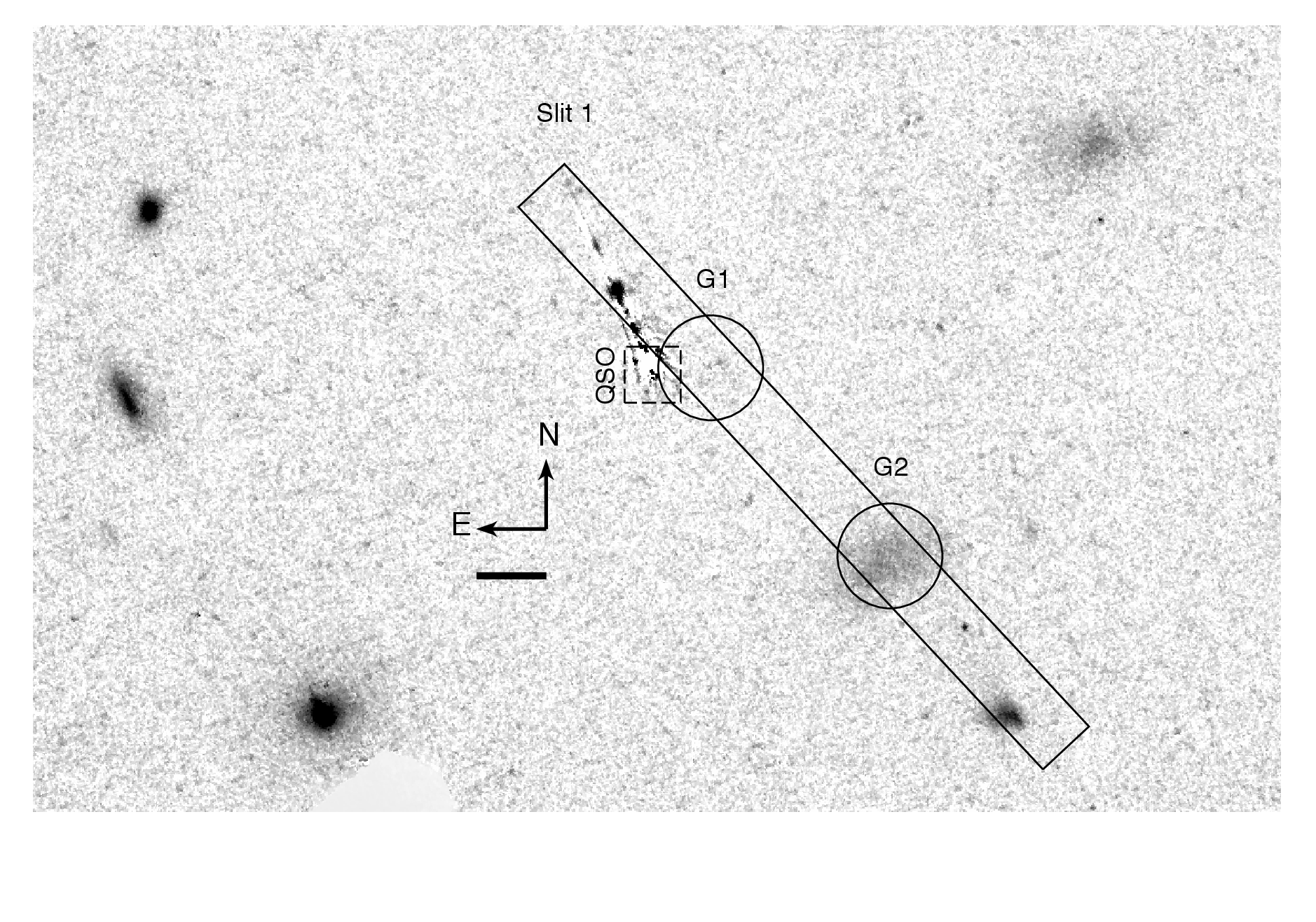}
	}
	\hbox{
		\includegraphics[width=0.55\hsize,bb=0 0 541 379,clip=,angle=0]{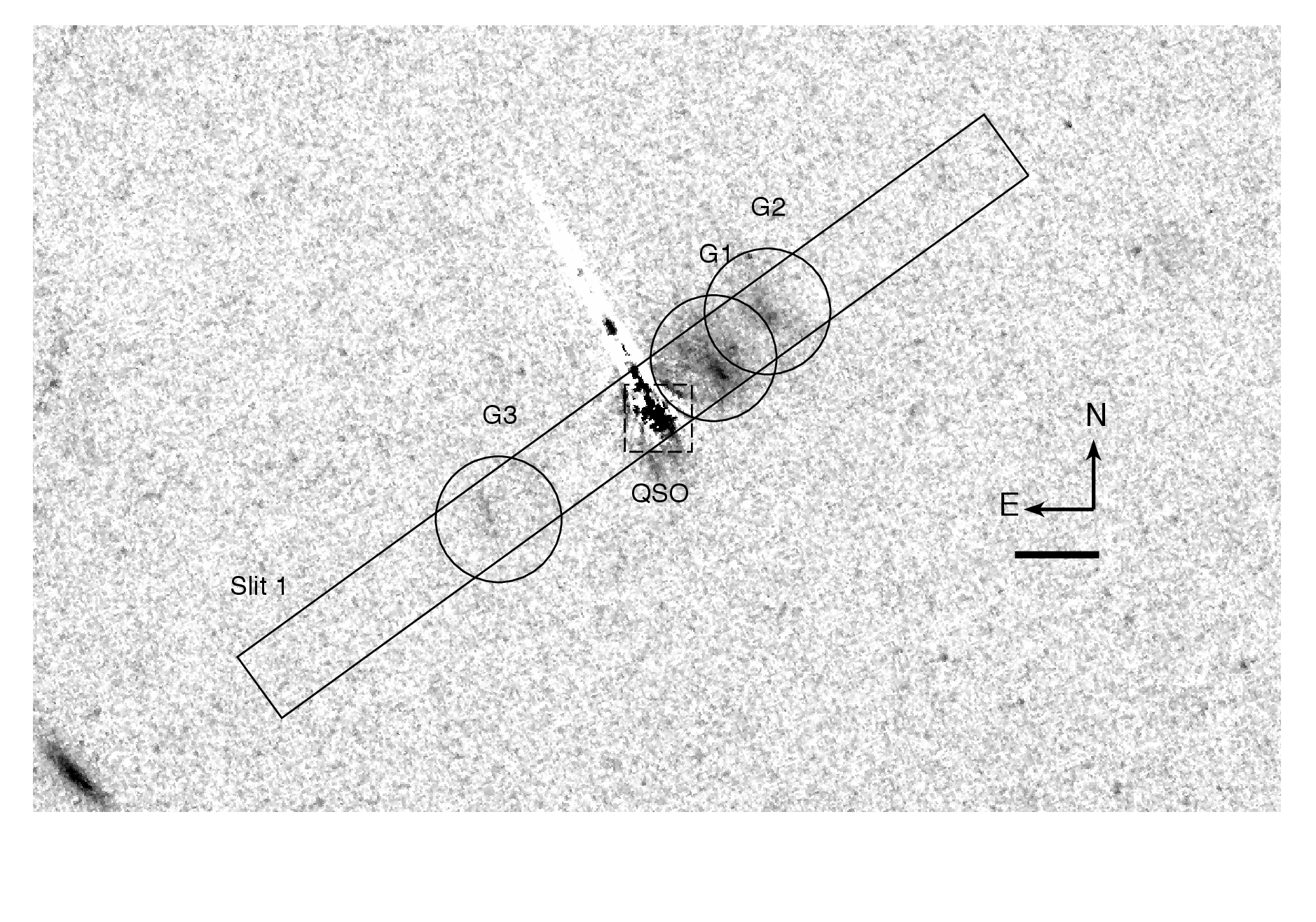}
		\includegraphics[width=0.55\hsize,bb=0 0 541 379,clip=,angle=0]{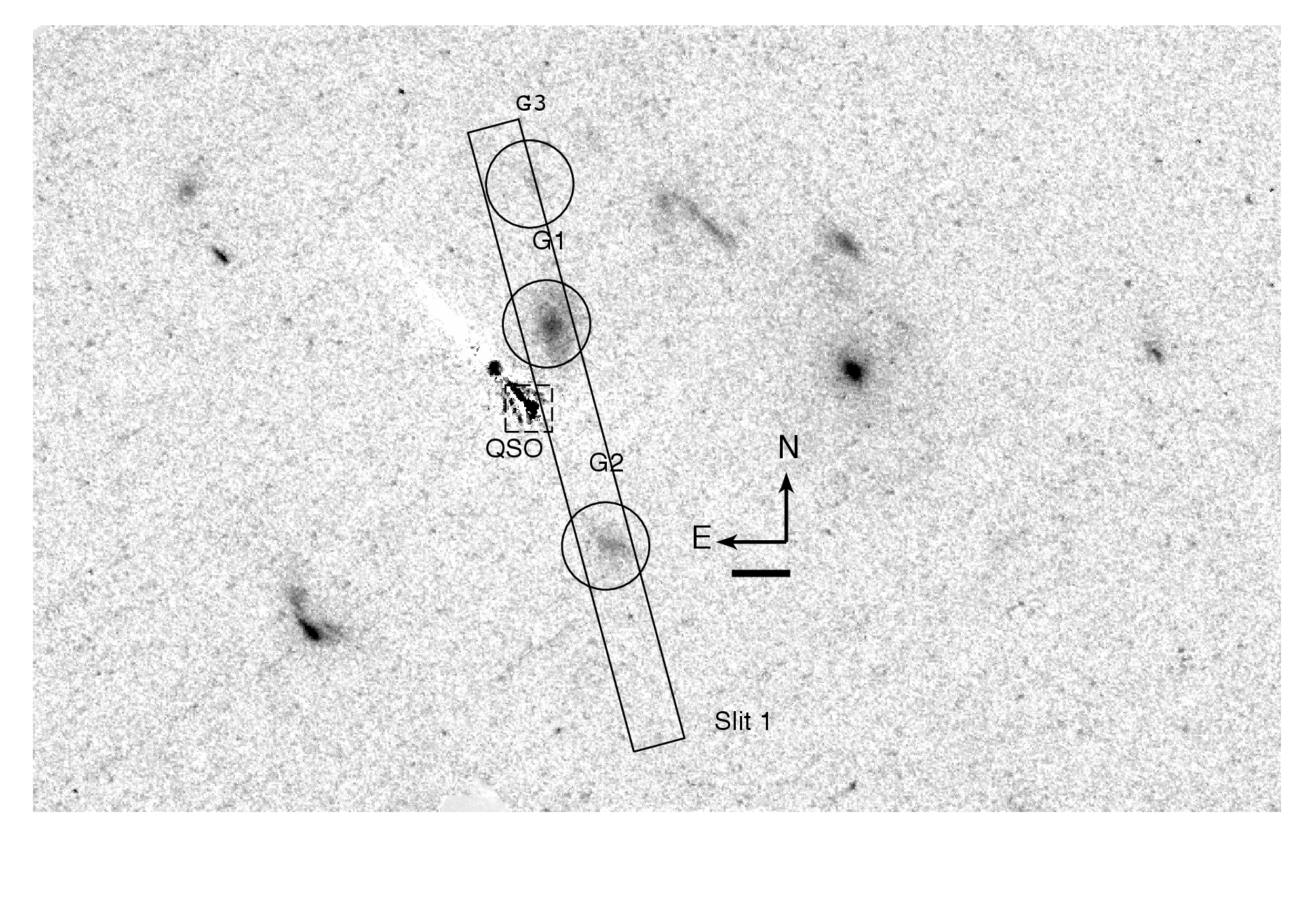}		
	}
	}
	\caption{Slit configurations for 9 QSO fields observed with X-Shooter. Panels are HST/ACS slitless prism images where QSOs PSF are subtracted. Different X-Shooter slit orientations used for observations are entitled ``Slit1`` and ``Slit2``. Circles and dashed square mark the sky position of candidate galaxies and the QSO of each field, respectively. North and East are towards up and left. The solid bar below the direction arrow in each plot is a 1$''$ scale. Longer (vertical or horizontal, in red) and shorter (inclined, in black) arrows in each panel mark detected galaxies associated with and unrelated to DLAs, respectively.}
	\vskip -.25cm
	\begin{picture}(0,0)(0,0)
	\put( -260,630){\bf \large 0218$-$0832}  \put( 25,630){\bf \large 0957$-$0807}
	\put( -260,435){\bf \large 0958$+$0549} \put( 25,435){\bf \large 1012$+$0739}
	\put( -260,240){\bf \large 1138$+$0139} \put( 25,240){\bf \large 1204$+$0953}
	\thicklines
	\put( -160,512){\textcolor{red}{\vector(0,1){25}}}
	\put( -120,600){\vector(-1,-1){10}}
	\put( -120,520){\vector(-1,1){10}}
	\put( 183,555){\vector(-1,-1){10}}	
	\put( -95,382){\textcolor{red}{\vector(-1,0){25}}}	
	\put( 150,335){\textcolor{red}{\vector(1,0){25}}}
	\put( 185,305){\textcolor{red}{\vector(1,0){25}}}
	\put( 200,295){\scriptsize G3}
	\put( -110,230){\textcolor{red}{\vector(0,-1){25}}}
	\put( 80,193){\textcolor{red}{\vector(1,0){25}}}
	\put( 115,130){\vector(1,1){10}}	
	\end{picture}
	\label{fig_FC_1}
\end{figure*}
\begin{figure*}
	\addtocounter{figure}{-1}
	\centering	
	\hspace*{-1cm}
	\vbox{
		\hbox{
			\includegraphics[width=0.55\hsize,bb=0 0 541 379,clip=,angle=0]{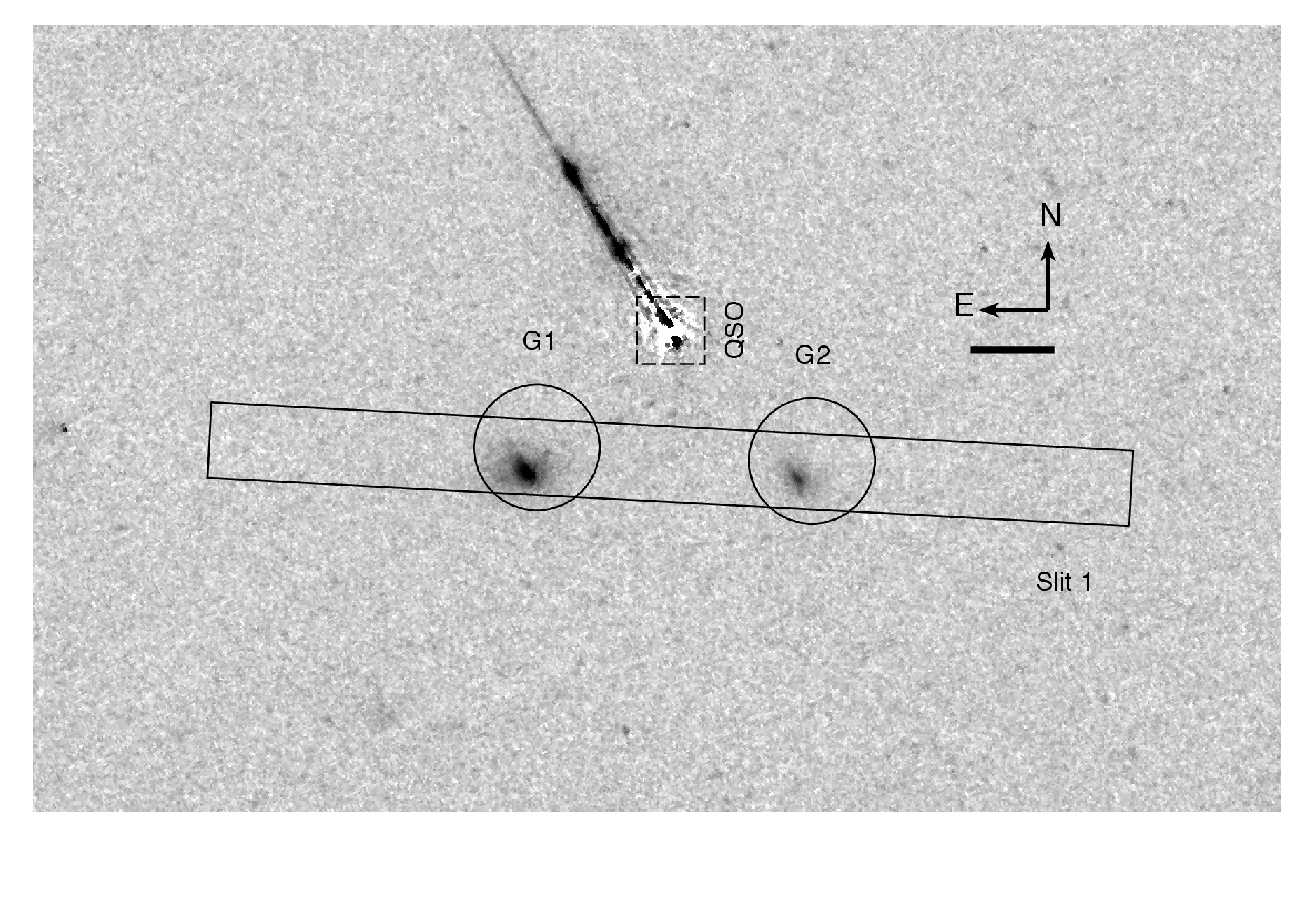}		
			\includegraphics[width=0.55\hsize,bb=0 0 541 379,clip=,angle=0]{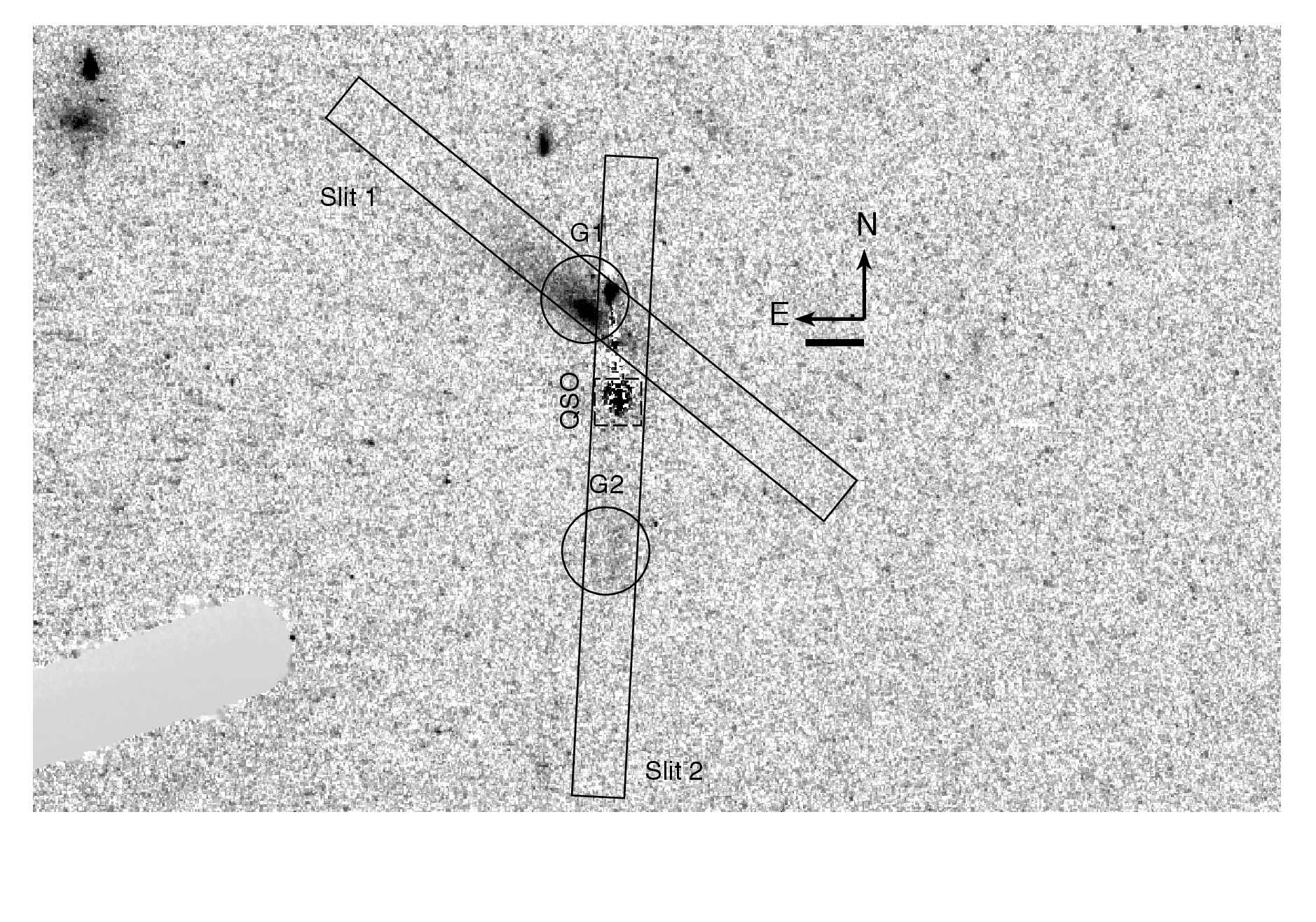}		
		}
	\hbox{
			\includegraphics[width=0.55\hsize,bb=0 0 541 379,clip=,angle=0]{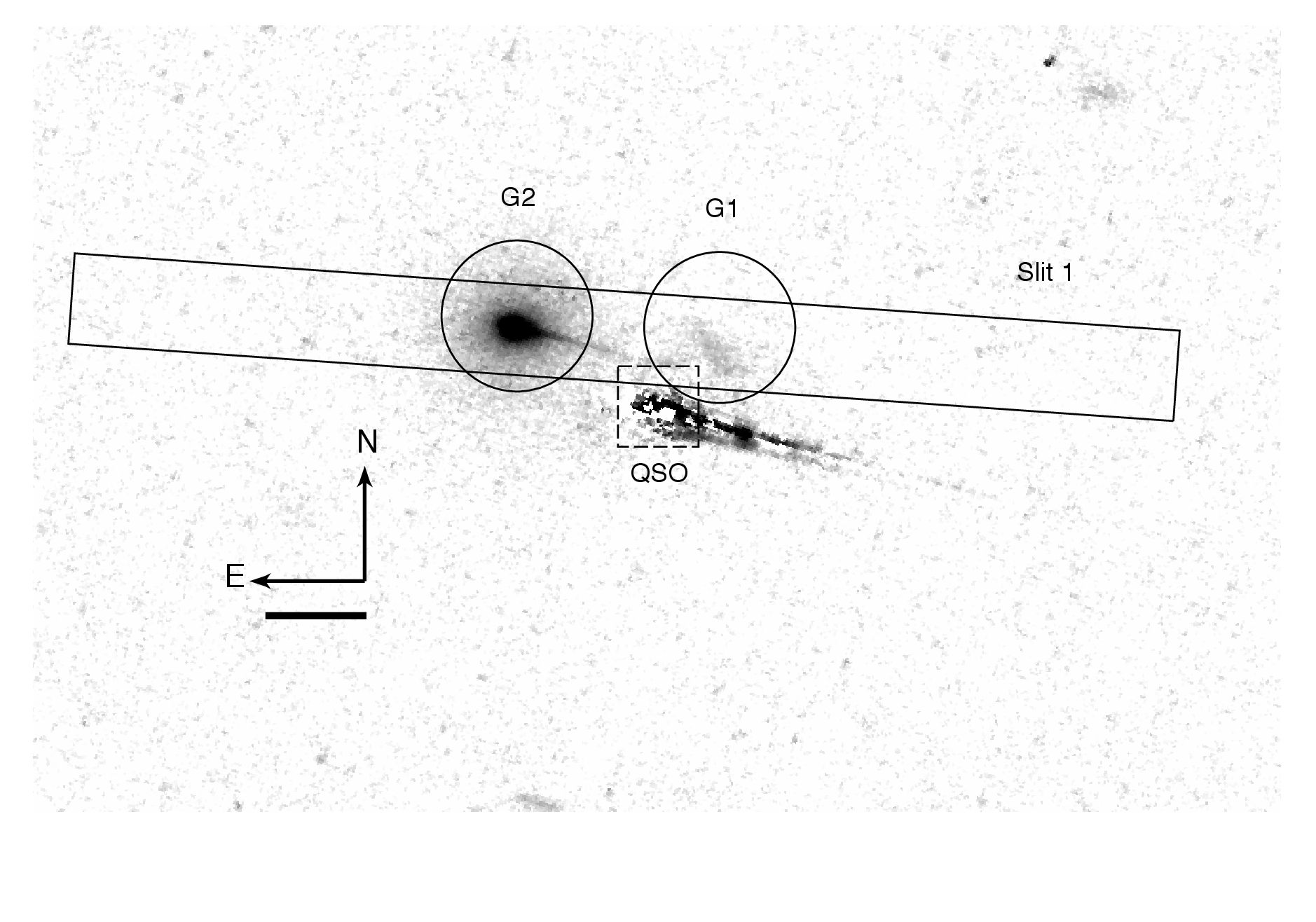}					
	}
	}
	\caption{continued.}
	\begin{picture}(0,0)(0,0)
	\put( -260,410){\bf \large 1217$+$0500}  \put( 25,410){\bf \large 1357$+$0525}
	\put( -260,210){\bf \large 1515$+$0410}
	\thicklines
	\put( -165,295){\textcolor{red}{\vector(0,1){25}}}	
	\put( -170,125){\textcolor{red}{\vector(0,1){25}}}
	\end{picture}
	\label{fig_FC_2}
\end{figure*}
We observed the DLA-galaxy candidates using VLT/X-Shooter \citep{Vernet11} at the European Southern Observatory (ESO) in service mode [programme 092.A-0690(A), Peroux, PI]. The X-Shooter spectrograph covers a wavelength range of 0.3 $\mu$m to 2.3 $\mu$m at medium resolution in a simultaneous use of three arms in UVB, VIS and NIR. Therefore, not only does it cover the interesting nebular emission lines it also covers the UV absorption lines in the spectrum of the QSO that are dominantly in the UVB arm. Having known the position of the DLA-galaxy candidates with respect to the QSOs we have observed all our targets by setting specified slit PAs. Furthermore, to have a robust sky subtraction, specially in NIR, the nodding mode was used following an ABBA scheme. We have not used the default nod lengths (2.5$''$) of the X-Shooter as it may smear out  the signal from our targeted faint galaxies by the negative trace of the bright QSOs while subtracting the sky off images. Instead, knowing the impact parameter of candidate galaxies, we fine tune the nodding length using the ``GenericOffset`` template.  Table \ref{log_XSH} presents a summary of the details of our X-Shooter observations. Slit widths of 1.0$''$,0.9$''$ and 0.9$''$ are used for respectively UVB, VIS and NIR arms of X-Shooter throughout of our observations.  This choice of slit widths results in formal spectral resolutions of 5100, 8800 and 5600 for the UVB, VIS and NIR respectively. 

We made use of the X-Shooter common Pipeline Library (CPL)  \citep{Goldoni06} release 6.5.1\footnote{http://www.eso.org/sci/facilities/paranal/instruments/xshooter/doc/} for reducing the science raw images and produce the final 2D spectra. All our science data are taken in \textsc{nodding} mode and hence we follow a standard procedure of data reduction as follows. We first compute an initial guess  for the wavelength solution and position of the center and edges of the orders. Then we trace the accurate position of the center of each  order and follow this step by generating the master flat frame out of five individual lamp flat exposures. Next we find a 2D wavelength solution and modify it by applying a flexure correction to correct for the shifts that can be of the order of the size of a pixel. Finally, having generated the required calibration tables we reduce each pair of science frames to obtain the  flat-fielded, wavelength calibrated and sky subtracted 2D spectrum. For the flux calibration of each science frame we have chosen the standard star from the X-Shooter archives which has the minimum airmass difference, observed within a maximum 7 days time gap from that science frame. However, we note that recalibration using the standard stars of the same nights results in fluxes consistent within 5\%. 

To extract the 1D flux of the QSO or other objects from a final reduced 2D frame we carry on a spectral point spread function  (SPSF) subtraction through the following steps: (1) dividing the 2D image in chunks of sizes $\sim$ 200 pixels along the wavelength; (2) estimating the mean spatial profile of each chunk by averaging the flux along the wavelength at each spatial pixel; (3) modeling this profile using a Moffat function to estimate the $\sigma$ and central position of each chunk; (4) fitting these parameters using a low order polynomial to extract their values for each pixel (wavelength bin). Having done so we know the center of the QSO ($y_\lambda$) and its spatial broadening parameter ($\sigma_\lambda$) at each wavelength ($\lambda$). (5) Now, we find the emission profile, P$_\lambda$($y$),  at each $\lambda$. We model P$_\lambda$($y$) with a Moffat, to obtain the amplitude, while keeping $y_\lambda$ and $\sigma_\lambda$ fixed based on what obtained at stage (4). We further integrate over the best fitted P$_\lambda$($y$) to calculate the total flux at each $\lambda$.  Therefore, we simultaneously find the 1D and 2D spectrum of the QSO or other bright sources. We then subtract the 2D modeled spectrum to obtain the residual image. We include masks in cases where we expect emission lines from the galaxy candidate at low impact parameters (b $\lesssim1.5''$). In such cases we mask a region of $\sim100$ \kms\ around the expected wavelengths of nebular emission lines and interpolate the parameters at these ranges using both sides to find the QSO flux. QSO's spectra obtained from the SPSF subtraction are used to study the UVB absorption lines and the residual images are searched for the expected DLA galaxies. 
\begin{table*}
		\small
		\caption{Log of the VLT/X-Shooter observations.}
		\begin{tabular}{lcccccccccccc}
			\hline
			
		 QSO/field (I)& $z_{\rm em}$ (II) & $z_{\rm abs}$ (III)& cumulative nodding (IV) & Position angle (V) & N$_{\rm gal}$ (VI) & N$_{\rm suit}\times 2\times$ EXPTIME (VII)  \\
			\hline
			   0218$-$0832(1)  & 1.218 & 0.5899  & +3.0,$-$3.0   & 4  & 2  & 2$\times$2$\times$1200 \\ 
			   0218$-$0832(2)  & 1.218 & 0.5899  & +3.0,$-$6.0  & 82 & 2 & 1$\times$2$\times$1200 \\ 
			   0957$+$0807(1)  & 0.870 & 0.6975  & +3.5,$-$7.0 &  160 & 2 & 2$\times$2$\times$1440 \\ 
			   0957$+$0807(2)  & 0.870 & 0.6975  & +2.0,$-$3.0 &  175   & 2 & 2$\times$2$\times$1200 \\ 
		       0958$+$0549(1)  & 0.730 & 0.6557  & +2.0,$-$3.0 & 151   & 2 & 2$\times$2$\times$1200 \\ 
		       0958$+$0549(2)  & 0.730 & 0.6557  & +2.0,$-$5.0  & 19    & 1 & 2$\times$2$\times$1200 \\ 
			   1012$+$0739   & 1.030 & 0.6164  & +3.0,$-$6.0 & 43      & 2 & 2$\times$2$\times$1200 \\ 
			   1138$+$0139   & 1.042 & 0.6130  & $-$2.5,+6.0  & 126   & 3 & 1$\times$2$\times$1200   \\ 
			   1204$+$0953   & 1.276 & 0.6401  & $-$3.0,+6.0 & 15   & 3 & 2$\times$2$\times$1200 \\ 
			   1217$+$0500   & 0.632 & 0.5413  & +3.0,$-$6.0  & 87   & 2 & 1$\times$2$\times$1200 \\ 
		       1357$+$0525(1)  & 0.740 & 0.6327  & +3.5,$-$6.0 & 51   & 1 & 3$\times$2$\times$1200 \\ 
		       1357$+$0525(2)  & 0.740 & 0.6327  & +3.0,$-$6.0  & 177  & 2 & 2$\times$2$\times$1200 \\ 
		       1515$+$0410   & 1.272 & 0.5592  & +3.5,$-$6.0 & 86  & 2 & 2$\times$2$\times$1200 \\ 
			\hline
		\end{tabular}
		\begin{flushleft}
			(I) QSO field: (1) and (2) define the same QSO field with different configurations of the slit position; (II) redshift of the QSO; (III) redshift of the DLA; (IV) cumulative nodding length from blind offset in arcsec; (V) position angle of the slit measured from North of East in degrees; (VI) Number of covered galaxy candidates in the X-Shooter slit; (VII) number of suitable exposures multiplied by the exposure time in second of individuals (a factor of 2 included for all 
			as all exposures are in nodding mode).  
		\end{flushleft}
\label{log_XSH}
\end{table*}
\begin{figure*}
	\centering	
	\vbox{
		\includegraphics[width=0.25\hsize,bb=206 -1 404 792,clip=,angle=90]{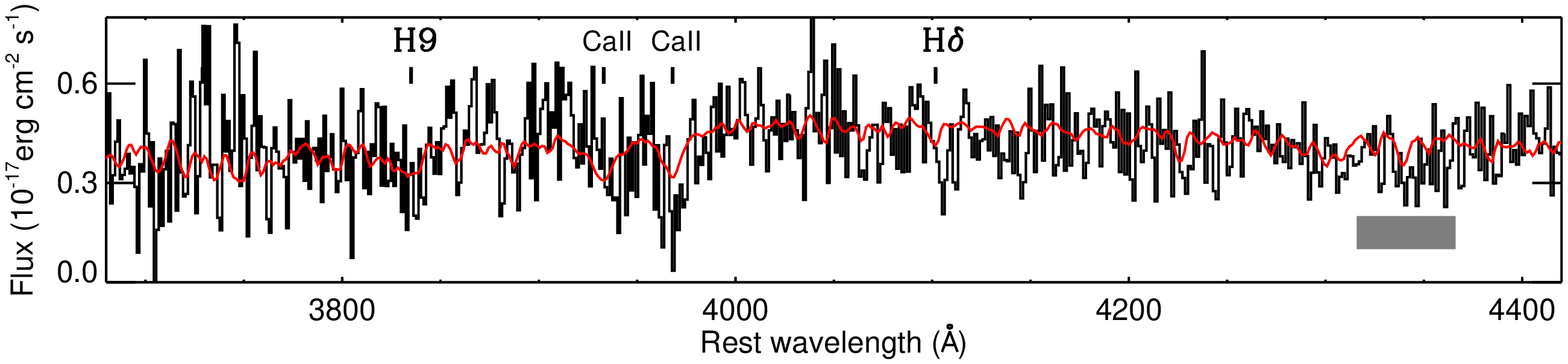}
		\includegraphics[width=0.50\hsize,bb=107 -1 503 792,clip=,angle=90]{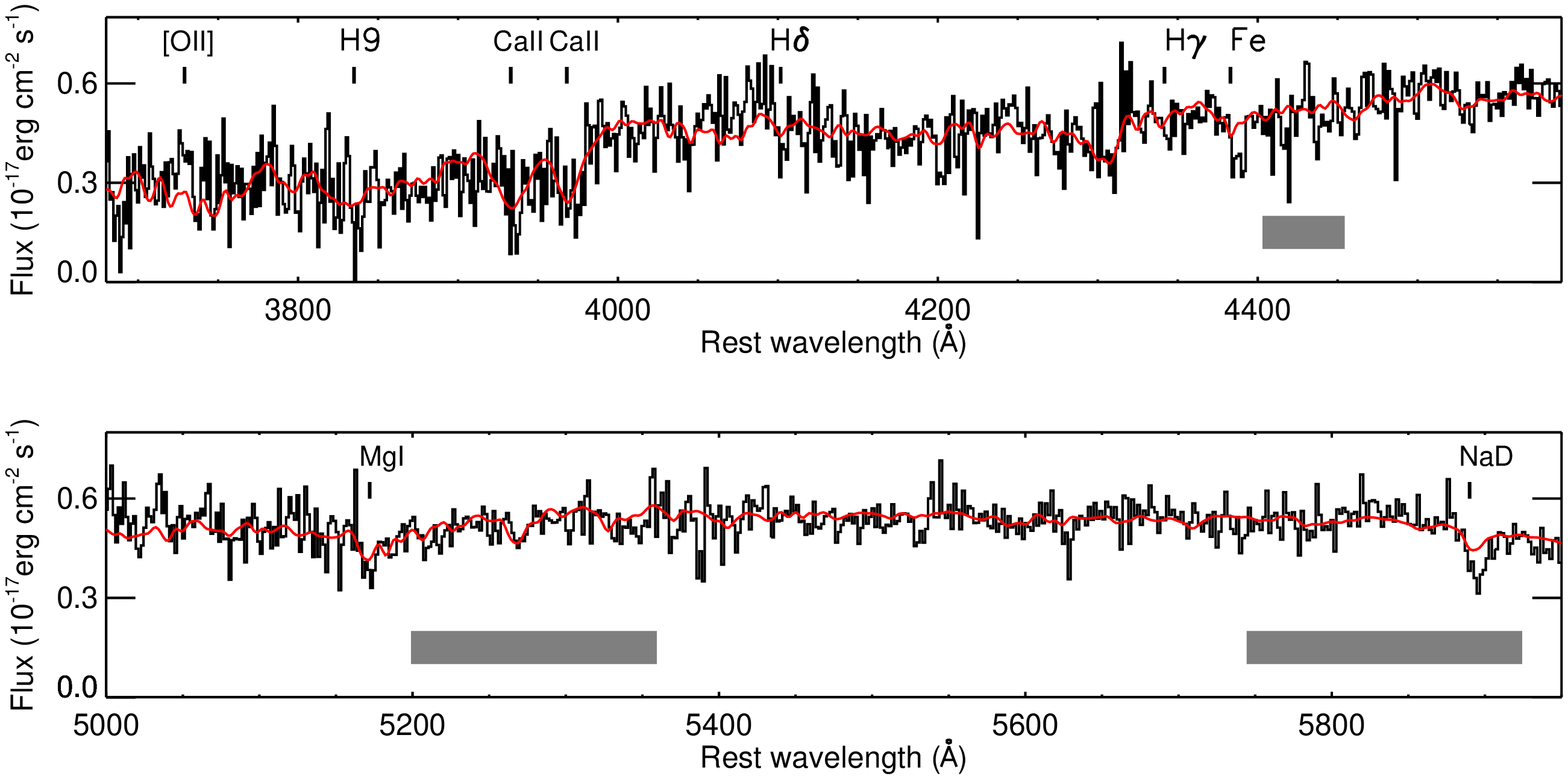}
	}
	\caption{{\it top:} part of the X-Shooter spectrum of the DLA-galaxy towards J0218$-$0832. Black and red lines are rest frame spectra of the DLA-galaxy (binned over 100 \kms), at $z_{\rm G2}=0.5895$, and an early type galaxy with similar absorption lines, respectively. Grey rectangles demonstrate the wavelength regions that are affected by telluric sky absorption lines. We have corrected the spectrum for the telluric absorption features using \textsc{Molecfit} \citep{Smette15,Kausch15}. \CaII\ H\&K, H$\delta$ and H9 absorption features are detected and labeled. {\it middle and bottom:} same as {\it top} but for the DLA-galaxy towards J1515$+$0410 at $z_{\rm G2}=0.5580$. \CaII\ H\&K, H$\delta$ and H$\gamma$ are tentatively and 4000 \AA\ break and NaD are clearly detected.}
	\begin{picture}(0,0)(0,0)
    \put( 130,438){\bf \large 0218$-$0832 (G2)}
    \put( 130,312){\bf \large 1515$+$0410 (G2)}
    \put( 130,185){\bf \large 1515$+$0410 (G2)}

	\end{picture}
	\label{fig_cont_0218_1515}
\end{figure*}
\begin{figure*}
	\centering	
	\vbox{
	\includegraphics[width=0.25\hsize,bb=206 -1 404 792,clip=,angle=90]{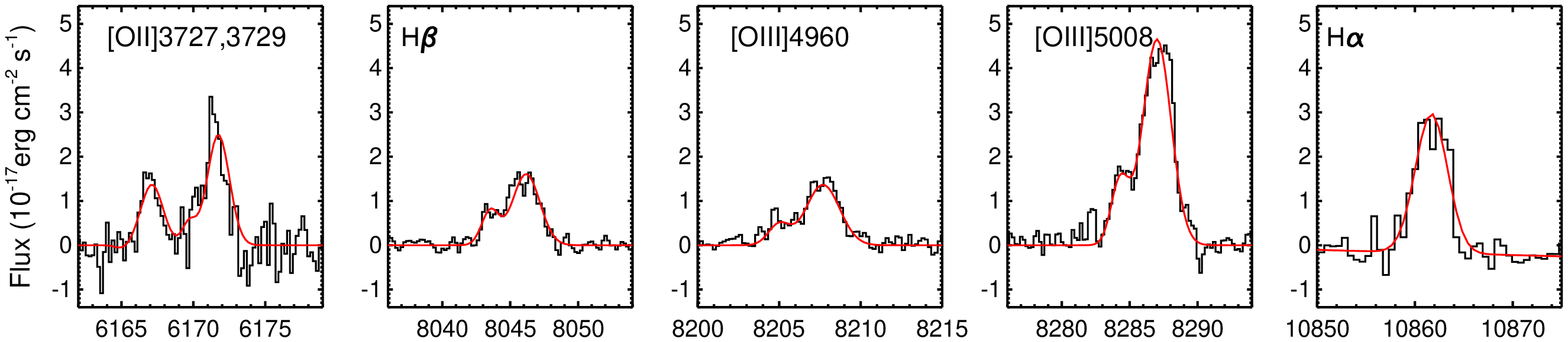}
	\includegraphics[width=0.25\hsize,bb=206 -1 404 792,clip=,angle=90]{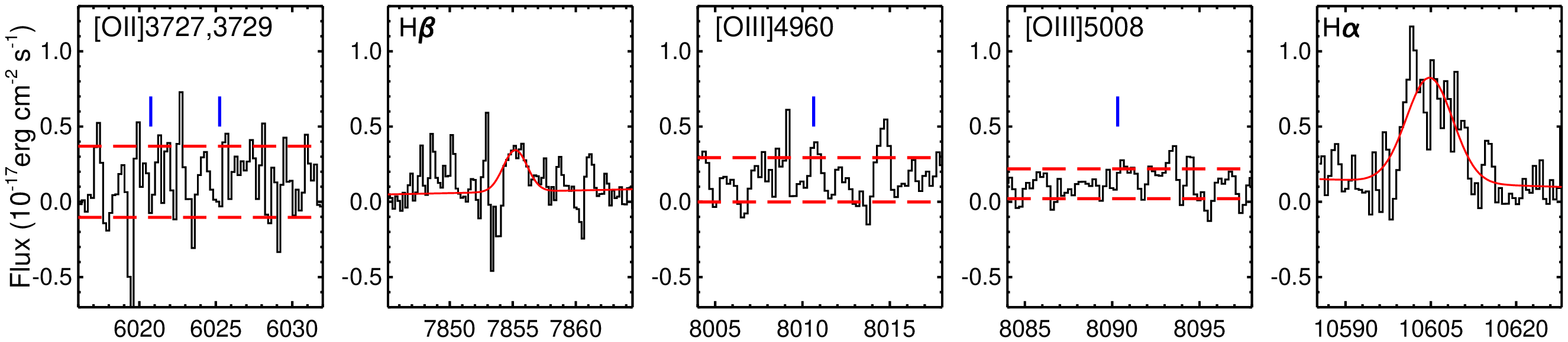}
	\includegraphics[width=0.25\hsize,bb=206 -1 404 792,clip=,angle=90]{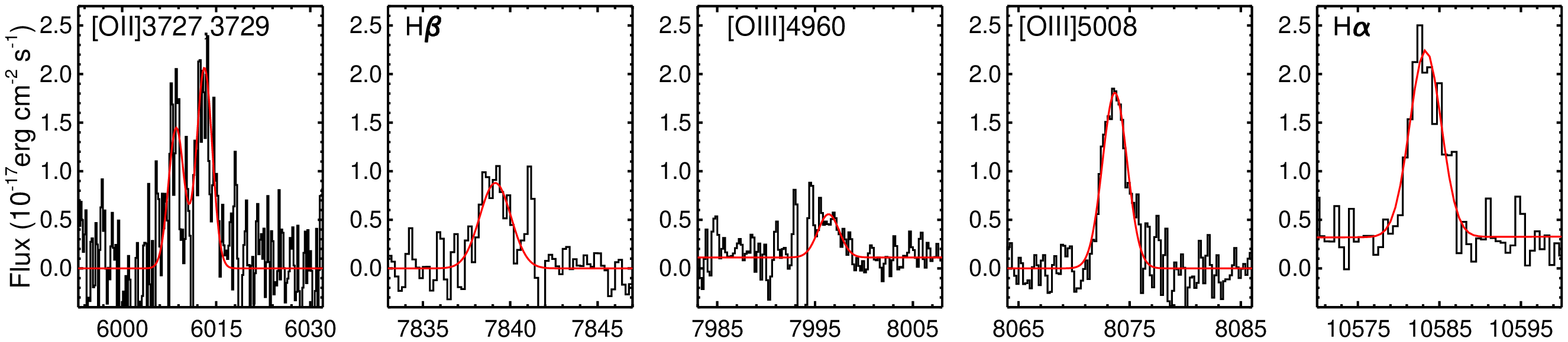}	
	\includegraphics[width=0.25\hsize,bb=206 -1 404 792,clip=,angle=90]{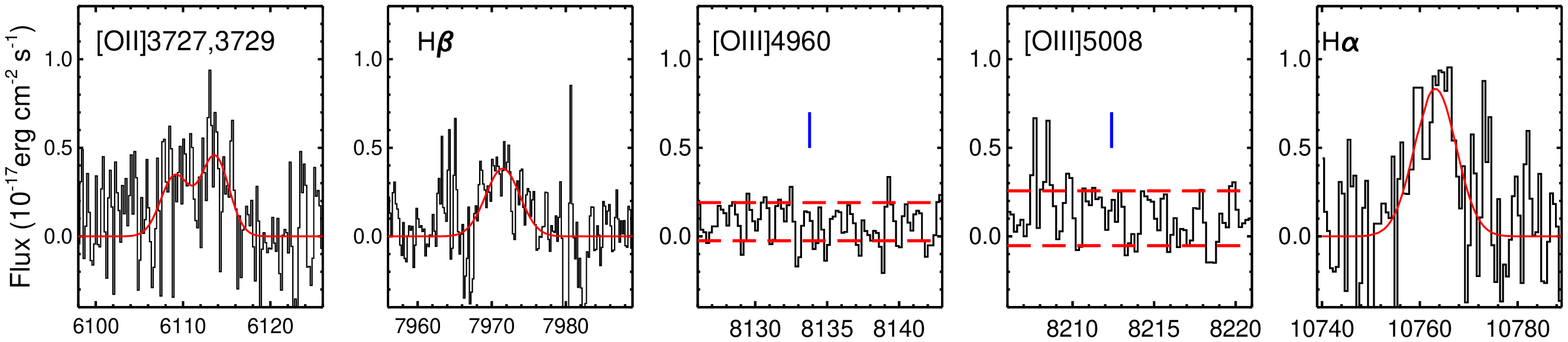}
	\includegraphics[width=0.25\hsize,bb=206 -1 404 792,clip=,angle=90]{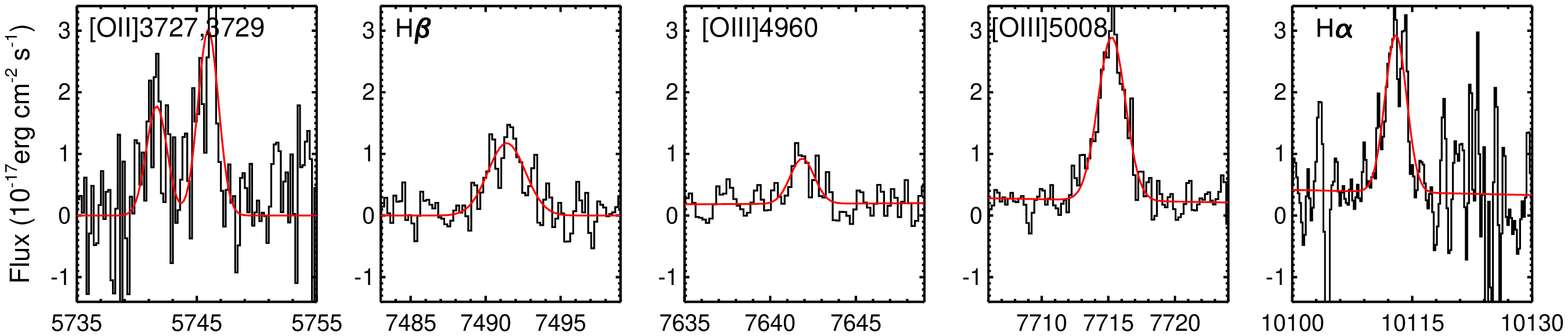}
	}
	\caption{The emission lines from the DLA-galaxies detected by X-Shooter (black histogram) and the best fit Gaussian models (red continuous lines).  Standard deviation of the flux ($\pm1\sigma$) for non-detected emission lines are presented with long-dashed lines. The vertical lines in panels with non-detected emission mark the expected position of the expected line. The QSO fields and DLA-galaxies are marked in top-left of each panel.}
\begin{picture}(400,400)(0,0)
\put( 160,445){\bf Observed wavelength (\AA)}
\end{picture}
\begin{picture}(0,0)(0,0)
    \put( -420,1070){\bf \large 0958$+$0549 (G2)}  
    \put( -420,944){\bf \large 1012$+$0739 (G2)}
    \put( -420,817){\bf \large 1138$+$0139 (G2)}  
	\put( -420,690){\bf \large 1204$+$0953 (G1)}  
	\put( -420,563){\bf \large 1217$+$0500 (G1)}
\end{picture}
	\label{fig_emit_all}
\end{figure*}

We have detected wavelength offsets of $\sim$ 10 -- 20 \kms\ in the VIS arm of the X-Shooter for some of our exposures.  
Such systematic errors in the wavelength calibration of the X-Shooter have been  
already reported by number of authors \citep[e.g.][]{Noterdaeme12}. 
To quantify and correct such shifts we use the  sky absorption lines obtained from Ultraviolet and Visual 
Echelle Spectrograph on VLT (VLT/UVES) as a reference and compare it with our X-Shooter sky lines.  To do so we use a cross-correlation technique described in \citet{Rahmani13}. We cross-correlate the sky absorption lines in the VIS 
arm of the X-Shooter with that of VLT/UVES  to measure and correct the offset for each spectrum. 

The shortcomings of the X-Shooter flux calibration has been reported in the literature  \citep[][]{Fynbo10,Schonebeck14,Japelj15}. Some of the potential problems include different slit widths and sky condition between the object and the standard star, slitloss and atmospheric dispersion.  
For example \citet[][]{Fynbo10} found that their QSOs fluxes were on average 30\% higher than that of SDSS fiber spectra while the spectral shapes were the same. Hence, they rescaled their fluxes to that of SDSS. However, \citet{Peroux13} \citep[see also][]{Fynbo11} found their X-Shooter fluxes to be consistent with that of SDSS. As QSOs are known to show flux variations over different time scales, such a comparison should be carried out very carefully. 
Our observing strategy is to cover the HST/ACS DLA-galaxy candidates within the X-Shooter slit. As we have multiple candidates in the field of each QSO we have always applied a blind offset from sky position of QSOs. As a result 8 out of 9 QSOs are partially covered by the X-Shooter slit and in one case (J1217$+$0500) we have not covered the QSO. Hence, while nodding we  cover only part of  QSOs fluxes. Therefore,  rescaling the fluxes based on a comparison between X-Shooter and SDSS fluxes might be inappropriate. 
%
The possible systematic errors in our measured fluxes which are dominated by the slitloss, depend on the observing conditions mainly the airmass and the parallactic angle. Hence, a comparison of measured fluxes of the same object between different OBs with different airmass and parallactic angles provides hints towards understanding the possible error budget in the flux measurements. We observe that the fluxes of the emission lines do not change more than 20\% between OBs. Therefore, we expect the error budget introduced by the slitloss will not exceed 20\% of our estimated fluxes. In practice to obtain the flux errors for the emission lines we find that fluctuations of the flux adjacent to the emission lines dominates the Poisson noise. This is mainly due to the residuals from subtracting the QSO spectrum. Thus, in modeling the emission lines we make use of the error estimated from the fluctuations of the flux adjacent to the emission lines. However, to understand the effect of using Poisson noise we re-fitted the emission profiles of some objects using Poisson noise. We find the total fluxes are consistent with our reported values while the estimated errors based on Poisson noise are smaller. Therefore, statistical flux errors based on the fluctuations of the background flux are conservative.
%
%
\section[]{DLA-galaxy candidates in X-Shooter}\label{dla_emission}
In this section we present the results of our search for DLA-galaxies using X-Shooter spectra. For galaxies detected in emission we recover the emission profiles and measure the total fluxes. To do so, we first integrate the flux over the spatial axis to obtain the emission profile. Then we model the resulting profile with a Gaussian function and integrate it to calculate the total flux. We estimate the 1$\sigma$ error of the total flux using the fluctuation of the background level close to the emission line. In cases of non-detection we estimate the 3$\sigma$ upper limits using the fluctuation of the flux at the expected wavelength position of the emission line. We estimate the SFR of galaxies based on their H$\alpha$ flux and using a \citet{Kennicutt98} relation which is corrected to a \citet{Chabrier03} initial mass function. Details of the detected galaxies and non-detections are summarized in Table \ref{xsh_flux}.
\subsection{0218$-$0832}\label{dlagal_0218}
There are four DLA-galaxy candidates in this QSO field. We have chosen two different observing configurations ``1`` and ``2`` to cover all four targets. In configuration ``1`` we detect the nebular emission lines from two galaxies at $z$ = 2.441 and 1.613 (indicated by black arrows in panel 0218$-0832$ of Fig. \ref{fig_FC_1}). These two galaxies are at higher redshifts than the QSO at $z_{\rm em}$ = 1.218 and are not seen in absorption in the quasar spectrum. Interestingly, both of these emitters are at the expected sky positions of G1 and G3 obtained from HST/ACS PSF-subtracted image. The fluxes of different nebular lines from these two high-$z$ galaxies are measured and quoted in Table \ref{xsh_flux}. For G3 where the H$\alpha$ line is also detected we further measure a SFR=4.8$\pm$0.4 M$_\odot$ yr$^{-1}$. The observed wavelength for Ly$\alpha$ line at the redshift of G1, $z = $ 2.441, is at $\lambda_{\rm obs}\sim4183$ \AA\ which falls in the UVB arm of X-Shooter. To measure the possible Ly$\alpha$ flux of this galaxy we masked a $\pm$150 \kms\ region centered on 4183 \AA\ and subtracted the QSO spectrum by applying a SPSF. No Ly$\alpha$ emission was detected for G2 and we put a stringent 3$\sigma$ upper limit of 2.4$\times 10^{-17}$ erg s$^{-1}$ cm$^{-2}$ on the Ly$\alpha$ flux. 

In configuration ``2`` we find a continuum emitting object with no emission line at $\sim2''$ from the QSO in our X-Shooter 2D image. This object is at the expected position of G2. A careful inspection of G2 in HST/ACS image indicates two very closely spaced galaxies, probably merging, surrounded in an extended lower surface brightness halo. We do not find any signature from G4 in configuration ``2`` of our X-Shooter spectrum. 

The 1D spectrum of G2 was extracted from the final reduced 2D frame of X-Shooter following the SPSF method described in section \ref{xsh_obs}. Top panel of Fig. \ref{fig_cont_0218_1515} presents a portion of the spectrum of G2. To increase the SNR we have binned the spectrum in wavelength bins of 100 \kms. By cross-correlating the spectrum of this DLA-galaxy with the spectrum of a typical early type galaxy (overplotted in red) we obtained $z_{\rm G2}=0.5895$. In Fig. \ref{fig_cont_0218_1515} the wavelength scale is converted to the rest frame corresponding to $z_{\rm G2}$. \CaII\ H\&K absorption features and two Hydrogen Balmer absorption lines are marked. However, the SNR of this spectrum is not good enough for modeling the continuum and absorption lines. Integrating the X-Shooter 1D spectrum of G2 having convolved it with SDSS i-band filter we find m$^i_{\rm ab}$ = 21.7$\pm$0.2 where the error was estimated based on 30\% variation of the flux level. Similarly by shifting the spectrum to its rest frame, based on $z_{\rm G2}$, we find the rest frame r-band and i-band absolute magnitudes of  $-20.1\pm0.2$ and $-21.4\pm0.2$ which indicate the host DLA is probably a sub-L$^\star$ galaxy.
\subsection{0957$-$0807}
There are four DLA-galaxy candidates in the field of this QSO at impact parameters $<2.7''$. We have observed this field with two slit configurations to suitably cover all the DLA-galaxy candidates. We note that G1 and G2 are low  surface brightness galaxies that are not visible in Fig. \ref{fig_FC_1} due to the chosen contrast. The brightest candidate in this field is G3 that looks like a bulgeless late-type disk galaxy. In the case of configuration ``1``, where G1 and G2 are covered, neither expected DLA-galaxies nor any other emitters are detected in our X-Shooter spectra. One of the exposures in this field has got a PA = 169$^{\circ}$ which differs from what was requested. We do not detect any emitter in this exposure as well. In case of configuration ``2`` we do not detect emission lines from the expected DLA-galaxies but we do find multiple emission lines at $z = 0.9289$ at the expected position of G3 which is at a higher redshift than  QSO J0957$-$0807 at $z_{\rm em}$ = 0.870. To be complete we have quoted total fluxes of the detected emission lines associated with this galaxy in Table \ref{xsh_flux}.  
\subsection{0958$+$0549}
As shown in Fig. \ref{fig_FC_1} three DLA-galaxy candidates at $b \lesssim3.0''$ are targeted in this QSO field using two slit configurations, though further objects are visible at larger impact parameters. Unfortunately, G3 fell out of slit in configuration ``2`` due to an oversight in PA calculation. Therefore, only G1 and G2 are expected to fall into slit in configuration ``1``. As can be inferred from Fig. \ref{fig_FC_1}, G2 presents an extended nature having brighter knots in both up and down sides. Multiple emission lines are detected in the X-Shooter 2D spectrum at the position of G2 having $z_{\rm em}$ = 0.655 that matches with that of DLA absorption lines. However, we do not detect emission lines at the expected position of G1. We also inspected X-Shooter configuration ``2`` 2D spectrum but found no emitter. 

\begin{figure}
	\centering	
	\vspace*{-2cm}
	\includegraphics[width=1.1\hsize,bb=30 97 580 693,clip=,angle=0]{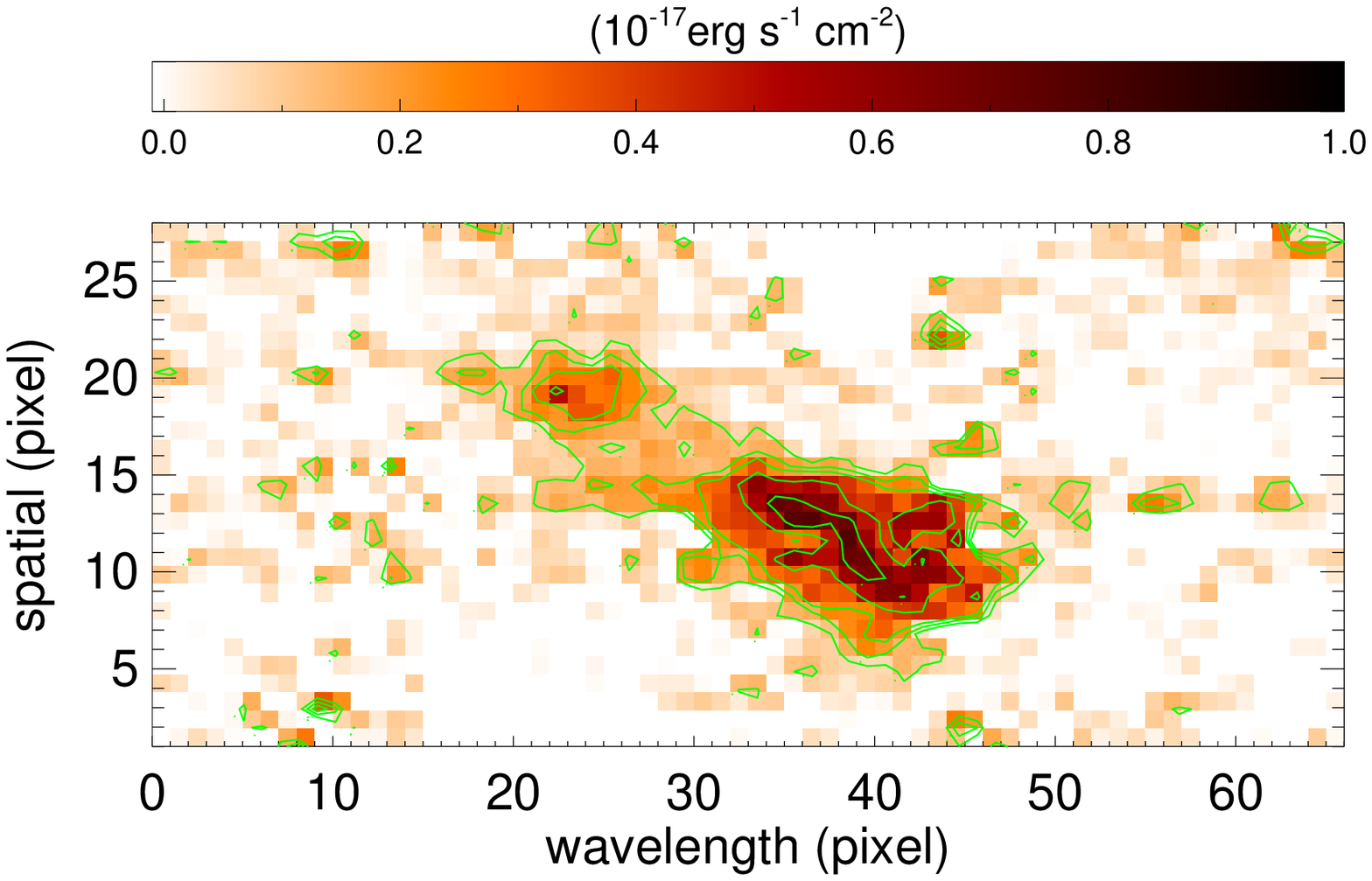}
	\vspace*{-2cm}
	\caption{The [OIII]5008 emission of G2 towards 0958$+$0549. The wavelength range is from 8279.6 to 8292.6 \AA\ and the spatial extent is $\sim4.3''$ ($\sim30$ kpc).}
	\label{fig_0958_2d}
\end{figure}

G2 has the highest  H$\alpha$ flux amongst the detected DLA-galaxies in our sample (see Table \ref{xsh_flux}). Moreover, from the 2D spectrum we notice the extended nature of emission lines from G2. Such an extended feature is more pronounced in [OIII]5008 emission line, as presented in Fig. \ref{fig_0958_2d}, where the spectrum reaches its highest SNR. The [OIII]5008 appears in two cores at a projected separation of $1.4''$ ($\sim$ 10 kpc) that are connected by a faint diffuse emission. The 1D spectra of emission lines also show a strong dominant component along with a weaker secondary one (see Fig. \ref{fig_emit_all}). We model the [\OII], H$\beta$ and [\OIII] emission lines using 2-component Gaussian profiles where the $z$ and $\sigma$ of  each component is same for all the emission lines. Based on such a fit we find a velocity separation of $\sim$ 100 \kms\ between the two cores where the fainter one contributes $\sim$ 20\% to the total flux. In Table \ref{xsh_flux} we have quoted the total flux for [OIII]5008 and other emission lines without decomposing the two emitters. Estimates of the SFR for each core is not possible as all the H$\alpha$ flux is concentrated in a single core. 
%
%
\begin{landscape}
	\begin{table}
		\centering
		\caption{Flux measurements of detected galaxies in our X-Shooter data (in unit of 10$^{-17}$ erg s$^{-1}$ cm$^{-2}$). \textbf{Highlighted} DLA-galaxies are those detected in emission at the expected redshift of DLAs and {\it italics} are those detected only in continuum. The 3$\sigma$ flux upper limits are also given in case of non-detections.}
		\begin{tabular}{lccccccccccccccccccccc}
			\hline
			
			QSO/config &  $z_{\rm abs}$&  DLA-galaxy&  $z_{\rm gal}^{a}$ &  [OII] & [OII] & H$\beta$ & [OIII]& [OIII]  & [NII] &  [NII] & H$\alpha$ & E(B$-$V)& SFR$^{\star}$  & SFR$_{\rm cor}^{\star\star}$\\
			&&(impact parameter$^\dagger$)&&3727&3729&&4960&5008&6549&6585&&&(M$_\odot$ yr$^{-1}$)&(M$_\odot$ yr$^{-1}$)\\
			\hline
			\multirow{ 2}{*}{0218$-$0832$^1$}     &      & G1 ($0.7''$:5.8 kpc)            & 2.441  & $<$6.3 & $<$5.1 & $<$4.5 & 1.8$\pm$0.3 & 5.3$\pm$0.6 & ---       &  ---        & ---  &---& --- & ---\\
			&0.5899$$ & G3 ($3.8''$:32.4 kpc)            & 1.613      & $<$1.8  &  $<$1.8      & $<$9.6    &  3.7$\pm$0.5    & 10.2$\pm$0.7         & $<$1.5      &  $<$2.7        & 6.2$\pm$0.5   &--- & 4.8$\pm$0.4 & 4.8$\pm$0.4    \\ 
			\\
			\multirow{ 2}{*}{0218$-$0832$^2$}      &      & \textit{\textbf{G2 ( 2.2$''$:14.8 kpc)}}            & 0.5895$^{b}$  & \boldmath$<$\textbf{\textit{2.0}} & \boldmath$<$\textbf{\textit{2.1}} & \boldmath$<$\textbf{\textit{1.3}}  & \boldmath$<$\textbf{\textit{2.1}} & \boldmath$<$\textbf{\textit{1.7}} & \boldmath$<$\textbf{\textit{6.4} } &  \boldmath$<$\textbf{\textit{3.9}} & \boldmath$<$\textbf{\textit{2.8}} & ---&\boldmath$<$\textbf{\textit{0.18}} & ---\\
			&0.5899$$ & G4 ($3.3''$:22.1 kpc)            & ---       & $<$1.8 &  $<$2.4      & $<$1.2    &  $<$1.8    & $<$1.5    & $<$3.3       &  $<$2.4       & $<$1.8 & ---   &$<$0.12     & ---      \\ 
			\\
			\multirow{ 2}{*}{0957$+$0807$^1$}      &      & G1 (1.1$''$:8.3 kpc)            & --- & $<$2.1 & $<$2.1 & $<$1.5 & $<$1.8 & $<$2.1 & $<$2.4      &  $<$3.9        & $<$2.7& --- & $<$0.26 & ---\\
			&0.6975 & G2 ($1.0''$:7.6 kpc)            & ---       & $<$2.7  &  $<$2.1      & $<$1.5    &  $<$2.7    & $<$1.5        & $<$3.3     &  $<$3.3        & $<$4.5   & --- & $<$0.43  & ---    \\ 
			\\
			\multirow{ 2}{*}{0957$+$0807$^2$}      &      & G3 (2.3$''$:16.9 kpc)            & 0.9289       & 6.0$\pm$2.3$^c$  &  6.0$\pm$2.3$^c$      & $<$8.3    &  $<$7.0    & $<$2.4        & $<$3.0     &  6.1$\pm$4.0        & 11.8$\pm$2.4  & --- & 2.3$\pm$0.5 & 2.3$\pm$0.5  \\
			&0.6975 &   G4 (2.7$''$:21.3 kpc)            & ---  & $<$2.7 & $<$4.8 & $<$1.5 & $<$3.6 & $<$1.8 & $<$3.6      &  $<$3.0       & $<$3.3& ---& $<$0.32  & ---  \\ 
			\\
			\multirow{ 2}{*}{0958$+$0549$^1$}      &      & G1 (1.1$''$:7.7 kpc)            & ---      & $<$3.3&  $<$3.3      & $<$1.8    &  $<$1.5    & $<$3.3        & $<$10.0     &  $<$8.6        & $<$8.1  & ---& $<$0.67 & ---\\
			&0.6557 &   {\bf G2 (2.9$''$:20.4 kpc) }           & \textbf{0.6547}  & \boldmath$2.7\pm0.8$ & \boldmath$5.5\pm0.8$ & \boldmath$5.3\pm0.3$ & \boldmath$4.3\pm0.4$ & \boldmath$14.5\pm0.4$ & \boldmath$<9.3$      &  \boldmath$<5.4$       & \boldmath$13.0\pm0.8$   & \boldmath$-0.15\pm0.08$& \boldmath$1.07\pm0.07$ & \boldmath$1.07\pm0.07$\\ 
			\\
			\\ 			
			
			\multirow{ 3}{*}{1012$+$0739}          &      & G1 (0.9$''$: 6.3 kpc)              & ---   & $<$1.7    & ---     & $<$2.4&   ---      & $<$1.2        & $<$2.3      &  $<$2.5       & $<$1.7 & ---& $<$0.12    & ---      \\
			&0.6164 & {\bf G2 (4.3$''$:29.8 kpc) }           & \textbf{0.6155}  & \boldmath$<1.8$    &\boldmath$<1.7$     & \boldmath$0.66\pm0.33$&  \boldmath$<1.7$       & \boldmath$<1.2$        & \boldmath$<3.7$      & \boldmath$2.3\pm1.7$ & \boldmath$7.2\pm1.6$& \boldmath$1.28\pm0.53$& \boldmath$0.5\pm0.1$  & \boldmath$6.7\pm1.3$ \\ 
			&      & {\bf G3}(7.2$''$:50.0 kpc)  & 0.6162  & 2.6$\pm$0.8& 3.9$\pm$0.8  & 1.9$\pm$0.4  & 1.3$\pm$0.2 &5.1$\pm$0.3 & $<$3.0       & $<$1.7         &  3.9$\pm$0.5& $-0.33\pm0.24$& 0.28$\pm$0.04   & 0.28$\pm$0.04\\ 
			\\
			\multirow{ 3}{*}{1138$+$0139}          &      & { G1} (1.0$''$: 7.0 kpc)                & ---   & $<$1.5   & $<$1.2      & $<$1.2  &   $<$1.5     & $<$1.2         & $<$2.4       &  $<$2.4       & $<$2.7& --- &$<$0.19    & ---      \\
			&0.6130 &  {\bf G2 (1.8$''$:12.4 kpc)}            & \textbf{0.6122} & \boldmath$4.5\pm1.1$ & \boldmath$6.4\pm1.1$& \boldmath$2.0\pm0.4$ & \boldmath$1.2\pm0.5$ & \boldmath$5.0\pm0.5$ & \boldmath$<2.7$       &  \boldmath$<3.6$  &  \boldmath$9.3\pm1.1$     & \boldmath$0.46\pm0.22$ &  \boldmath$0.65\pm0.08$    &   \boldmath$1.7\pm0.2$   \\ 
			&      & G3 (2.2$''$:15.2 kpc)            & ---    & $<$1.8   &     $<$1.8         & $<$1.2  &   $<$1.5    & $<$1.2        & $<$2.4      &  $<$2.4       & $<$1.8  & --- &$<$0.13  & ---\\ 
			\\
			\multirow{ 3}{*}{1204$+$0953}      &      & {\bf G1 (1.5$''$:10.0 kpc)}             & \textbf{0.6395}  & \boldmath$1.4\pm1.0$     &   \boldmath$1.8\pm1.0$  & \boldmath$2.2\pm1.0$   &    \boldmath$<1.9$   & \boldmath$<2.1$         & \boldmath$<10.3$      &  \boldmath$2.1\pm0.8$& \boldmath$8.9\pm3.5$ & \boldmath$0.33\pm0.58$&  \boldmath$0.69\pm0.27$ &\boldmath$1.3\pm0.5$ \\
			&0.6401 & G2 (2.7$''$:23.0 kpc)             & 1.2776     & 1.9$\pm$0.2  &  3.1$\pm$0.2  & 2.1$\pm$1.0 & $<$6.3     & 11.9$\pm$1.5        & $<$0.9       & $<$0.9        & 8.5$\pm$0.3   & 0.33$\pm$0.46 &  3.7$\pm$0.1  & 7.2$\pm$0.3\\ 
			&      & G3 (3.8$''$:27.2 kpc)             & ---      & $<$2.4   &    $<$2.4   & $<$1.5          & $<$1.2   &   $<$1.5   & $<$3.3       &  $<$2.4        & $<$3.6 & ---& $<$0.28  & ---         \\ 
			\\
			\multirow{ 2}{*}{1217$+$0500}      &      & {\bf G1 (2.3$''$:14.7 kpc)}            & \textbf{0.5405} & \boldmath$3.8\pm1.4$ & \boldmath$6.5\pm1.4$ & \boldmath$3.7\pm0.6$ & \boldmath$1.3\pm0.4$ & \boldmath$6.9\pm0.5$ & \boldmath$<7.3$       &  \boldmath$<9.3$        & \boldmath$8.6\pm3.0$ & \boldmath$-0.21\pm0.37$& \boldmath$0.44\pm0.15$   & \boldmath$0.44\pm0.15$ \\
			&0.5413 & G2 (2.4$''$:15.1 kpc)             & ---   & $<$7.5  &  $<$7.2      & $<$2.1    &  $<$2.4    & $<$3.0         & $<$7.8      &  $<$9.9        & $<$7.2    & --- & $<$0.37 & ---     \\ 
			\\
			1357$+$0525$^1$ &0.6327 & G1 (1.8$''$:12.7 kpc)            & ---     & $<$3.3 &  $<$3.0      & $<$2.4    &  $<$2.1    & $<$1.2        & $<$3.0     &  $<$4.5       & $<$5.7 & --- & $<$0.43   & ---     \\
			\\ 			
			1357$+$0525$^2$ &0.6327 & G2 (2.6$''$:17.7 kpc)            & ---      & $<$3.3 &  $<$3.0      & $<$1.5    &  $<$1.8    & $<$1.8        & $<$3.6     &  $<$7.2       & $<$5.7  & --- & $<$0.43   & ---    \\
			\\ 			
			\multirow{ 2}{*}{1515$+$0410}      &      &  G1 (0.9$''$:5.9 kpc)            & --- & $<$1.8 & $<$1.8 & $<$1.2 & $<$0.9 & $<$1.2 & $<$6.9      &  $<$3.6       & $<$3.6  & ---& $<$0.20& ---\\
			&0.5592 & \textbf{\textit{G2 (1.6$''$:10.2 kpc)}}               & 0.5580$^{b}$   & \boldmath$<$\textbf{\textit{2.6}} &  \boldmath$<$\textit{\textbf{2.5}}      & \boldmath$<$\textit{\textbf{1.6}}    &  \boldmath$<$\textit{\textbf{1.6}}    & \boldmath$<$\textit{\textbf{1.2}}        & \boldmath$<$\textit{\textbf{12.3}}     &  \boldmath$<$\textit{\textbf{5.2}}       & \boldmath$<$\textit{\textbf{6.6}}       & --- & \boldmath$<$\textit{\textbf{0.37 }} & ---\\ 
			\\			 
			\hline
		\end{tabular}%
		\begin{flushleft}
			 $^\dagger$ Impact parameters (in kpc) are calculated based on redshifts of the detected galaxies. For non-detections we assume the galaxy candidates are at the redshift of the DLAs. \\
			$^{\star}$ SFR estimated from H$\alpha$ flux assuming a \citet{Kennicutt98} conversion corrected to a \citet{Chabrier03} initial mass function.\\
			$^{\star\star}$ SFR corrected for dust extinction in the host galaxy.\\ 
			$^{a}$ average redshift of those emission lines in the VIS arm\\
			$^{b}$ absorption redshifts of the DLA-galaxies obtained from cross-correlation analyses (see Section \ref{dlagal_0218} and \ref{dlagal_1515})\\
			$^c$ Separate fluxes of the two [OII] lines can not be measured. \\ 
		\end{flushleft}
		\label{xsh_flux}
	\end{table}
\end{landscape}
\subsection{1012$+$0739}
The HST/ACS image of this QSO presents a crowd with several possible DLA-galaxy candidates. However, a single X-Shooter slit configuration was chosen to cover the two closest objects, called G1 and G2, at impact parameters of respectively $0.9''$ and $4.3''$. Given the selected configuration we incidentally cover a third object in the slit at an impact parameter of $7.2''$ which is dubbed G3. No emission is detected at the expected position of G1 in our X-Shooter spectrum. G2 is detected in H$\alpha$ emission with high significance at a redshift matching $z_{\rm abs}$ along with a faint continuum. The continuum emission from G2 is visible in the entire range from $\lambda_{rest} \sim$ 4300 -- 12600 \AA\ but does not have a high enough SNR to enable the analysis of galaxy absorption lines. The H$\alpha$ emission from G2 with $\sigma=146$ \kms\ is the broadest line associated to DLA-galaxies in our sample. Apart from a tentative H$\beta$ and an [\NII]$\lambda$6585 none of other emission lines are significantly detected from G2. Interestingly, multiple emission lines are detected from G3 at $z=0.6162$ that also matches with the redshift of the sub-DLA. A velocity separation of $\sim140$ \kms\ persists between G2 and G3. The impact parameter of G2 is $\sim$ 20 kpc smaller than that of G3. Hence, we will consider G2 as the galaxy responsible for the sub-DLA in this sightline.

Due to a high clustering possibility of DLA-galaxies in this field we inspected the SDSS database of this QSO sightline. Two objects presented on the left side of HST/ACS image are detected as a faint elongated one having m$_i \sim$ 21 mag with a photometric redshift $z_{\rm phot}$ = 0.58$\pm$0.21 at $\sim7.7''$ distance to the QSO. There further exists a faint, m$_i \sim$ 21 mag, object at $12''$ south-east of this QSO (not covered in HST/ACS image) with $z_{\rm phot}$ = 0.57$\pm$0.10. Photometric redshifts of both of these faint objects are consistent with $z_{\rm abs}$ of the associated sub-DLA. Therefore, a clustering of faint galaxies may exists in this field where further deep spectroscopic observations will be required to confirm or infirm this hypothesis.
%
%
\begin{figure*}
	\centering	
	\vspace*{-0.6cm}
	\includegraphics[width=0.95\hsize,bb=18 17 594 773,clip=,angle=0]{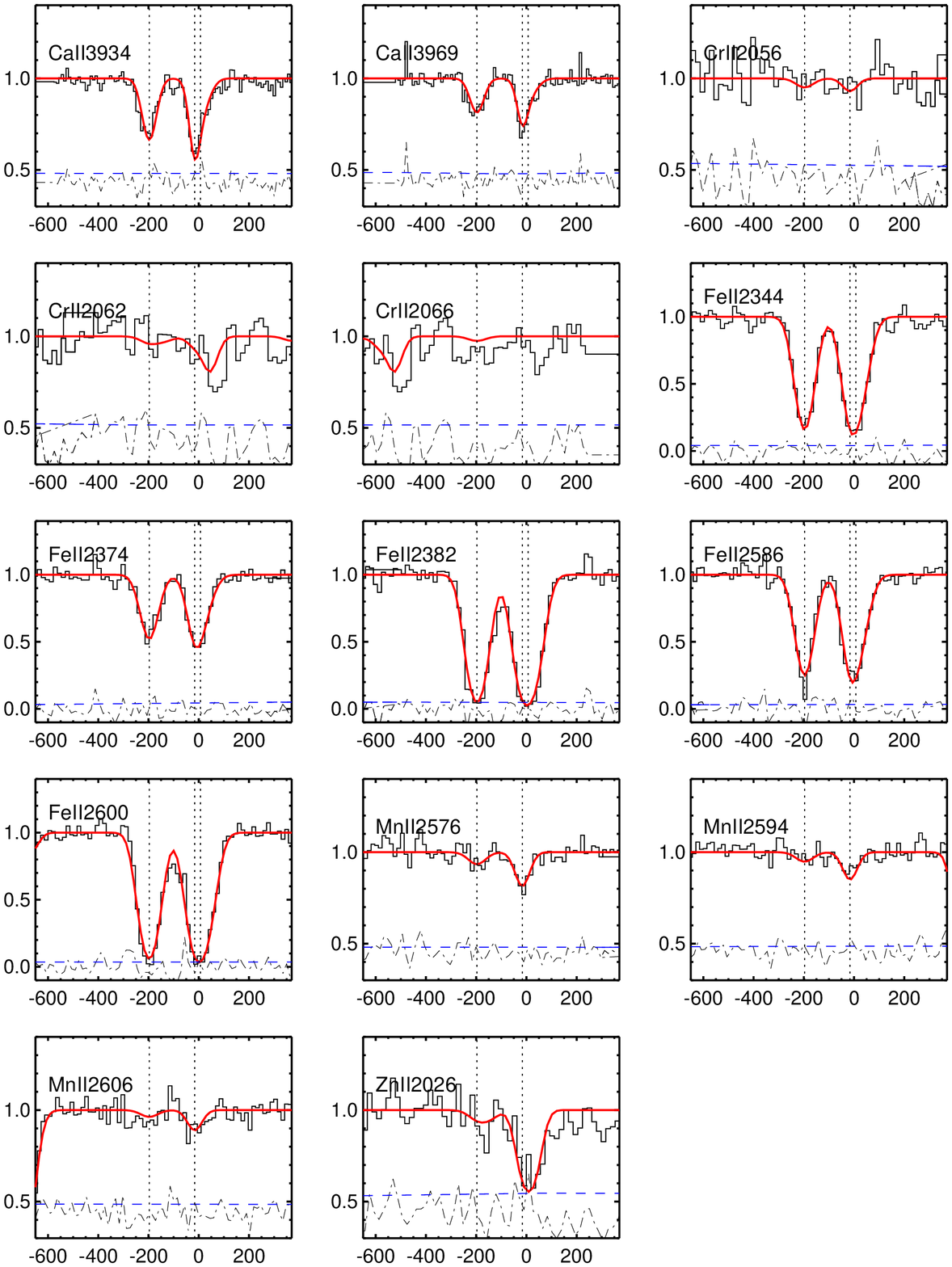}
	\vskip 0.5cm	
	\caption{ Metal absorption lines towards J0218$-$0832. The zero velocity is set at the redshift of the DLA-galaxy, $z_{\rm G2}=0.5895$. Vertical dashed lines denote the velocity of different absorbing components. The histogram and continuous lines present the observed normalized QSO flux and the best obtained Voigt profile model, respectively. The error spectrum is shown as dashed line and the residual of the fit ([data]-[model]) as the dashed-dotted line (both shifted with a constant offset of 0.45 for panels not covering zero flux on y-axis). Similar figures for the remaining absorbers in our sample are presented in Appendix \ref{app_abs_profile}.  }
	\begin{picture}(0,0)(0,0)
	\put( -260,360){\rotatebox{90}{\large Normalized Flux}}
	\put( -30,59){\large Velocity (\kms)}
	\end{picture}
	\label{fig_0218_abs}
\end{figure*}
\subsection{1138$+$0139}
There are three DLA-galaxy candidates towards this QSO sightline in the HST/ACS image at impact parameters $<2.2''$. All three candidates show some UV fluxes, though at different levels. A single X-Shooter configuration covers these 3 candidates within the slit and partially the QSO. However, the nebular emission lines are detected only from the candidate at an impact parameter of $1.8''$ (12 kpc). Inspecting the rest of the X-Shooter 2D spectrum we do not find any other emitter. Therefore, we estimate the emission line fluxes for G2 while for G1 and G3 we calculate 3$\sigma$ upper limits from non-detections. Fig. \ref{fig_emit_all} presents detected emission profiles of the nebular lines. The [OIII]4960 line is  contaminated by the residuals from a sky emission line since it is close to a sky emission line. Hence, to model the emission profile we fix the $\sigma$ and the $z_{\rm em}$ to that obtained from [OIII]5008. The total measured flux of the [OIII]4960 is significant at a level lower than 3$\sigma$ (see Table \ref{xsh_flux}). 

This QSO sightline has another strong \MgII\ absorber at $z = $ 0.7658 \citep{Quider11}. Using QSO spectrum obtained from our X-Shooter observations we estimate the rest frame equivalent widths (EW) of \MgII$\lambda$2796 and \FeII$\lambda$2600 to be respectively 0.93$\pm$0.01 \AA\ and 0.34$\pm$0.01 \AA. \MgII\ systems at such EWs have high chances to be associated with sub-DLAs or DLAs \citep{Rao00,Rao06}. However, none of the expected emission lines at $z = $ 0.7658 are detected in our X-Shooter spectrum.   
\subsection{1204$+$0953}
The PSF subtracted HST/ACS image of this field presents several objects at impact parameters of $<6''$. We have chosen a single slit configuration for this field in which we cover three of such galaxy candidates and partially the QSO. Multiple nebular emission lines are detected in X-Shooter 2D spectrum at impact parameters of $1.5''$ and $2.7''$. The redshift of the former, $z$ = 0.6395, matches that of the DLA but the latter,  $z$ = 1.2776, is $\sim$ 450 \kms\ higher than that of the emission redshift of the QSO. Furthermore, no absorption signature is detected to be associated with this high redshift galaxy. Hence, most likely this galaxy is in the background with respect to the QSO. As no emission line is detected at impact parameter of $3.8''$ as expected from G3 we only place upper limits for fluxes of nebular emission lines assuming the same redshift as that of the DLA. We also notice the detection of [NII]6585 emission line from the DLA-galaxy in this sightline at more than 2$\sigma$ significance. 

Two further \MgII\ absorption systems are detected towards this QSO at $z \sim$ 0.830 and 0.837 with rest-frame EW$^{2796}$ of respectively 0.14$\pm$0.01 \AA\ and 0.46$\pm$0.03 \AA. The expected N(\HI) for such \MgII\ absorbers are in the range of LLS but lower than that of sub-DLAs \citep{Rao06}. We inspected the 2D spectrum searching for nebular emission lines at such redshifts. A weak emission line was detected at an observed wavelength of $\lambda \sim$  9146.7 \AA\ and an impact parameter of $7.8''$. In case of assuming it is [OIII]5008 such a line is at $z = 0.8263$ or $\sim$ 635 \kms\ with respect to the weaker \MgII\ absorber with the impact parameter of 60 kpc. However, further deeper observations are required to conform this hypothesis. 
\subsection{1217$+$0500}
Two DLA-galaxy candidates are present in the HST/ACS image of this field at impact parameters of $2.3''$ and $2.4''$. We are able to cover both of them in the slit using a single X-Shooter slit configuration at the cost of not covering the QSO. Hence, in the case of this sightline we neither have details of the metal absorption lines nor measure the absorption metallicity. We detect nebular emission lines from one of the candidates at an impact parameter of $2.3''$ at the redshift of the DLA. Apart from the detected emission lines tabulated in Table \ref{xsh_flux} we also detect with a high significance [SIII]9533 from this DLA-galaxy. No emission line is detected for G2 at the DLA's redshift or other redshifts. Hence, upper limits were estimated for the fluxes of this galaxy assuming it lies at the redshift of this sub-DLA. 
\subsection{1357$+$0525}\label{1357_em}
%
%
%
Our original analysis of this prism field suggested that there were two DLA-galaxy candidates towards this QSO at impact parameters of $1.8''$ and $2.6''$. G1 looked like a disk galaxy with a bright bulge in its center. Two different X-Shooter configurations were used to observe G1 and G2 in two different slits (Table \ref{log_XSH}). However, no significant continuum or emission lines were detected from any objects at these locations. Hence, we have used the observations at the positions of G1 and G2 to estimate upper limits from non-detections of nebular emission lines.

Our inability to detect apparently bright optical flux at the position of G1 is inconsistent with our original interpretation of this field, namely that, in addition to providing a dispersed image of a blue QSO, the prism image also provides a deep image of mildly-dispersed optical light from galaxies in the field (e.g., Section 2.1). Instead, as the X-Shooter observation showed, there is no optical light at the position of G1. The above two observational results indicate that the light seen in the prism image is UV light. Only if we assume that the object is a Lyman-$\alpha$ emitter (LAE) do the two above observations make sense; it explains why an object was not seen with X-Shooter, despite the fact that it appeared to be the brightest galaxy candidate in our sample. This leads to the interpretation that the significant flux seen in the prism image is actually due to a Ly$\alpha$ emitter (LAE) at the DLA redshift ($z=0.6327$). If so, the LAE has a total extended Ly$\alpha$ emission flux of $\sim 7.6 \times 10^{-14}$ ergs cm$^{-2}$ s$^{-1}$, which translates to a total Ly$\alpha$ luminosity of $\sim 1.3 \times 10^{44}$ ergs s$^{-1}$. It has an impact parameter of $\sim 0.81''$ ($\sim 5.5$ kpc) relative to the QSO sightline and an oval-like extent of $\sim 22$ kpc by $\sim 6$ kpc. It has a bright central region $\sim 1$ kpc across which contains $\sim 24$\% of the total measured flux. These characteristics are similar to the many high-redshift LAE found in the literature \citep[e.g.][and references therein]{Bielby16}, but this example is rather remarkable since it has a low redshift and is identified with a DLA absorber. However, as we do not have a spectrum of this object we do not count it as a confirmed detection.
\subsection{1515$+$0410}\label{dlagal_1515}
There are two DLA-galaxy candidates at impact parameters of $0.9''$ and $1.6''$ in the field of this QSO. A high level of UV flux was detected at the core of G2 in the PSF subtracted HST/ACS image. In a single X-Shooter configuration we are able to cover both galaxy candidates and partially the QSO in the slit. No emitting object were detected at the expected position of G1. However, at the expected position of G2, a continuum flux (with no emission line) is detected over an observed wavelength range from 5500 \AA\ to 21000 \AA. The extracted 1D spectrum of G2 represents several absorption features (see middle and bottom panels of Fig. \ref{fig_cont_0218_1515}). We cross-correlate the spectrum of G2 with a typical early type galaxy to obtain $z_{\rm G2}=0.5580$. The absorption features and the break match those of a typical early type galaxy redshifted at the position of the DLA. The \CaII\ H\&K, H$\delta$ and \MgI$\lambda$5170 in particular are detected and labeled in Fig. \ref{fig_cont_0218_1515}. However, the SNR of this spectrum is not enough for modeling the continuum and absorption lines. Therefore, we confirm G2 to be the galaxy responsible for the sub-DLA absorption in this QSO sightline at an impact parameter of 10 kpc. In the absence of H$\alpha$ we put a stringent 3$\sigma$ upper limit of 0.4 M$_\odot$ yr$^{-1}$ on the SFR of this quenched galaxy. Integrating the X-Shooter spectrum of this galaxy when convolved with the SDSS i-band filter we find a m$^i_{\rm AB}$ = 21.4$\pm$0.2 mag. We further estimate the SDSS ``r`` and ``i`` band absolute magnitudes to be respectively $-$19.2$\pm$0.2 and $-$19.5$\pm$0.2 mag which given the uncertainties are similar to those estimated for G2 towards 0218$-$0832. This finding is in line with their rest frame flux levels (see Fig. \ref{fig_cont_0218_1515}). 
\section{Absorption line properties}
\begin{table*}
	\centering
		 \begin{adjustwidth}{-1cm}{}
	\caption{Extracted properties of DLAs in our sample from absorption line analysis. Column densities are in units of log10(cm$^{-2}$). 3$\sigma$ upper limits are estimated for ions that are not detected. \ZnII\ abundances are used to estimate the [Zn/H] with respect to solar \citep[Zn$_\odot=4.56\pm0.05$,][]{Asplund09}. $\Delta v_{90}$ is estimated from the \FeII\ absorption line.}
	\begin{tabular}{cccccccccccccccccccccc}
		\hline
		QSO & $z_{\rm abs}$ & N(\HI)& N(\CaII)& N(\CrII)& N(\FeII)  & N(\MnII) & N(\TiII) & N(\ZnII) & [Zn/H]& Z$_{\rm cor}^{\star}$&$\Delta~v_{90}$ \\
		& &&&&&&&&&&(\kms)\\
		\hline
		J0218& 0.5899 &  20.84$\pm$0.12 &  13.02$\pm$0.02& 13.02$\pm$0.42 & 14.84$\pm$0.02 & 12.95$\pm$0.04 & $<$12.50 & 13.21$\pm$0.08 &$-0.19\pm$0.15&$-0.20\pm0.31$&261 \\
		J0957& 0.6975 &  21.44$\pm$0.05 &  12.49$\pm$0.05& 13.75$\pm$0.19 & 15.65$\pm$0.17 & 13.64$\pm$0.09 & 13.09$\pm$0.05 & 12.71$\pm$0.31 & $-1.30\pm$ 0.31 &$-1.42\pm0.28$& 25\\ 
		J0958& 0.6557 &  20.54$\pm$0.15 &  12.18$\pm$0.04& 12.97$\pm$0.10 & 14.78$\pm$0.03 & 12.58$\pm$0.04 & 12.07$\pm$0.06 & 11.96$\pm$0.10 & $-1.14\pm$0.18 &$-1.33\pm0.23$& 112\\
		J1012& 0.6164 &  20.18$\pm$0.12 &  $<$12.53  & $<$12.83 & 14.50$\pm$0.10 & $<$11.95 & $<$11.67 & $<$12.16 & $<-$0.58&---&177\\
		J1138& 0.6130 &  21.25$\pm$0.10 &  12.84$\pm$0.03& 13.97$\pm$0.05 & 15.64$\pm$0.04 & 13.59$\pm$0.03 & 12.99$\pm$0.05 & 13.23$\pm$0.06 & $-0.58\pm$0.11 &$-0.78\pm0.16$& 105\\           
		J1204& 0.6401 &  21.04$\pm$0.08 &  12.60$\pm$0.05& $<$13.58 & 15.16$\pm$0.06 & 13.03$\pm$0.04 & 12.37$\pm$0.07 & $<$13.24 & $<-$0.36 &$-0.72\pm0.16$& 163\\           
		J1357& 0.6327 &  20.81$\pm$0.05 &  12.57$\pm$0.04& 13.47$\pm$0.10 & 14.83$\pm$0.02 & 12.89$\pm$0.04 & $<$12.34 & 12.75$\pm$0.09 & $-0.62\pm$0.11&$-0.86\pm0.13$&150\\ 
		J1515&  0.5592&  20.20$\pm$0.19 &  $<$12.26 & $<$13.68  & 13.92$\pm$0.13 & $<$12.19 & $<$11.57 & $<$13.40 &$<+$0.64 &---& 302\\
		\hline 
	\end{tabular}%
		\label{xsh_abs}
	\begin{flushleft}
		$^{\star}$ The estimated metallicities based on \citet{Jenkins09} analysis.
	\end{flushleft}
\end{adjustwidth}	
\end{table*}			
For the analysis of the metal absorption lines associated with a DLA we first model the QSO continuum. We find the continuum by fitting low order polynomials to spectral chunks that are free from absorption lines. However, working with narrow metal absorption lines, the uncertainty of different estimated parameters are dominated by flux noise rather by the continuum placement. We  model the UV resonance absorption lines associated with the DLAs with Voigt profiles by using \textsc{VPFIT}\footnote{http://www.ast.cam.ac.uk/~rfc/vpfit.html} v10.0. \textsc{VPFIT} is a code developed for fitting multiple Voigt profiles (convolved with instrumental broadening) to spectroscopic data while minimizing the $\chi^{2}$. In our multicomponent Voigt profile analysis we have assumed that each absorption component has the same redshift for all ions in the fit. While the thermal broadening will be different for different elements, we assume the same broadening parameter for all ions since the turbulent motion may dominate. From the best fit model to metal absorption lines we calculate the total column density of different elements and their associated uncertainties. For the non-detected elements we estimate the 3$\sigma$ upper limit of the column density from the flux fluctuation at the expected position of those absorption lines. In table \ref{xsh_abs} we list the total column density for each detected element and the upper limits for non-detections. Furthermore, for absorption systems with multiple detected metal lines we follow the prescription proposed by \citet{Jenkins09} to estimate the dust free metallicity \citep[see][for example]{Quiret16}

Kinematics of DLAs and sub-DLAs have been studied using the velocity width of low ionization metal absorption lines  \citep{Nestor03,Peroux03,Ledoux06a,Neeleman13,Som15,Quiret16}. Such a velocity width is usually represented by $\Delta v_{90}$ defined as the velocity range over which the integrated optical depth ($\tau_{tot}~ = ~ \int_{-\infty}^{+\infty} \tau(v) ~ dv$) reaches from 5\% to 95\% of its total value \citep{Prochaska97,Ledoux06a}. A mild correlation holds between $\Delta v_{90}$ and metallicity of DLA and sub-DLAs spread over more than two orders of magnitude in metallicities \citep{Ledoux06a,Quiret16}. Dark matter simulations have shown that there is a direct relation between velocity width of DLAs and the virial mass of dark matter halos hosting DLA galaxies, though with large scatters \citep{Pontzen08,Bird15}. Furthermore, stellar mass of galaxies present a direct correlation with their metallicities \citep{Tremonti04,Savaglio05,Erb06,Maiolino08}. Therefore, the metallicity-velocity relation is sometimes interpreted as a consequence of the mass-metallicity relation of the DLA host galaxies \citep{Moller13}. However more complex processes such as gas accretion and galactic winds are also known to explain the width of typical absorption line profiles \citep[][]{Cen12}. To estimate $\Delta v_{90}$ we implement the method described in \citet{Quiret16} which uses the best fit Voigt profile model instead of the observed spectrum. Integrating over the noise and blend free Voigt profile obtained from simultaneous fit of several absorption lines in this approach leads to a more robust estimate of $\Delta v_{90}$. \FeII\ absorption lines are the only lines observed to be associated with all DLAs and sub-DLAs in our sample. Hence, we make use of \FeII\ absorption to measure $\Delta v_{90}$ while knowing it is prone to dust depletion in the ISM \citep{Vladilo98,Ledoux02a,Khare04,Wolfe05,Jenkins09}. We notice that the measured $\Delta v_{90}$ in our approach will be same for all transitions of a given ion. The $\Delta v_{90}$ values derived for individual systems are presented in the last column of Table \ref{xsh_abs}.

In the following we describe the absorption line properties of the quasar absorbers in our sample. For QSO fields with detected DLA-galaxies we set the zero velocity at the systemic redshift of the galaxy. For non-detections (J0957 and J1357) we set the zero velocity at the redshift of absorbing components with highest N(\FeII) obtained from the best Voigt profile fit. 
\subsection{0218$-$0832  [$z_{\rm abs}$=0.590, N(\HI)=20.84] }
Fig. \ref{fig_0218_abs} presents the metal absorption lines towards J0218$-$0832. The absorption profiles present two main clumps. The redder clump contains a sub-component at $v\sim+20$ \kms\ seen in \CaII, where the (VIS) spectrum has a factor of $\sim$ 1.7 higher resolution compared to the other arms. We tried to incorporate this component in our Voigt profile fit by initiating the fit with three Voigt profile components. However, apart from \CaII\ and \FeII, the remaining lines can be modeled only with two components corresponding to the two clumps. We further modeled all the absorption profiles, including \CaII, with only two components and noticed the total column density of the ions remain unchanged in such a case. The \ZnII\ abundance is [Zn/H]=$-0.19\pm0.15$ which is higher than average population of DLAs at similar redshifts. A ratio of [\FeII/\ZnII] = -1.32 indicates that \FeII\ is highly  depleted onto dust grains. 
Estimated value of $\Delta v_{90}$ = 261 \kms\ for this DLA is the second largest amongst the systems in our sample. A large  $\Delta v_{90}$ was also expected for this DLA given its high metallicity and the metallicity-$\Delta v_{90}$ correlation discussed earlier.  
\subsection{0957$-$0807 [$z_{\rm abs}$=0.697, N(\HI)=21.44]}
The DLA towards this QSO has a N(\HI) = 21.44$\pm$0.05 which is the highest column density of neutral hydrogen amongst the DLAs in our sample. Fig. \ref{fig_0957_abs} presents a set of metal absorption lines associated with this DLA. The best model is composed of two Voigt profile components  in which the bluer one is not detected in \CrII, \ZnII\ and \MnII. In this system we have [Zn/H]=$-1.30 \pm0.31$ which is consistent with the average DLA metallicity at similar redshifts. The [Fe/Zn]=0.0$\pm$0.35 is consistent with zero for this DLA which can be interpreted as the lack of dust depletion of metals associated with this neutral gas. A $\Delta v_{90}$ = 25 \kms\ for this system, is the smallest measured amongst systems in our sample. Moreover, a few DLAs have already been reported having $\Delta v_{90}$ smaller than 25 \kms. It is worth noting that much smaller metallicity is expected for this DLA based on a metallicity-$\Delta v_{90}$ relation \citep{Quiret16}. From Fig. \ref{fig_0957_abs} it appears that \FeII\ absorption profiles have quite broader lines compared to our estimated $\Delta v_{90}$ for this system (25 \kms). However, our Voigt profile analysis indicates the existence of a \FeII\ component with N(\FeII)=15.6$\pm$0.2 that produces such saturated profiles broaden due to the intermediate resolution of the X-Shooter. To confirm the high column density of \FeII\ we carried out a curve of growth analysis from which we obtained N(\FeII)=15.3$\pm$0.1 which is marginally consistent with the result of our Voigt profile analysis and confirms the existence of a saturated \FeII\ component. Hence, the intermediate resolution of X-Shooter can hamper the $\Delta v_{90}$ measurements if using the apparent optical depth method \citep[also see][]{Srianand16}. It is worth noting that we obtained a $\Delta v_{90}$ = 24 \kms\ based on \MnII\ lines which is well in agreement with that obtained from \FeII\ lines.
\subsection{0958$+$0549 [$z_{\rm abs}$=0.655, N(\HI)=20.54]}
Fig. \ref{fig_0958_abs} presents the absorption profiles of the metal lines along with the best Voigt profile fit associated with the DLA towards QSO J0958$+$0549. Apart from \ZnII, the remaining absorption lines can be modeled with two components. There exists a further weaker component ($\sim2$ dex less column density of metals) which is seen in \MgII\ and the strongest \FeII\ transitions at $\sim218$ \kms. This DLA has a [Zn/H]=$-1.14\pm0.18$ which is in line with the average population of DLAs at this redshift. We further find an [Fe/Zn] = $-0.1\pm0.1$ which indicates iron is not depleted much onto dust. Redshift of the absorbing galaxy in this field matches well that of the stronger absorption component associated with this DLA. There exists a weaker emitter in this field at $\sim - 100$ \kms\ which is blueward of the rest of Voigt profile components. 
\subsection{1012$+$0739 [$z_{\rm abs}$=0.616, N(\HI)=20.18]}
Fig. \ref{fig_1012_abs} presents the absorption profiles and the best obtained Voigt fit model for the \FeII\ lines associated with the sub-DLA towards QSO J1012$+$0739. As we have only partially covered this QSO in the slit the resulting spectrum has a low SNR over most of the UVB arm where the UV absorption lines fall. Hence, we can model only the strongest \FeII\  transitions  and obtain upper limits for the rest of metal absorption lines. The upper limit on the \ZnII\ column density translates into a metallicity upper limit of [Zn/H] $<-0.58$. We further obtain a Fe abundance of [Fe/H]=$-1.16 \pm0.17$ that can be also considered as a lower limit for the metallicity. As shown in Fig. \ref{fig_1012_abs}, velocity of the bluest absorption component matches well that of the detected emission lines in this system. In other words, most of the neutral gas is seen redwards of the nebular emission lines for this galaxy. There is another emitter (G3) in this field at $z = 0.6154$ (or at $v\sim +140$ \kms) though at a larger impact parameter of $\sim$ 50 kpc. The redshift of the G3 matches well the reddest absorbing component shown as dashed-dotted line in Fig. \ref{fig_1012_abs}. 
\subsection{1138$+$0139 [$z_{\rm abs}$=0.613, N(\HI)=21.25]}
Fig. \ref{fig_1138_abs} presents the absorption lines associated with the DLA towards QSO J1138$+$0139. Having a high SNR spectrum of this QSO we can obtain accurate Voigt profile model for the metal absorption lines. The best fit Voigt profile model is obtained with three components. However, not all the components are detected in all metal absorption lines. All absorbing components fall redward of  detected emission lines where the closest component is at $+12$ \kms. This DLA has a metallicity of [Zn/H]=$-0.58\pm0.11$ which is larger than the average metallicity of DLAs at similar redshift. From our best fit we obtain a [Fe/Zn] = $-0.5\pm0.14$ that shows a mild dust depletion of Fe in this system. The estimated $\Delta v_{90}$ = 105 \kms\ is quite smaller than the DLAs at similar metallicities.

We searched for the possible DLA associated with the \MgII\ system in this sightline at $z=0.7658$ in our HST/ACS data. However, the quality of our HST/ACS spectrum close to this Lyman-$\alpha$ line is not good enough for constraining the N(\HI) of this system. Therefore, to estimate the effect of this {\it possible} DLA on the N(\HI) measurement of the DLA at $z=0.6130$ we generated 6 DLA absorption profiles at $z=0.7658$ having $\log$ N(\HI) = 20.30 -- 21.80 convolved and noise added to have similar resolution and SNR to that of HST/ACS. We then fitted the DLA profile at $z=0.6130$ to measure N(\HI). We notice that for all cases we recover an N(\HI) consistent with $21.25\pm0.10$. As a result the N(\HI) of the low $z$ system is not overestimated due the existence of the {\it possible} DLA at $z=0.7658$.
\subsection{1204$+$0953 [$z_{\rm abs}$=0.639, N(\HI)=21.04]}
Fig. \ref{fig_1204_abs} presents the metal absorption lines associated with the DLA towards QSO J1204$+$0953. The X-Shooter spectrum of this QSO has a poor SNR close to the \ZnII\ features. As a result, we can only put upper limit for the \ZnII\ column density, N(\ZnII) $<$ 13.24 corresponding to an upper limit of [Zn/H] $<-0.36$ for the metallicity. However, the best obtained Voigt profile fit involves three components based on which we obtain [Fe/H] = $-1.38\pm0.10$. 
The systematic redshift of the DLA-galaxy in this field is $\sim 30$ \kms\ redwards of the strongest absorbing component and bluewards of the other two weaker components.  
\subsection{1357$+$0525 [$z_{\rm abs}$=0.632, N(\HI)=20.81]}
Fig. \ref{fig_1357_abs} presents the absorption lines associated with the DLA towards QSO J1357$+$0525. We measure a metallicity of [Zn/H]=$-0.62\pm0.11$ for this DLA. A high depletion of \FeII\ is derived in this system given [Fe/H] = $-1.48 \pm0.05$. 
\subsection{1515$+$0410 [$z_{\rm abs}$=0.558, N(\HI)=20.20]}
As presented in Fig. \ref{fig_1515_abs} even the strongest \FeII\ transitions are hardly detected in X-Shooter spectrum of this QSO. We can only measure the column density of \FeII\ associated with this DLA (log [N(\FeII) cm$^{-2}$]= 13.92$\pm$0.13) and place upper limits for the remaining elements. The upper limit on \ZnII\ column density (log [N(\ZnII) cm$^{-2}$] $<$ 13.40) translates into an upper limit of [Zn/H] $< +0.64$ for the metallicity of gas associated to this DLA. However, for \FeII\ we find a relative abundance of [Fe/H]=$-1.78\pm0.23$. Taking into account typical Fe depletion in sub-DLAs indicates that most likely this system does not have a high metallicity. 
%
%
\begin{table*}
	\centering
	\caption{Properties of the detected DLA and sub-DLA host galaxies in this study and those from the literature. The two galaxies detected in continuum are in italic.}
	\begin{tabular}{lccccccccccccccccccccc}
		\hline
		QSO (1) & $z_{\rm gal}$ (2)& impact parameter (3) & SFR$^a$ (4)& Z$_{\rm em}$ (5) &N(\HI) (6) &Z$_{\rm abs}$ (7)&  $\Delta v_{90}$  (8) & References (9)\\
		\hline
J0005$+$0524& 0.8518 & 26  & $6.5^{\rm F}$ & --- & 19.08 $\pm$0.04&--- &--- & [15,36]\\
J0021$+$0043& 0.9419 &  85 & 3.6$\pm$2.2$^A$& $<+$0.4&19.38$\pm$0.15&$+$0.57$\pm$0.23&126& [1]\\ 
J0051$+$0041& 0.7397 &  24 &  5.3$^C$                 & ---  & 20.4$\pm$0.1   & ---                  &   ---& [2]\\    %
J0100$+$0211& 0.6120 &   8  & 0.25$\pm$0.01$^C$& ---  & 20.0$\pm$0.1  & $+$0.08$\pm$0.01&   ---& [3,4,5]\\
J0120$+$2133 & 0.5763&  7    & --- & --- & $19.15\pm0.07$& --- & ---& [15,40]\\ 
J0139$-$0824& 2.677   &  13 &  ---                    & ---  & 20.70$\pm$0.15 & $-$1.15$\pm$0.15  &  110& [6,7] \\ 
J0154$+$0448& 0.16    &  18 & 0.50$\pm$0.15$^B$ & +0.33$\pm$0.14  &19.70$\pm$0.17    & ---    &     ---& [8,9]\\            
{\it \textbf{J0218$-$0832}}& ---   & \textit{\textbf{15}} & \textit{\textbf{\boldmath$<$0.18$^A$}}      &  ---                     &\textit{\textbf{20.84\boldmath$\pm$0.12} }&  \textit{\textbf{\boldmath$-0.20\pm0.32$ }} &  \textit{\textbf{261}}& \textit{\textbf{this work}}\\
J0238$+$1636& 0.5253 &  7   & 1.24$\pm$0.01$^B$&  $-$0.3$\pm$0.2 &21.70$\pm$0.09 &$-$0.6$\pm$0.4& ---& [4]\\ 
J0304$-$2211& 1.0095 &  25 & 1.8$\pm$0.6$^A$&$+$0.04$\pm$0.2 & 20.36$\pm$0.11  & $-$0.51$\pm$0.12 &   ---& [1,10]\\  
J0310$+$0055& 3.115  & 28 & 0.97$\pm$0.13$^E$ & --- & 20.05$\pm$0.05 & ---& ---& [31]\\
J0338$-$0005& 2.223    &    4  & ---          & ---  & 21.05$\pm$0.05   & $-$1.25$\pm$0.10 &   ---& [7]\\    
J0441$-$4313& 0.1010  &    7  & 0.14$\pm$0.01$^{A\dagger}$& $+$0.50$\pm$0.16 & 19.63$\pm$0.15 & $+$0.10$\pm$0.15&  275& [4,5,11]\\ 
J0452$-$1640& 1.007   &16 & 3.5$\pm$1.0$^A$   & $-$0.46$\pm$0.20   & 20.98$\pm$0.07  &$-$0.96$\pm$0.10&---& [12]\\
J0456$+$0400 & 0.8596&  5   & --- & --- & $20.67\pm0.03$& --- & ---& [15,40] \\
J0501$-$0159& 2.040   &   3 & ---            & ---  &  21.65$\pm$0.09  & $-$1.2$\pm$0.1  &    84 & [6,7] \\  
J0530$-$2503& 2.811   &     9 & ---           & ---  &  21.35$\pm$0.07   & $-$0.91$\pm$0.07&  304 & [7,13]\\ 
J0741$+$3112& 0.2222&  20 & $<$0.002$^C$  & ---  &  20.90$\pm$0.08 &$-$0.70$\pm$0.17 &  ---& [4,14,39]\\  
J0813$+$4813& 0.4374 &    8 & 1.09$\pm$0.02$^B$ &   $>+$0.4  & 20.80$\pm$0.2  & ---& --- & [4] \\ 
J0830$+$2410& 0.5263 &  38 & 0.35$\pm$0.01$^B$ & $>+$0.1  &20.30$\pm$0.05  & $-$0.49$\pm$0.30&---& [4,15]\\
J0853$+$4349& 0.1635 &  25 & ---          & --- & 19.81$\pm$0.04  & ---& --- & [16,17]\\ 
J0918$+$1636& 2.413 &  $< 2$ & ---  & --- & 21.26$\pm$0.06 & $-$0.60$\pm$0.2&  350 & [35]\\ 
J0918$+$1636& 2.583 &  16 & 22$\pm$7.0$^{A\dagger}$  & $+$0.04$\pm$ 0.2 & 20.96$\pm$0.05 & $-$0.12$\pm$0.05&  295 & [18,35]\\ 
J0956$+$4734& 3.40   &     3 & ---        & --- & 21.15$\pm$0.15  & $-$1.8$\pm$0.3& --- & [7]\\  
{\bf J0958$+$0549}& \textbf{0.6546} & \textbf{20} & \textbf{1.07\boldmath$\pm$0.07$^{A\dagger}$} & \textbf{\boldmath$-0.81\pm$0.06} &\textbf{20.54\boldmath$\pm$0.15} &  \textbf{\boldmath$-1.33\pm0.23$} &  \textbf{112}&  \textbf{this work} \\
J1009$-$0026& 0.8864&  39  & 2.9$\pm$1.0$^A$ &  $+$0.04$\pm$0.8 & 19.48$\pm$0.05 &  $+$0.25$\pm$0.06 & 94& [1,33]\\
{\bf J1012$+$0739$^{\star}$}& \textbf{0.6154 }& \textbf{30} & \textbf{\boldmath 6.7$\pm$1.3$^{A\dagger}$} & \textbf{\boldmath $-0.07\pm0.19$} &\textbf{\boldmath 20.18$\pm$0.12} &  \textbf{\boldmath $<-$0.58}            &  \textbf{177}&  \textbf{this work} \\
J1130$-$1449& 0.3132 & 18 & --- & --- & 21.71$\pm$0.08 & --- &---  &[15,36$^{\star\star}$]\\
J1135$-$0010& 2.21    &   1  & 25$\pm$6$^A$ &  --- & 22.10$\pm$0.05  & $-$1.1$\pm$0.1& 180 & [19]\\   
J1137$+$3907&0.7185 &  14 & ---         & --- & 21.10$\pm$0.1  & $-$0.3$\pm$0.15& --- & ---\\ 
{\bf J1138$+$0139}& \textbf{0.6126} & \textbf{12} & \textbf{\boldmath 1.7$\pm$0.2$^{A\dagger}$ }& \textbf{\boldmath $-0.41\pm0.18$} &\textbf{\boldmath 21.25$\pm$0.10} &  \textbf{\boldmath $-0.78\pm0.16$ }&  \textbf{105}& \textbf{this work} \\           
{\bf J1204$+$0953}& \textbf{0.6390} & \textbf{10} & \textbf{\boldmath1.3$\pm$0.5$^{A\dagger}$} & \textbf{\boldmath$-0.15\pm0.14$}            &\textbf{\boldmath21.04$\pm$0.08} &  \textbf{\boldmath$-0.72\pm0.16$}             &  \textbf{163}&  \textbf{this work} \\           
J1211$+$1030& 0.3922 &  37& 1.02$\pm$0.01$^B$ & --- & 19.46$\pm$0.08  & $+$0.04$\pm$0.2& --- & [5,15]\\
{\bf J1217$+$0500}& \textbf{0.5405 }& \textbf{15} & \textbf{\boldmath0.44$\pm$0.15$^{A\dagger}$} & \textbf{\boldmath$-0.80\pm0.18$} &\textbf{\boldmath20.00$\pm$0.17} &  ---                       & ---& \textbf{ this work }\\           
J1436$-$0051&  0.7390& 50  & 48$^C$         &   $+$0.33$\pm$0.30 & 20.08$\pm$0.11 & $-$0.05$\pm$0.12 & ---& [8,32] \\  
J1422$-$0001&  0.9096& 12  & $4.7\pm2.0^{A\dagger}$   &   $<-0.06$ & 20.4$\pm$0.4 & $-$0.4$\pm$0.4 & ---& [43] \\  
{\it \textbf{J1515$+$0410}}& ---  & \textit{\textbf{10}} & \textit{\boldmath$<$\textbf{0.37}$^A$}            & ---                       & \textit{\boldmath\textbf{20.20$\pm$0.19}} &  \textit{\boldmath$<+$\textbf{0.64}}          &  \textit{\textbf{302}}& \textit{\textbf{this work}} \\
J1544$+$5912& 0.0102   &   1   & 0.006$^B$ & $-$0.4$\pm$0.30 & 20.35$\pm$0.01 & $-$0.50$\pm$0.33& --- & [20,34]\\
J1624$+$2345& 0.6561 & 100 & --- &  --- & 20.36$\pm$0.08 & --- & ---& [15,21,41]\\
J1624$+$2345& 0.8915 & 28 & --- &  --- & 19.23$\pm$0.03 & $>-$0.40 & ---& [15,21]\\ 
J2059$-$0528 & 2.210   & $<6.3$ & $>0.40^{\rm E}$ & --- & $21.00\pm0.05$ &  $-0.91\pm0.06$ & --- & [37,38]\\
J2131$-$1207& 0.4299   &   48 &  2.26$^B$   & --- & 19.18$\pm$0.03  & ---& --- & [15,22] \\ 
J2142$-$4420& 2.3796 &   182 & ---         & --- & 19.7$\pm$0.1   & $< -$1.1& --- & [23] \\    
J2208$-$1944& 1.9220 &  13   & 9.5$\pm$1.2$^D$  & --- & 20.65$\pm$0.07  & $-$0.38$\pm$0.07& 136 & [13,24]\\ 
J2222$-$0946& 2.354    &  6.3    &  17.1$\pm$5.1$^A$ & $-0.30\pm0.13$ & 20.65$\pm$0.05  & $-$0.49$\pm$0.05& 185&[12,25,26,44]\\  
\hline
	\end{tabular}%
	\begin{flushleft}
		(2) systematic redshift of the DLA-galaxy\\
		(3) impact parameter in kpc\\
		(4) SFR in M$_\odot$ yr$^{-1}$ obtained from H$\alpha$, A; H$\beta$, B; \OII, C; \OIII, D; Ly$\alpha$, E; UV continuum F; and corrected for dust extinction, $\dagger$\\
		(5) emission metallicity with respect to solar metallicity. \citet{Kobulnicky99} and \citet{Pettini04} calibrations are used to derive Z$_{\rm em}$ for all DLA-galaxies.\\
		(6) \HI\ column density in log N(\HI) [cm$^{-2}$]\\
		(7) absorption metallicity with respect to solar\\
		(8) $\Delta v_{90}$ in km s$^{-1}$ \\
		(9) References: [1] \citet{Peroux16}, [2] \citet{Lacy03}, [3] \citet{Pettini00}, [4] \citet{Chen05}, [5] \citet{Peroux11a}, [6] \citet{Wolfe08}, [7] \citet{Krogager12}, [8] \citet{Rao00}, [9] \citet{Christensen05}, [10] \citet{Peroux11b}, [11] \citet{Som15}, [12] \citet{Peroux12}, [13] \citet{Ledoux06a}, [14] \citet{Kulkarni05}, [15]\citet{Rao06}, [16] \citet{Lanzetta95}, [17] \citet{Lanzetta97}, [18] \citet{Fynbo11}, [19] \citet{Noterdaeme12}, [20] \citet{Schulte-Ladbeck05}, [21] \citet{Steidel97}, [22] \citet{Bergeron86}, [23] \citet{Francis04_c}, [24] \citet{Weatherley05}, [25] \citet{Fynbo10}, [26] \citet{Neeleman13}, [27] \citet{Djorgovski96}, [28] \citet{Christensen04}, [29] \citet{Lu97}, [30] \citet{Bouche13}, [31] \citet{Kashikawa14}, [32] \citet{Straka16}, [33] \citet{Meiring09}, [34] \citet{Schulte-Ladbeck04}, [35] \citet{Fynbo13}, [36] \citet{Kacprzak10}, [37] \citet{Hartoog15}, [38] \citet{Herbert-Fort06}, [39] \citet{Turnshek01}, [40] \citet{Churchill96}, [41] \citet{Steidel97}, [42] \citet{Srianand16}, [43] \citet{Bouche16}, [44] \citet{Krogager13}\\
		$^{\star}$ The second galaxy at $z_{\rm em}$ =0.6162 and b = 50 kpc has a Z$_{\rm em}$=$-0.6\pm0.2$ with SFR=0.28$\pm$0.04.\\
		$^{\star\star}$ \citet{Kacprzak10} found 3 brighter galaxies matching \zabs\ at $b\sim46$ -- 90 kpc. \\
		$^{\star\star\star}$ \citet{Steidel97} commented this galaxy is farther than to be responsible for this DLA.  
	\end{flushleft}	
	\label{tab_res}
\end{table*}			
\begin{table*}
	\addtocounter{table}{-1}
	\centering
	\caption{continued}
	\begin{tabular}{lccccccccccccccccccccc}
		\hline
		QSO (1) & $z_{\rm gal}$ (2)& impact parameter (3) & SFR$^a$ (4)& Z$_{\rm em}$ (5) &N(\HI) (6) &Z$_{\rm abs}$ (7)&  $\Delta v_{90}$  (8) & References (9)\\
		\hline
		J2236$+$1326& 3.1530  & 18 & 19$\pm$10$^E$& --- & 20  & $-$0.8$\pm$0.24 &---&[24,27,28,29]\\ 
		J2247$-$6015& 2.33 & 23 &  36$\pm$10$^{A\dagger}$ & $-$0.6$\pm$20 & 20.67$\pm$0.05  & $-$0.72$\pm$0.05&  289& [7,13,30]\\ 
		J2352$-$0028   & 1.0318  &   12 & 1.3$\pm$0.6$^A$  & $-$0.26$\pm$0.03 & 19.81$\pm$0.14 & $<-$0.51& 164& [1,12,33]\\
		J2358$+$0149  & 2.9791  &   12 & $<2$$^D$  & --- & 21.69$\pm$0.10 &$-1.83\pm0.18$& 168& [42]\\
		J2358$+$0149  & 3.2477  &   $<5$ & 0.2--1.7$^E$  & --- & 21.12$\pm$0.10 &$-0.97\pm0.13$& 149& [42]\\
		\hline
	\end{tabular}%
	\label{tab_res}
\end{table*}		
\section{Results}
\subsection{Detection rate}
We detect DLA-galaxies using X-Shooter in seven out of nine QSO fields that translates to a high success rate of 78\%. Such a high success rate is mainly achieved due to the fact that we had set the slit orientations to cover the galaxy candidates based on prior information we obtained from HST/ACS observation. In five out of seven fields, multiple nebular emission lines are detected from the DLA-galaxies. Only one of these five DLA-galaxies (J1012$+$0739) is also detected in continuum emission. There are two fields (J0958$+$0549 and J1012$+$0739) in which we detect two emitters both at the redshifts of the DLAs. This can be a signature of clustering of DLA-galaxies as already reported in the literature \citep{Chen03,Rao03,Kacprzak10,Rao11}. 

Two out of seven DLA-galaxies are quiescent early-type galaxies detected only in continuum. We confirm their redshifts using several absorption signatures seen in their spectra (see Fig. \ref{fig_cont_0218_1515}) and cross-correlating their spectra with that of a typical early type galaxy. In addition to the detected \CaII\ H\&K doublets along with some Balmer absorption that confirm these galaxies are at the redshift of the DLAs, the DLA-galaxy towards J1515$+0410$ also presents a strong 4000 \AA\ break. These are amongst the first quiescent DLA-galaxies ever discovered \citep[see][for the only detection of such objects before this work]{Chen05}. The quenched SFR observed for these galaxies may be due to the lack of neutral gas reservoirs in these objects. As a result, smaller covering fraction of \HI\ ($f_{cor}(\HI)$) is expected for these sort of objects. Furthermore, recent simulations demonstrate that $f_{cor}(\HI)$ sharply decreases at large impact parameters \citep{Rahmati15,Suresh15}. Therefore, it is less likely for a DLA to be produced where a background QSO passes through large impact parameters from such foreground galaxies. This may be one reason for the small impact parameters of these two DLA-galaxies. Moreover, owing small impact parameters the detection of such foreground galaxies is observationally challenging because of the bright background quasar next to the faint galaxy. While this may explain the few number of detected quiescent DLA-galaxies no stringent conclusion can be drawn given the small size of the sample. 

\begin{figure}
	\centering	
	\includegraphics[width=0.8\hsize,bb=18 18 594 774,clip=,angle=90]{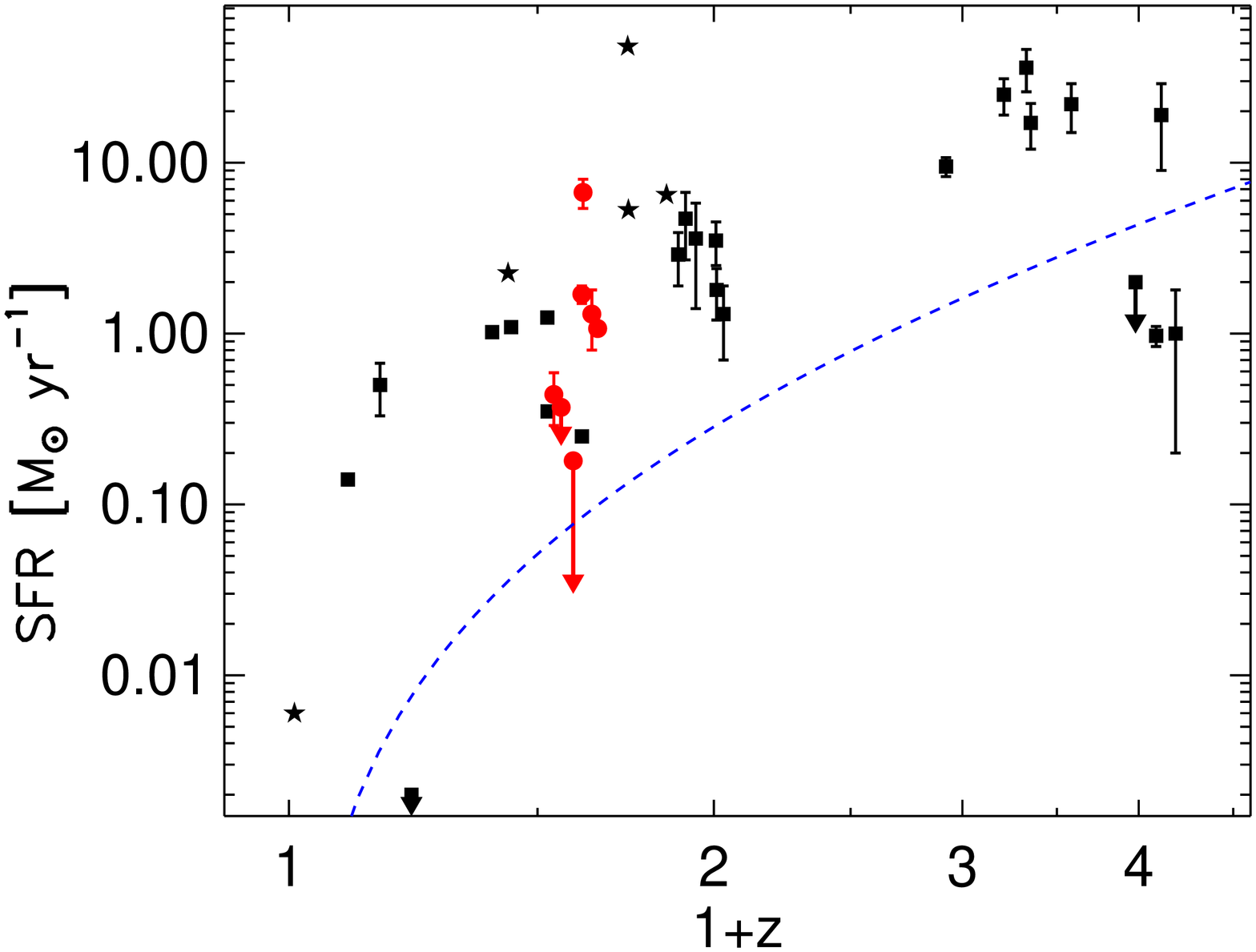}
	\caption{The star formation rate vs redshift for the host galaxies of strong \HI\ absorbers having N(\HI) $> 10^{19}$ cm$^{-2}$. The red filled circles present our measured extinction corrected SFR obtained from the H$\alpha$ flux assuming Chabrier IMF. The black squares are the reported SFR for \HI\ absorber galaxies in the literature (see Table \ref{tab_res}). Stars mark reported measurements without estimated errors. The dashed line is the SFR sensitivity of the SINFONI survey \citep{Peroux12} extrapolated at different redshifts.}
	\label{fig_sfr_halpha}
\end{figure}

\citet{Fumagalli15} could detect DLA-galaxies at $z\sim2.7$, in the rest frame UV, with a success rate of $\sim10$\% (3 out of 33) in an imaging survey. While they are sensitive to SFR $\sim0.1$ M$_\odot$yr$^{-1}$, the dust attenuation of the DLA-galaxies can be one reason for such a low detection rate. \citet{Peroux12} (P12 hereafter) reported success rates of 25\% (4 out of 16) and 10\%  (1 out of 12) in a VLT/SINFONI search for DLA-galaxies at $z\sim$1 and $z\sim$ 2, respectively. 
 The faint nature of DLA-galaxies and the depths of the $z \sim $ 2 observations may partly explain the differences between the success rates in this study and those of P12. However, the flux sensitivity of $z \sim$ 1 sample of P12 (indicated using dashed line in Fig. \ref{fig_sfr_halpha}) is enough for the detection of our DLA-galaxies. The smaller success rate of P12 can be mainly due to two effects: (1) redshift evolution of the SFR of DLA-galaxies from $z\sim1$ to $z\sim0.6$; (2) Quiescent DLA-galaxies with no emission lines will not be detected with SINFONI which is insensitive to continuum light. 
 Excluding the two objects detected only in continuum decreases our success rate to 56\% which is still a factor of $\sim2$ higher than that of P12. 
 There have been suggestions for higher detection probability of host galaxies of (sub-)DLAs having higher metallicities \citep[e.g.][]{Moller02,Fynbo11,Peroux12}. However, the median metallicity of absorbers in this study which is [Zn/H]=$-0.62$ (and $-0.58$ after excluding detected systems in continuum) is not larger than those of P12 ([Zn/H]=$-0.46$). 
 %
 In summary this comparison provides evidences that indicate a redshift evolution of the properties of DLA-galaxies between $z \sim$ 1 and $z \sim$0.6 (5.9 Gyr to  7.9 Gyr after the big bang). 
\begin{figure}
	\centering	
	\vbox{
		\includegraphics[width=0.7\hsize,bb=18 18 594 774,clip=,angle=90]{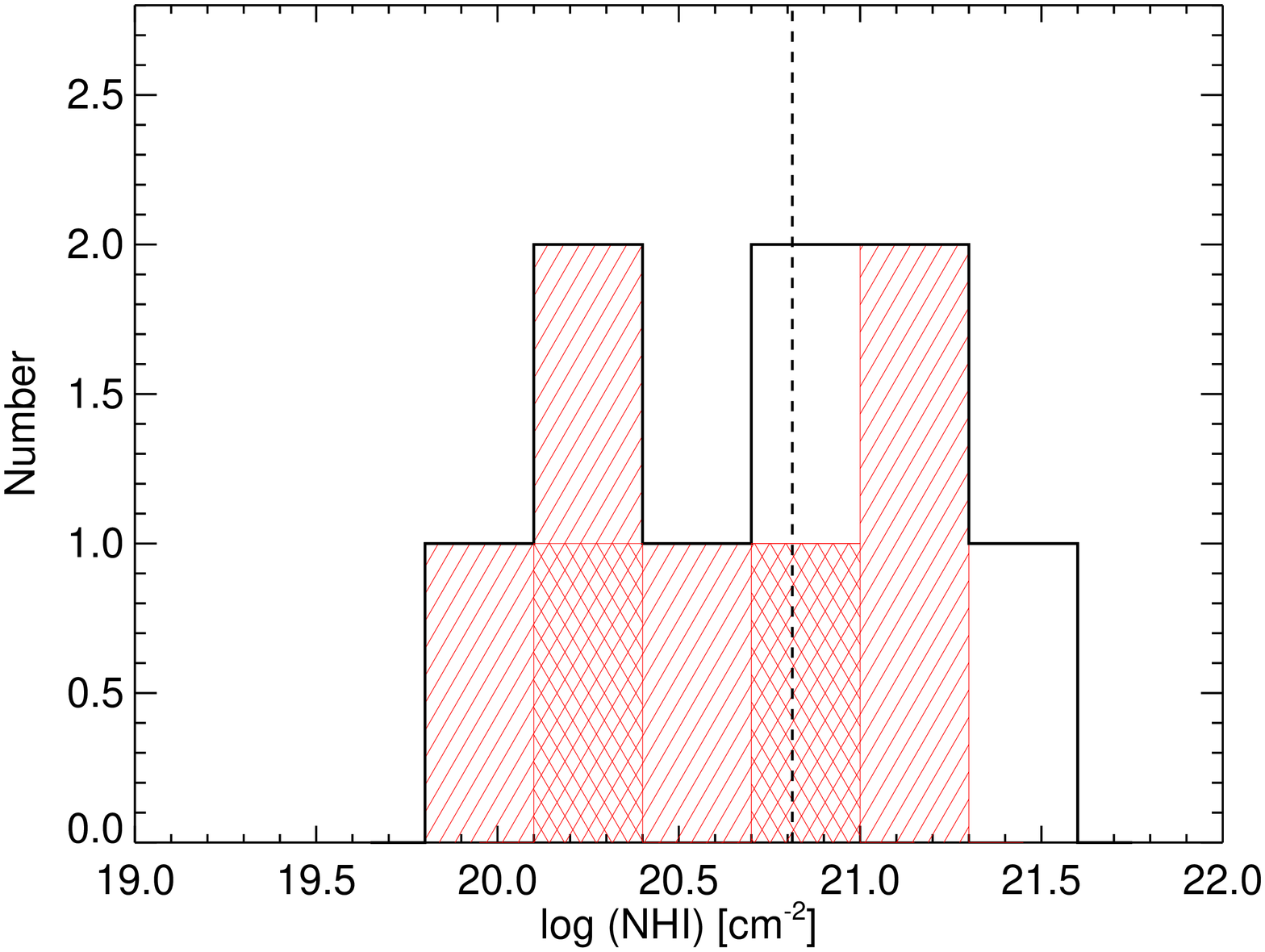}
	}
	\caption{N(\HI) distribution of the DLAs in our sample. Those filled area are associated with detected DLA-galaxies and the hashed ones present the two systems detected in continuum.
	}
	\label{fig_dist_nhi}
\end{figure}

Fig. \ref{fig_dist_nhi} presents the distribution of N(\HI) for absorbers in our sample where detected DLA-galaxies in emission are filled and those detected only in continuum shown with hashed area. The vertical dashed line marks the median of the N(\HI) distribution that is 10$^{20.81}$ cm$^{-2}$. 
The two non-detections of our sample fall in larger column density side of the distribution. There are evidences in favor of occurrence of higher N(\HI) DLAs at lower impact parameters \citep{Moller98a,Wolfe05,Christensen07,Monier09,Rao11,Peroux12,Krogager12,Noterdaeme14,Rahmati15} [also see Fig. \ref{fig_b_nhi} and corresponding discussion]. As a result detection of host galaxies of high-N(\HI) DLAs would be observationally more challenging. This may partly be one reason for the non-detection of two host galaxies but no robust conclusions can be drawn given the small sample size. 
\subsection{SFR and possible AGN contribution}
\begin{figure}
	\centering	
	\hskip -1.2cm
	\includegraphics[width=1\hsize,bb=0 0 576 432,clip=,angle=0]{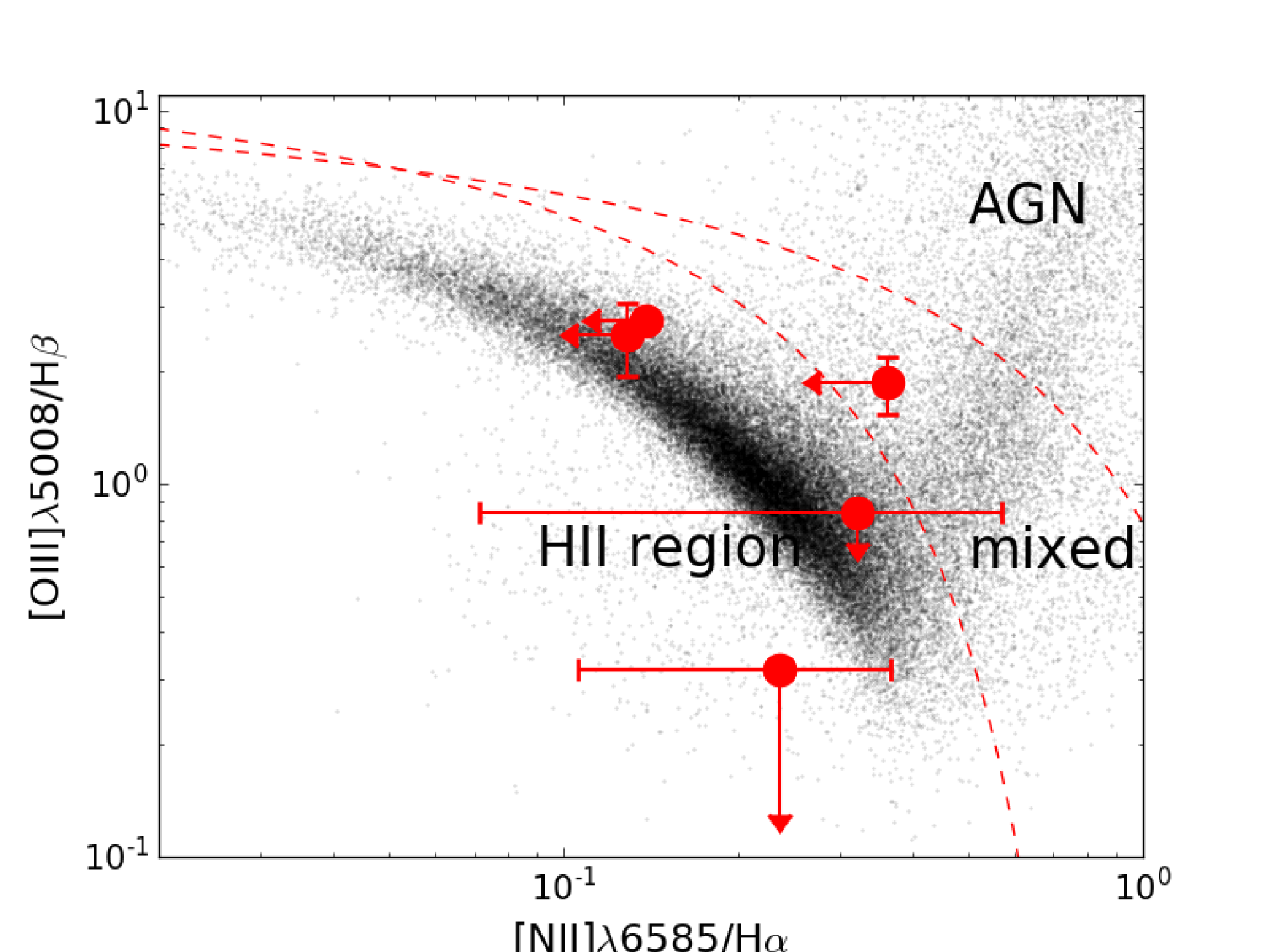}
	\caption{The BPT \citep*{Baldwin81} diagram to distinguish the nature of ionizing sources. Dashed lines are obtained from theoretical calculation \citep{Kewley01,Kauffmann03} and separate the position of AGN dominated from those of star forming regions. Red circles present the position of DLA galaxies from this study where they are dominantly presenting emission lines originating from star forming regions. The black points are SDSS galaxies \citep{Aihara11}.  }
	\label{fig_bpt}
\end{figure}
%
%
We convert the total H$\alpha$ fluxes of galaxies to SFR assuming a \citet{Kennicutt98} conversion corrected to a \citet{Chabrier03} initial mass function. We further utilize the Balmer decrement between H$\alpha$ and H$\beta$ to estimate the dust extinction of galaxies detected in our X-Shooter data. For case B recombination at T=10$^4$ K and $n_e$ = 10$^2$--10$^4$ cm$^{-3}$ the intrinsic value of Balmer decrement is 2.88 \citep{Osterbrock89book} which is typical for \HII\ regions. To estimate the colour excess we adopt a Small Magellanic Cloud-type (SMC-type) extinction law and utilize the following parametrization
\begin{equation}
E(B-V) = \frac{1.086}{k({\rm H\beta})-k({\rm H\alpha})} \ln \left(\frac{\rm H\alpha}{\rm 2.88 H\beta}\right)
\end{equation}
where $k(\rm H\alpha)$ and $k(\rm H\beta)$ are respectively the extinction curve values at $\lambda$ = 6564 \AA\ and 4862 \AA\ taken from \citet{Pei92}. Estimated values of $E(B-V)$ and extinction curves are then used to evaluate the intrinsic flux of H$\alpha$ from which we derive the SFR corrected for dust extinction. The Balmer decrement for DLA-galaxies towards 0958$+$0549 and 1217$+$0500 are found to be respectively 2.5$\pm$0.3 and 2.3$\pm$0.4 that are $<$2.88 (though consistent with 2.88 within 2$\sigma$) which seems not physical. For such cases we retain the dust uncorrected SFRs. Our SFR estimates are presented in the last two columns of Table \ref{xsh_flux}. We have reached SFR sensitivities down to 0.1 M$_\odot$ yr$^{-1}$ (1$\sigma$) which are amongst the most sensitive detection limits to date for DLA-galaxy searches. Fig. \ref{fig_sfr_halpha} presents the SFR vs redshifts for absorbing galaxies in our sample and those spectroscopically confirmed in the literature with N(\HI) $>$ 10$^{19}$ cm$^{-2}$ (see Table \ref{tab_res}). The range of SFR measured for our DLA-galaxies are comparable with those at similar redshifts. There is an overall trend of increasing SFR with redshift that can be partly due to the lower SFR sensitivities at higher redshifts. However, \citet{Kulkarni06} (and also \citet{Gharanfoli07}) reported a consistent redshift evolution of SFR for emitters associated with metal absorbing systems (which may be at lower column densities of N(\HI)). 

The H$\alpha$ to SFR conversion relies on the assumption that the source of ionizing photons are young stellar objects and there is no contribution from the active galactic nuclei (AGNs). 
 We utilize the BPT digram \citep*[][]{Baldwin81} to investigate the possible contribution of AGN to the fluxes of emission lines. 
Fig. \ref{fig_bpt} presents the BPT diagram where abscissa and ordinate are respectively flux ratios of [\OIII]5008/H$\beta$ and [\NII]6853/H$\alpha$. 
Dashed lines obtained from theoretical calculations \citep{Kewley01,Kauffmann03} and broadly define the separation between emission lines originating from star forming regions and AGN activities. A mixed nature of ionizing spectrum is expected from objects occupying the region in between the two dashed lines. Filled circles are DLA-galaxies obtained by our X-Shooter observations where one of the them (1138$+$0139) indicates possible signature of AGN contamination. The black dots present the positions of galaxies and AGNs obtained from SDSS DR8 \citep{Aihara11}. Clearly the majority of our points are well within the star forming region of the BPT diagram. Hence, systematic errors in the estimated SFRs introduced by AGN contamination is thought to be negligible.
\subsection{Emission metallicity}
\begin{figure}
	\centering	
	\includegraphics[width=0.8\hsize,bb=18 18 594 774,clip=,angle=90]{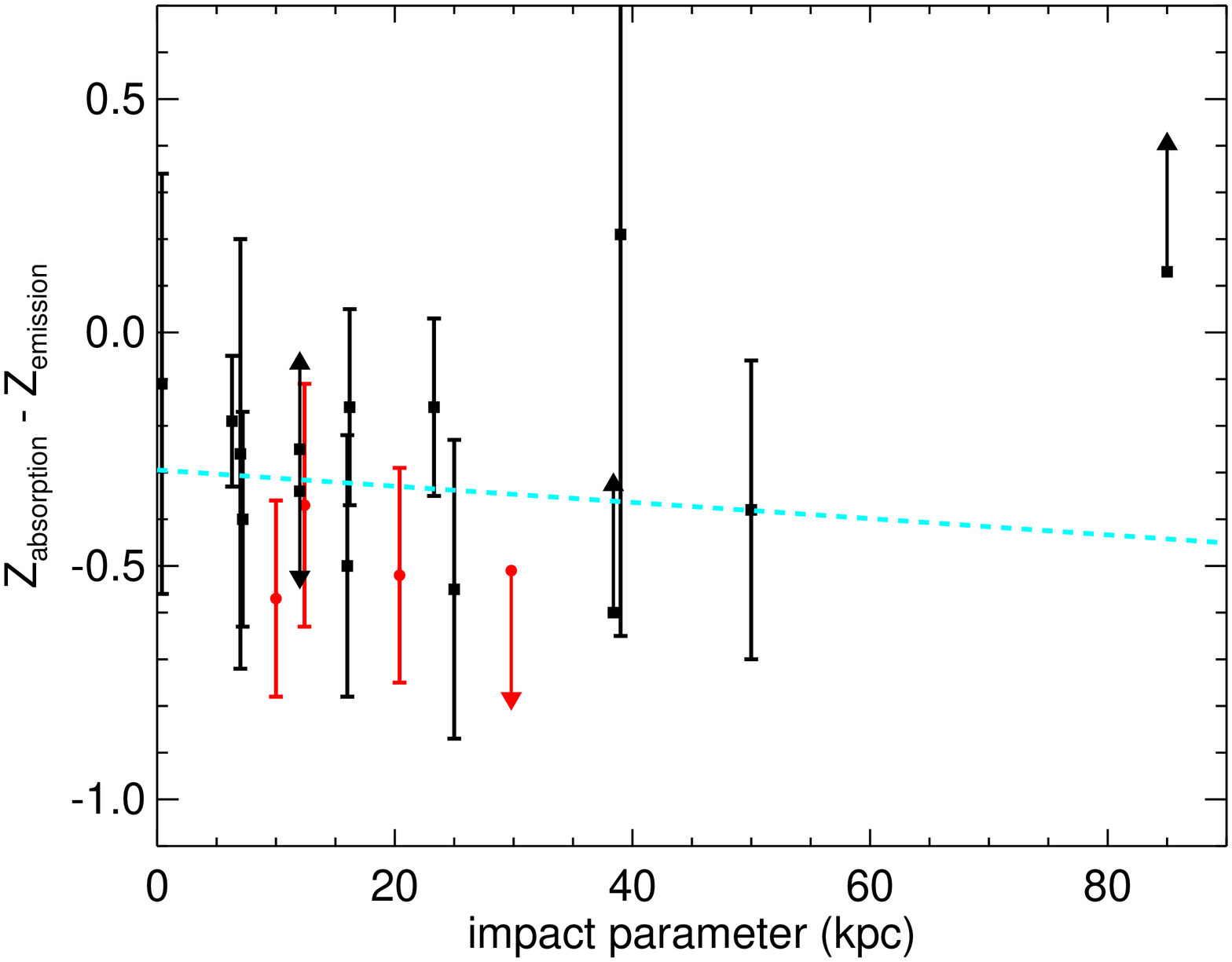}
	\caption{The emission and absorption metallicity difference as a function of the impact parameter. Filled circles present measurements obtained in this study and squares those in the literature. Dashed line presents the best fitted line to all measurements (ignoring the limits) resulting in a slope of $-$0.002$\pm$0.007 dex kpc$^{-1}$.
	}
	\label{fig_met_em_b}
\end{figure}
$N2$, $O3N2$ and $R23$ are combinations of flux ratios of nebular emission lines from [\OII], [\OIII], [\NII], H$\alpha$ and H$\beta$  
that can be utilized to estimate the metallicity of the ionized gas, Z$_{\rm em}$ \citep{Pagel79,Pettini04,Kobulnicky99}. [\NII]$\lambda$6585 is tentatively detected for DLA-galaxies towards 1012$+$0739 and 1204$+$0953. Hence we derive the Z$_{\rm em}$ of these two galaxies based on $N2$ and the calibration given by \citet{Pettini04}. For the rest of DLA-galaxies we have an upper limit for the flux of [\NII]$\lambda$6853 that translates to an upper limit for the metallicity of ionized gas, Z$_{\rm em}^{\rm up}$. However, using $R23$ \citep[equations (8) and (9) of ][]{Kobulnicky99} the lower and upper branch metallicities are calculated. From these two estimates the one consistent with Z$_{\rm em}^{\rm up}$ is then chosen as Z$_{\rm em}$. Our DLA-galaxies have sub-solar \citep[O$_\odot=8.69\pm0.05$,][]{Asplund09} emission metallicities in the range of 0.2 -- 0.9 Z$_\odot$ (see Table \ref{tab_res}).

The clumpy distribution of metals has proven to be fundamental in understanding the role of interaction, mergers, accretion and gas flows in galaxy formation and evolution \citep{Hou00,Cescutti07,DiMatteo09,Kewley10,Rupke10,Pilkington12}. We measure the difference between emission and absorption metallicities towards J0958$+$0549, J1138$+$0139 and J1204$+$0953 to be respectively $-0.025\pm0.009$, $-0.030\pm0.017$ and $-0.057\pm0.021$ dex kpc$^{-1}$. These are consistent with those reported in \citet{Peroux16} with a median value of $-0.022$ dex kpc$^{-1}$ at $z\sim1$ and also those of  \citet{Christensen14} with an average of $-0.022$ dex kpc$^{-1}$ at $0.1<z<3.2$. Fig. \ref{fig_met_em_b} presents the metallicity difference of DLA-galaxies and the absorbing gas as a function of impact parameter. Taking into account all measurements (ignoring the limits) we find a best fitted slope of $-0.002\pm0.007$ dex kpc$^{-1}$ which implies an insignificant variation of metallicity difference as a function of impact parameter for the average population of DLA-galaxies. The average value of metallicity gradient for the low luminosity sub-sample of spiral galaxies in local universe is $-0.063\pm0.011$ dex kpc$^{-1}$ \citep{Ho15}. We notice such metallicity gradients are measured for scales $<10$ kpc which are limited to local galaxy disks while our measurements compares mainly CGM neutral phase metallicity at scales usually $>10$ kpc. A shallow metallicity difference measured for DLA-galaxies can be due to interactions that can make the accretion of IGM pristine gas more efficient \citep{Cresci10,Queyrel12}. Evidences for such interactions have been observed for two DLA-galaxies in our sample and some more in the literature \citep[see][for other examples]{Kacprzak10,Straka16}. Disentangling the possible scenarios producing the observed metallicity difference in DLA-galaxies and the nature of absorbing gas requires detailed structural analysis of foreground galaxies \citep[see for example][]{Bouche13,Schroetter15,Bouche16}.
\begin{figure}
	\centering	
	\includegraphics[width=0.8\hsize,bb=18 18 594 774,clip=,angle=90]{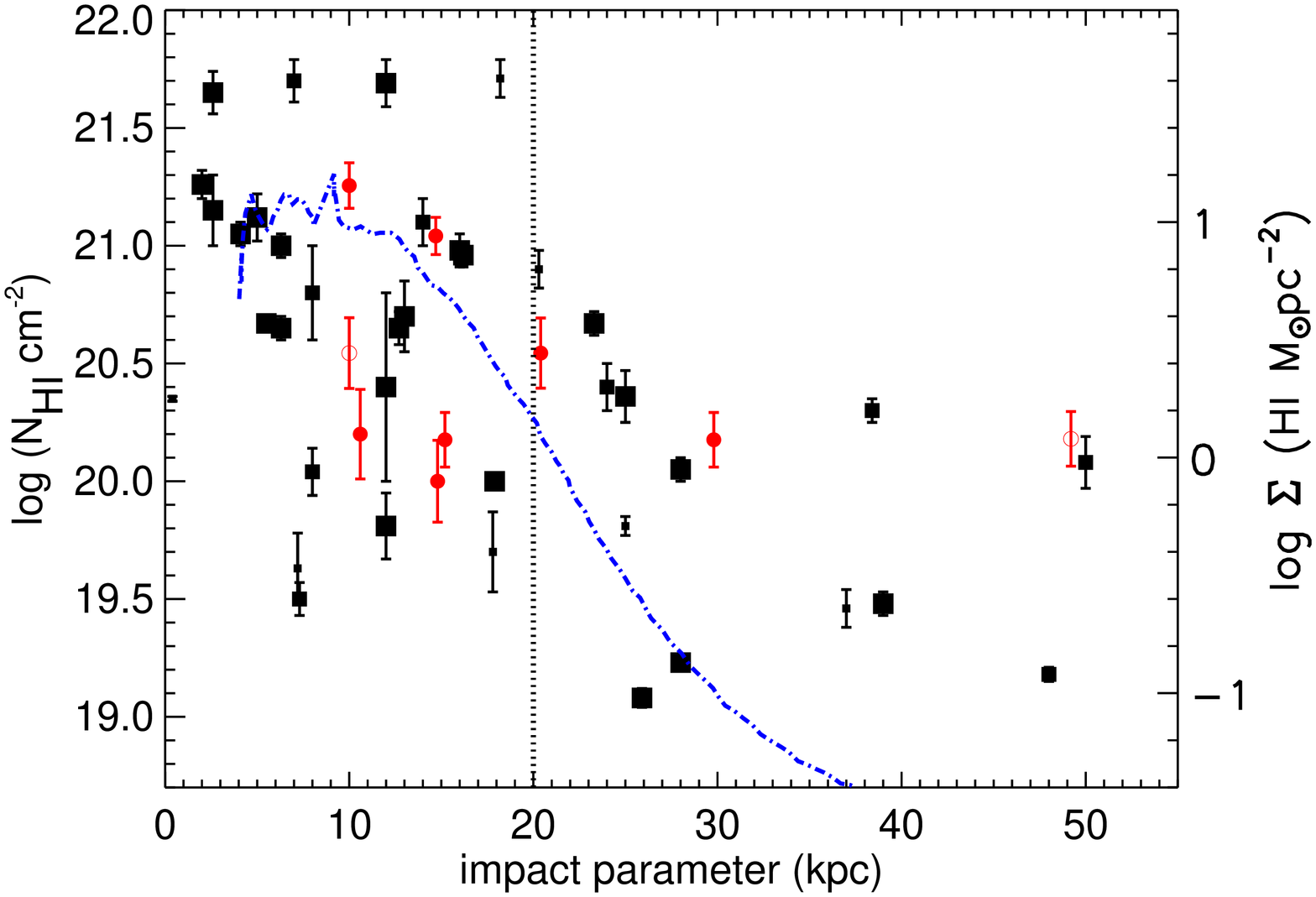}
	\caption{N(\HI) as a function of impact parameter. Filled circles are obtained in this study. Two empty circles present the second galaxies detected in fields of 0958$+$0549 and 1012$+$0739. Squares show those measurements in the literature where they have three different sizes: smaller squares are at  $z <$ 0.6 while the larger ones are at higher $z >$ 0.6. The vertical line at b = 20 kpc is chosen by eye and marks a transition where at larger impact parameters no DLA-galaxy with N(\HI) $>$ 21.0 were detected. Note that we have added 6 new points in the low impact parameter regime where only 15 measurements were known.	The dashed-dotted curve presents the \HI\ surface density for the disk of Milky Way \citep{Kalberla09}.
	}
	\label{fig_b_nhi}
\end{figure}
\subsection{Neutral gas in absorption}
The established correlation between $\Delta v_{90}$ and metallicity for DLAs and sub-DLAs is usually interpreted as a consequence of the underlined mass-metallicity relation observed in galaxies \citep{Ledoux06a,Neeleman13}. Detail studies of a handful of host galaxies of such absorbers (that allow mass estimates) have supported this interpretation \citep{Peroux11a,Peroux14,Christensen14}. The measured $\Delta v_{90}$ for systems, in this work, with detected DLA-galaxies range from 105--302 \kms. The two largest $\Delta v_{90}$ are measured towards J0218$-$0832 and J1515$+$0410 where DLA-galaxies are detected in continuum. We notice the detection of a faint continuum from the DLA-galaxy towards J1012$+$0739 where we have the third rank $\Delta v_{90}$. Being detected in continuum we expect higher underlined masses for these three galaxies which is in favor of the usual interpretation of $\Delta v_{90}$-metallicity correlation. However, this finding is limited by small number statistics. The realistic physics behind $\Delta v_{90}$ is more complicated as some fraction of DLAs and sub-DLAs belong to the CGM where the width of absorption lines is dominated by bulk flows rather than the gravitational potential well \citep[see][]{Bouche07,Lundgren09,Gauthier14}. 
%
%
%
%

Surface density maps of neutral hydrogen obtained from 21-cm emission of spiral galaxies in the local universe show a smooth decrease of N(\HI) with increasing impact parameter \citep{Braun04,Heald11,Wolfe_s15}. We can obtain an average picture for the N(\HI) distribution of DLA host galaxies by studying the N(\HI) vs impact parameters for a population of DLAs. For example, from a sample of absorbing galaxies identified using primarily photometric redshifts, \citet{Rao11} find a statistically significant trend of decreasing impact parameter with N(\HI). Their DLA galaxy sample has a median impact parameter of 17.4 kpc in comparison to 33.3 kpc obtained for their sub-DLA sample. In Fig. \ref{fig_b_nhi} we present N(\HI) as a function of impact parameter of sub-DLAs and DLAs for the spectroscopically confirmed absorbing galaxies in Table \ref{tab_res}. Circles present our results and squares those in the literature  where they have three different sizes: smaller squares are at lower $z$ than our survey (i.e. $z$ $\lesssim$ 0.6) while the larger ones are at higher $z$. For two fields of 0958$+$0549 and 1012$+$0739 where two galaxies are detected the second galaxies are shown using open circles.  In this work we have contributed to 6 new measurements at $b <$ 20 kpc where previously only 15 were known. The vertical dashed line at $b$ = 20 kpc is chosen by eye where at higher impact parameters $\log$ N(\HI) $>$ 21.0 DLAs were not detected. There are some high-$z$ DLA-galaxies in Fig. \ref{fig_b_nhi} detected at very small impact parameters of few kpc \citep{Moller93,Noterdaeme12,Krogager12}. Such DLA-galaxies are at $z > $ 2 so that the Lyman-$\alpha$ emission falls in the wavelength range where the QSO light is totally filtered by the DLA absorption of the foreground galaxy. Therefore, in principle there is no lower limit on the scale of impact parameters of such DLA-galaxies. We note the scatter decreases at $b > $ 20 kpc due to lack of high N(\HI) systems but this can be partly a bias introduced from the triangulation technique. In such a technique \citep[see][]{Fynbo10} one observes the QSO field using three different slit PA centered on the QSO separated by 120$^\circ$ optimized to cover the DLA-galaxy within the slit. However, the total area covered by the slit in three PAs decreases with increasing impact parameter resulting in a higher detection probability at lower impact parameters. This may be one reason for having fewer measurements at larger $b$ values. The real picture can be more complicated if DLA-galaxies are not randomly distributed around QSO sightlines \citep[See][]{Fumagalli14}. 
\begin{figure}
	\centering
	\includegraphics[width=0.8\hsize,bb=18 18 594 774,clip=,angle=90]{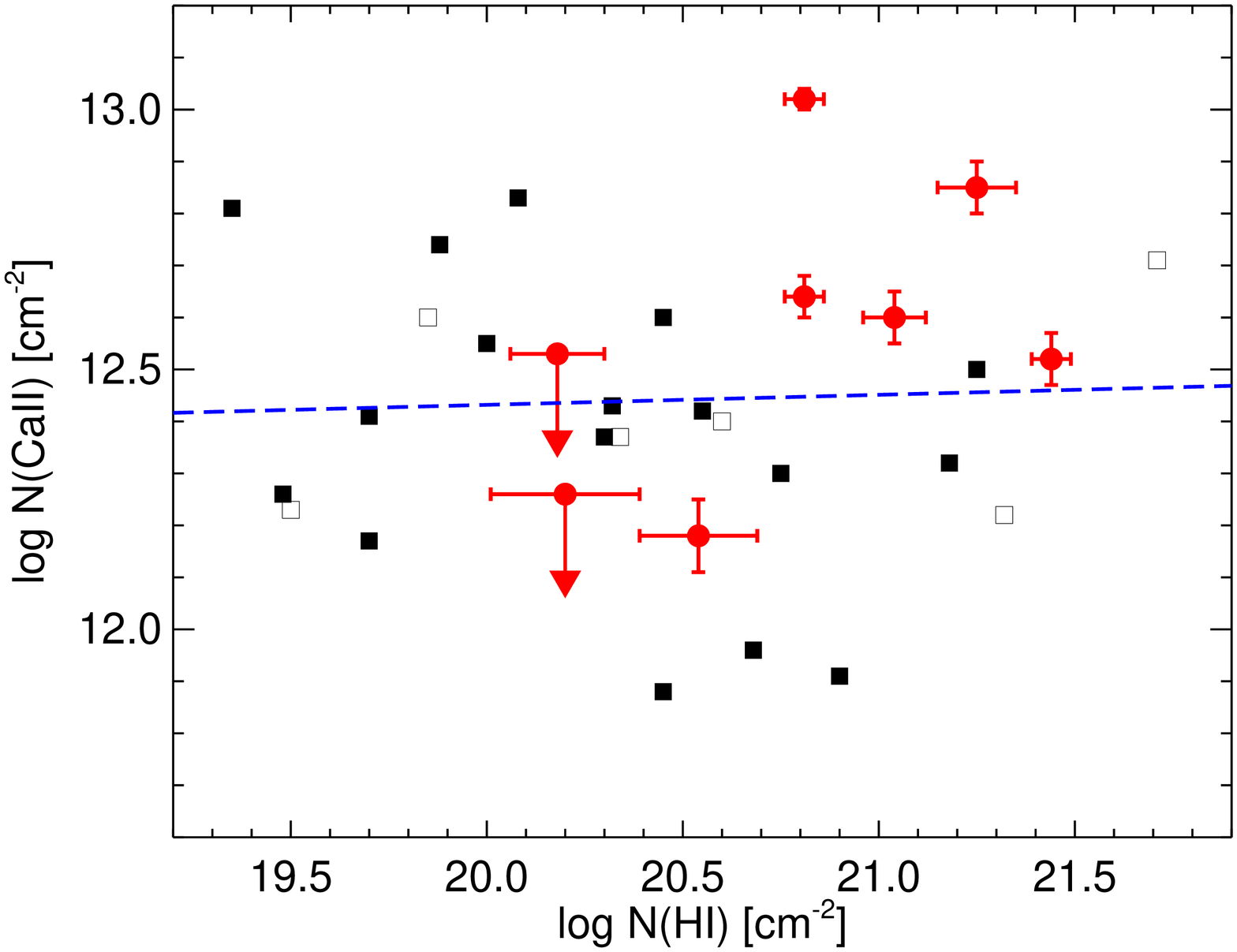}
	\caption{ N(\CaII) as a function of N(\HI). Filled circles present our measurements and squares those in the literature \citep{Boisse98,Ledoux02,Kulkarni05,Meiring06,Peroux06,Meiring07,Peroux08,Nestor08,Meiring09,Battisti12,Guber16}. The empty squares are from \citet{Richter11}. Dashed lines was obtained from a regression analysis to all points with $\log$ N(\HI) $>$ 19.0 with a slope of 0.02$\pm$0.08. A non-varying N(\CaII) over 2 orders of magnitude increase of N(\HI) may trace a competition between ionization and depletion of Calcium at respectively low and high N(\HI). 
	}
	\label{fig_ca_hi}
\end{figure}

Fig. \ref{fig_ca_hi} presents the column density of \CaII\ as a function of N(\HI) where our results are plotted as circles along with errorbars and those in the literature as squares. The filled and open squares are from the compilation of \citet{Quiret16} and \citet{Richter11}, respectively. \citet{Richter11} reported a mild correlation for their 7 systems with both \CaII\ had N(\HI) measurements at $z < $ 0.5. Similar trends have been also reported in studies of large number of sightlines within the Galaxy widespread over 4 orders of magnitude for N(\HI) \citep{Wakker00,Murga15}. However, no meaningful correlation is apparent for the overall population of sub-DLAs and DLAs presented in Fig. \ref{fig_ca_hi}. The dashed line obtained from a least square analysis has a slope of 0.04$\pm$0.09. 
This means the column density of \CaII\ do not change significantly over 2 orders of magnitude increase in N(\HI) though there exists a large scatter. We find a median value of 12.4 for the log [N(\CaII) cm$^{-2}$] with a 1-$\sigma$ standard deviation of 0.3 dex for all systems in Fig. \ref{fig_ca_hi}. Given the fact that Calcium is strongly depleted onto dust in the ISM this may imply a much higher Calcium depletion for higher N(\HI). However, it is worth noting that the ionization potential of \CaII\ is 11.87 eV, less than that of \HI. Therefore, even in DLAs, \CaII\ may not be the dominant ionization stage of Calcium and certain fraction of this element is expected to be in \ion{Ca}{iii}. This ionization effect can increase the scatter particularly at lower N(\HI). To test this we carried out a regression analysis for systems with N(\HI) $\geq$ 20.3 to obtain $\log$ N(\CaII) = 7.1$\pm$3.6 + (0.25$\pm$0.17) $\log$ N(\HI). Carrying out a correlation test using Spearman's $\rho$ (and Kendall's $\tau$) test we find $\rho$ = 0.36 and probability of no correlation P($\rho$) $<$ 0.13 ($\tau$=0.24, P($\tau$)=0.16). As a result, a possible correlation between N(\HI) and N(\CaII) can not be rejected for DLA systems. In summary two contrasting effects of dust depletion and ionization play dominant roles in the density of \CaII\ in high and low N(\HI) sides \citep{Nestor08,Zych09}. 

Given the fact that Calcium is highly depleted into dust a correlation between the strength of the \CaII\ absorption and the SFR in the host halo of the absorbers might be expected \citep{Wild07,Straka15,Sardane15}. By stacking QSO spectra of a sample \CaII\ absorbers from SDSS, \citet{Wild07} found higher SFR for the subsample with higher EW(\CaII). \citet{Zych07} found the host galaxies of very strong \CaII\ (EW$_0^{3934}\sim$ 0.5--0.9 \AA) absorbers to be L $\sim$ L$^*$ galaxies at Z $\sim$ Z$_\odot$. However, in non of these work N(\HI) is known. Amongst 22 DLA and sub-DLAs with measured N(\CaII) in the literature there are only 3 cases where SFR can be measured for the absorbing galaxies. In this work we have obtained 6 new N(\CaII) measurements where in 3 cases we are able to measure the SFR of absorbing galaxies. However, combining our measurements and those in the literature we do not find any correlation between \CaII\ content and SFR in DLA-galaxies. 
\section{Summary and conclusions}
We have studied X-Shooter observations of a set of 9 QSO fields. These QSOs are selected based on HST/ACS observations that indicate they have a DLA or sub-DLA in their spectra at $<z>\sim0.6$. A PSF subtraction of high spatial resolution prism images obtained from these HST/ACS observations provide us with 24 candidate DLA-galaxies in 9 fields at low impact parameters ($b\gtrsim0.7''$ or 6.0 kpc at $z\sim0.6$). The main goal of our X-Shooter observations is to confirm such candidates and study galaxies responsible for DLA and sub-DLA systems. The wide wavelength range covered by the X-Shooter (0.3 -- 2.3 $\mu$m) allows us to search for several nebular emission lines from DLA-galaxy candidates while simultaneously obtaining QSOs' spectra to study the associated metal absorption lines. We detect a total number of 12 galaxies in our X-Shooter observations. Four out of 12 detections are high redshift line emitters unrelated to \HI\ absorbers. The remaining 8 detected galaxies are related to DLAs in 7 QSO fields at impact parameters from 10 to 50 kpc. Having detected galaxies responsible for very strong \HI\ absorption in 7 out of 9 QSO fields we reach a high detection rate of 78\%. The high rate in detection of DLA-galaxies obtained in this work indicates that a high spatial resolution imaging of QSOs followed by deep spectroscopic observations is an efficient strategy in finding galaxies responsible for DLAs and sub-DLAs.

In 2 of the fields (0957+0807 and 1357+0525) we could not detect the absorbing galaxies in our X-Shooter data. We postulate the following two reasons for such non-detections: (I) they are not detected because they lie in the PSF of HST/ACS images. This can happen for absorbing galaxies at (Ia) $b\lesssim0.5''$ or (Ib) in the tail of the PSF where the QSO subtraction leaves larger residuals. (II) They have fainter emission lines or continuum than can be observed with our X-Shooter observations. We ignore the possibility of the absorbing galaxies to be at larger distances than that covered by HST/ACS field of view as the two non-detections have high N(\HI) of 10$^{21.44}$ and 10$^{20.81}$ cm$^{-2}$ and hence are unlikely to lie at $b\gtrsim20$ kpc ($\sim3''$) (see Fig. \ref{fig_b_nhi}). While the high N(\HI) of these two systems favors the reason (Ia) for the non-detections, still the small sample size prevents us to draw a concrete conclusion. 
 
We estimate the extinction corrected SFR of the 5 DLA-galaxies detected in emission, to be in the range of 0.4 to 6.7 M$_\odot$yr$^{-1}$ with a median of 1.07 M$_\odot$yr$^{-1}$. We also discover two sub-L$^\star$ (M$_i\sim-19.5$ mag) DLA-galaxies detected only in continuum with 3$\sigma$ SFR upper limits of 0.2 and 0.4 M$_\odot$yr$^{-1}$. Overall, such findings demonstrate that the average population of DLA-galaxies at $z\sim0.6$ are low luminosity dwarf galaxies at maximum impact parameter of 30 kpc. A comparison between our results and those reported in the \citep{Peroux12} but also $z\sim1$ indicate possible signatures of redshift evolution of SFR of DLA-galaxies.

In two of the QSO fields we detect two galaxies matching the absorption redshift. In particular in the case of 0958+0549 the appearing morphology resembles a merging configuration of two galaxies connected via a faint emitting stream. The absorption in such cases might be produced by the disrupted gas streams resulting from strong gravitational interactions of galaxies. Further 3D morpho-kinematic integral field unit (IFU) observations using instrument like Multi Unit Spectroscopic Explorer (MUSE) on VLT of such systems can unravel the physics of these galaxy absorbers.  

\section*{acknowledgement}
This work has been carried out thanks to the support of the OCEVU Labex (ANR-11-LABX-0060) and the A*MIDEX project (ANR-11-IDEX-0001-02) funded by the ``Investissements d'Avenir`` French government program managed by the ANR. CP thanks the ESO science visitor program for support. VPK was supported in part by  the NASA STScI grant GO/13733.02. Additional support from US National Science Foundation grant AST/1108830 is gratefully acknowledged.
\def\aj{AJ}%
\def\actaa{Acta Astron.}%
\def\araa{ARA\&A}%
\def\apj{ApJ}%
\def\apjl{ApJ}%
\def\apjs{ApJS}%
\def\ao{Appl.~Opt.}%
\def\apss{Ap\&SS}%
\def\aap{A\&A}%
\def\aapr{A\&A~Rev.}%
\def\aaps{A\&AS}%
\def\azh{AZh}%
\def\baas{BAAS}%
\def\bac{Bull. astr. Inst. Czechosl.}%
\def\caa{Chinese Astron. Astrophys.}%
\def\cjaa{Chinese J. Astron. Astrophys.}%
\def\icarus{Icarus}%
\def\jcap{J. Cosmology Astropart. Phys.}%
\def\jrasc{JRASC}%
\def\mnras{MNRAS}%
\def\memras{MmRAS}%
\def\na{New A}%
\def\nar{New A Rev.}%
\def\pasa{PASA}%
\def\pra{Phys.~Rev.~A}%
\def\prb{Phys.~Rev.~B}%
\def\prc{Phys.~Rev.~C}%
\def\prd{Phys.~Rev.~D}%
\def\pre{Phys.~Rev.~E}%
\def\prl{Phys.~Rev.~Lett.}%
\def\pasp{PASP}%
\def\pasj{PASJ}%
\def\qjras{QJRAS}%
\def\rmxaa{Rev. Mexicana Astron. Astrofis.}%
\def\skytel{S\&T}%
\def\solphys{Sol.~Phys.}%
\def\sovast{Soviet~Ast.}%
\def\ssr{Space~Sci.~Rev.}%
\def\zap{ZAp}%
\def\nat{Nature}%
\def\iaucirc{IAU~Circ.}%
\def\aplett{Astrophys.~Lett.}%
\def\apspr{Astrophys.~Space~Phys.~Res.}%
\def\bain{Bull.~Astron.~Inst.~Netherlands}%
\def\fcp{Fund.~Cosmic~Phys.}%
\def\gca{Geochim.~Cosmochim.~Acta}%
\def\grl{Geophys.~Res.~Lett.}%
\def\jcp{J.~Chem.~Phys.}%
\def\jgr{J.~Geophys.~Res.}%
\def\jqsrt{J.~Quant.~Spec.~Radiat.~Transf.}%
\def\memsai{Mem.~Soc.~Astron.~Italiana}%
\def\nphysa{Nucl.~Phys.~A}%
\def\physrep{Phys.~Rep.}%
\def\physscr{Phys.~Scr}%
\def\planss{Planet.~Space~Sci.}%
\def\procspie{Proc.~SPIE}%
\let\astap=\aap
\let\apjlett=\apjl
\let\apjsupp=\apjs
\let\applopt=\ao
\bibliographystyle{mnras}
\bibliography{/Users/hrahmani/work/IGM/files-ref/bib.bib}

\begin{thebibliography}{}
\makeatletter
\relax
\def\mn@urlcharsother{\let\do\@makeother \do\$\do\&\do\#\do\^\do\_\do\%\do\~}
\def\mn@doi{\begingroup\mn@urlcharsother \@ifnextchar [ {\mn@doi@}
  {\mn@doi@[]}}
\def\mn@doi@[#1]#2{\def\@tempa{#1}\ifx\@tempa\@empty \href
  {http://dx.doi.org/#2} {doi:#2}\else \href {http://dx.doi.org/#2} {#1}\fi
  \endgroup}
\def\mn@eprint#1#2{\mn@eprint@#1:#2::\@nil}
\def\mn@eprint@arXiv#1{\href {http://arxiv.org/abs/#1} {{\tt arXiv:#1}}}
\def\mn@eprint@dblp#1{\href {http://dblp.uni-trier.de/rec/bibtex/#1.xml}
  {dblp:#1}}
\def\mn@eprint@#1:#2:#3:#4\@nil{\def\@tempa {#1}\def\@tempb {#2}\def\@tempc
  {#3}\ifx \@tempc \@empty \let \@tempc \@tempb \let \@tempb \@tempa \fi \ifx
  \@tempb \@empty \def\@tempb {arXiv}\fi \@ifundefined
  {mn@eprint@\@tempb}{\@tempb:\@tempc}{\expandafter \expandafter \csname
  mn@eprint@\@tempb\endcsname \expandafter{\@tempc}}}

\bibitem[\protect\citeauthoryear{{Aihara} et~al.,}{{Aihara}
  et~al.}{2011}]{Aihara11}
{Aihara} H.,  et~al., 2011, \mn@doi [\apjs] {10.1088/0067-0049/193/2/29}, \href
  {http://adsabs.harvard.edu/abs/2011ApJS..193...29A} {193, 29}

\bibitem[\protect\citeauthoryear{{Asplund}, {Grevesse}, {Sauval}  \&
  {Scott}}{{Asplund} et~al.}{2009}]{Asplund09}
{Asplund} M.,  {Grevesse} N.,  {Sauval} A.~J.,   {Scott} P.,  2009, \araa,
  \href {http://adsabs.harvard.edu/abs/2009arXiv0909.0948A} {47, 481}

\bibitem[\protect\citeauthoryear{{Baldwin}, {Phillips}  \&
  {Terlevich}}{{Baldwin} et~al.}{1981}]{Baldwin81}
{Baldwin} J.~A.,  {Phillips} M.~M.,   {Terlevich} R.,  1981, \mn@doi [\pasp]
  {10.1086/130766}, \href {http://adsabs.harvard.edu/abs/1981PASP...93....5B}
  {93, 5}

\bibitem[\protect\citeauthoryear{{Battisti} et~al.,}{{Battisti}
  et~al.}{2012}]{Battisti12}
{Battisti} A.~J.,  et~al., 2012, \mn@doi [\apj] {10.1088/0004-637X/744/2/93},
  \href {http://adsabs.harvard.edu/abs/2012ApJ...744...93B} {744, 93}

\bibitem[\protect\citeauthoryear{{Bergeron}}{{Bergeron}}{1986}]{Bergeron86}
{Bergeron} J.,  1986, \aap, \href
  {http://adsabs.harvard.edu/abs/1986A%26A...155L...8B} {155, L8}

\bibitem[\protect\citeauthoryear{{Bielby} et~al.,}{{Bielby}
  et~al.}{2016}]{Bielby16}
{Bielby} R.~M.,  et~al., 2016, \mn@doi [\mnras] {10.1093/mnras/stv2914}, \href
  {http://adsabs.harvard.edu/abs/2016MNRAS.456.4061B} {456, 4061}

\bibitem[\protect\citeauthoryear{{Bird}, {Haehnelt}, {Neeleman}, {Genel},
  {Vogelsberger}  \& {Hernquist}}{{Bird} et~al.}{2015}]{Bird15}
{Bird} S.,  {Haehnelt} M.,  {Neeleman} M.,  {Genel} S.,  {Vogelsberger} M.,
  {Hernquist} L.,  2015, \mn@doi [\mnras] {10.1093/mnras/stu2542}, \href
  {http://adsabs.harvard.edu/abs/2015MNRAS.447.1834B} {447, 1834}

\bibitem[\protect\citeauthoryear{{Boiss\'e}, {Le Brun}, {Bergeron}  \&
  {Deharveng}}{{Boiss\'e} et~al.}{1998}]{Boisse98}
{Boiss\'e} P.,  {Le Brun} V.,  {Bergeron} J.,   {Deharveng} J.-M.,  1998, \aap,
  \href {http://adsabs.harvard.edu/abs/1998A%26A...333..841B} {333, 841}

\bibitem[\protect\citeauthoryear{{Booth} \& {Schaye}}{{Booth} \&
  {Schaye}}{2009}]{Booth09}
{Booth} C.~M.,  {Schaye} J.,  2009, \mn@doi [\mnras]
  {10.1111/j.1365-2966.2009.15043.x}, \href
  {http://adsabs.harvard.edu/abs/2009MNRAS.398...53B} {398, 53}

\bibitem[\protect\citeauthoryear{{Bouch{\'e}}, {Lehnert}, {Aguirre},
  {P{\'e}roux}  \& {Bergeron}}{{Bouch{\'e}} et~al.}{2007a}]{Bouche07}
{Bouch{\'e}} N.,  {Lehnert} M.~D.,  {Aguirre} A.,  {P{\'e}roux} C.,
  {Bergeron} J.,  2007a, \mn@doi [\mnras] {10.1111/j.1365-2966.2007.11740.x},
  \href {http://adsabs.harvard.edu/abs/2007MNRAS.378..525B} {378, 525}

\bibitem[\protect\citeauthoryear{{Bouch{\'e}}, {Murphy}, {P{\'e}roux},
  {Davies}, {Eisenhauer}, {F{\"o}rster Schreiber}  \& {Tacconi}}{{Bouch{\'e}}
  et~al.}{2007b}]{Bouche07_simple}
{Bouch{\'e}} N.,  {Murphy} M.~T.,  {P{\'e}roux} C.,  {Davies} R.,  {Eisenhauer}
  F.,  {F{\"o}rster Schreiber} N.~M.,   {Tacconi} L.,  2007b, \mn@doi [\apjl]
  {10.1086/523594}, \href {http://adsabs.harvard.edu/abs/2007ApJ...669L...5B}
  {669, L5}

\bibitem[\protect\citeauthoryear{{Bouch{\'e}} et~al.,}{{Bouch{\'e}}
  et~al.}{2012}]{Bouche12}
{Bouch{\'e}} N.,  et~al., 2012, \mn@doi [\mnras]
  {10.1111/j.1365-2966.2011.19500.x}, \href
  {http://adsabs.harvard.edu/abs/2012MNRAS.419....2B} {419, 2}

\bibitem[\protect\citeauthoryear{{Bouch{\'e}}, {Murphy}, {Kacprzak},
  {P{\'e}roux}, {Contini}, {Martin}  \& {Dessauges-Zavadsky}}{{Bouch{\'e}}
  et~al.}{2013}]{Bouche13}
{Bouch{\'e}} N.,  {Murphy} M.~T.,  {Kacprzak} G.~G.,  {P{\'e}roux} C.,
  {Contini} T.,  {Martin} C.~L.,   {Dessauges-Zavadsky} M.,  2013, \mn@doi
  [Science] {10.1126/science.1234209}, \href
  {http://adsabs.harvard.edu/abs/2013Sci...341...50B} {341, 50}

\bibitem[\protect\citeauthoryear{{Bouch{\'e}} et~al.,}{{Bouch{\'e}}
  et~al.}{2016}]{Bouche16}
{Bouch{\'e}} N.,  et~al., 2016, \mn@doi [\apj] {10.3847/0004-637X/820/2/121},
  \href {http://adsabs.harvard.edu/abs/2016ApJ...820..121B} {820, 121}

\bibitem[\protect\citeauthoryear{{Braun} \& {Thilker}}{{Braun} \&
  {Thilker}}{2004}]{Braun04}
{Braun} R.,  {Thilker} D.~A.,  2004, \mn@doi [\aap]
  {10.1051/0004-6361:20034423}, \href
  {http://adsabs.harvard.edu/abs/2004A%26A...417..421B} {417, 421}

\bibitem[\protect\citeauthoryear{{Cen}}{{Cen}}{2012}]{Cen12}
{Cen} R.,  2012, \mn@doi [\apj] {10.1088/0004-637X/748/2/121}, \href
  {http://adsabs.harvard.edu/abs/2012ApJ...748..121C} {748, 121}

\bibitem[\protect\citeauthoryear{{Cescutti}, {Matteucci}, {Fran{\c c}ois}  \&
  {Chiappini}}{{Cescutti} et~al.}{2007}]{Cescutti07}
{Cescutti} G.,  {Matteucci} F.,  {Fran{\c c}ois} P.,   {Chiappini} C.,  2007,
  \mn@doi [\aap] {10.1051/0004-6361:20065403}, \href
  {http://adsabs.harvard.edu/abs/2007A%26A...462..943C} {462, 943}

\bibitem[\protect\citeauthoryear{{Ceverino} \& {Klypin}}{{Ceverino} \&
  {Klypin}}{2009}]{Ceverino09}
{Ceverino} D.,  {Klypin} A.,  2009, \mn@doi [\apj]
  {10.1088/0004-637X/695/1/292}, \href
  {http://adsabs.harvard.edu/abs/2009ApJ...695..292C} {695, 292}

\bibitem[\protect\citeauthoryear{{Chabrier}}{{Chabrier}}{2003}]{Chabrier03}
{Chabrier} G.,  2003, \mn@doi [\pasp] {10.1086/376392}, \href
  {http://adsabs.harvard.edu/abs/2003PASP..115..763C} {115, 763}

\bibitem[\protect\citeauthoryear{{Chen} \& {Lanzetta}}{{Chen} \&
  {Lanzetta}}{2003}]{Chen03}
{Chen} H.-W.,  {Lanzetta} K.~M.,  2003, \mn@doi [\apj] {10.1086/378635}, \href
  {http://adsabs.harvard.edu/abs/2003ApJ...597..706C} {597, 706}

\bibitem[\protect\citeauthoryear{{Chen}, {Kennicutt}  \& {Rauch}}{{Chen}
  et~al.}{2005}]{Chen05}
{Chen} H.-W.,  {Kennicutt} Jr. R.~C.,   {Rauch} M.,  2005, \mn@doi [\apj]
  {10.1086/427088}, \href {http://adsabs.harvard.edu/abs/2005ApJ...620..703C}
  {620, 703}

\bibitem[\protect\citeauthoryear{{Christensen}, {S{\'a}nchez}, {Jahnke},
  {Becker}, {Wisotzki}, {Kelz}, {Popovi{\'c}}  \& {Roth}}{{Christensen}
  et~al.}{2004}]{Christensen04}
{Christensen} L.,  {S{\'a}nchez} S.~F.,  {Jahnke} K.,  {Becker} T.,  {Wisotzki}
  L.,  {Kelz} A.,  {Popovi{\'c}} L.~{\v C}.,   {Roth} M.~M.,  2004, \mn@doi
  [\aap] {10.1051/0004-6361:20034371}, \href
  {http://adsabs.harvard.edu/abs/2004A%26A...417..487C} {417, 487}

\bibitem[\protect\citeauthoryear{{Christensen}, {Schulte-Ladbeck},
  {S{\'a}nchez}, {Becker}, {Jahnke}, {Kelz}, {Roth}  \&
  {Wisotzki}}{{Christensen} et~al.}{2005}]{Christensen05}
{Christensen} L.,  {Schulte-Ladbeck} R.~E.,  {S{\'a}nchez} S.~F.,  {Becker} T.,
   {Jahnke} K.,  {Kelz} A.,  {Roth} M.~M.,   {Wisotzki} L.,  2005, \mn@doi
  [\aap] {10.1051/0004-6361:20041591}, \href
  {http://adsabs.harvard.edu/abs/2005A%26A...429..477C} {429, 477}

\bibitem[\protect\citeauthoryear{{Christensen}, {Wisotzki}, {Roth},
  {S{\'a}nchez}, {Kelz}  \& {Jahnke}}{{Christensen}
  et~al.}{2007}]{Christensen07}
{Christensen} L.,  {Wisotzki} L.,  {Roth} M.~M.,  {S{\'a}nchez} S.~F.,  {Kelz}
  A.,   {Jahnke} K.,  2007, \mn@doi [\aap] {10.1051/0004-6361:20066410}, \href
  {http://adsabs.harvard.edu/abs/2007A%26A...468..587C} {468, 587}

\bibitem[\protect\citeauthoryear{{Christensen}, {M{\o}ller}, {Fynbo}  \&
  {Zafar}}{{Christensen} et~al.}{2014}]{Christensen14}
{Christensen} L.,  {M{\o}ller} P.,  {Fynbo} J.~P.~U.,   {Zafar} T.,  2014,
  \mn@doi [\mnras] {10.1093/mnras/stu1726}, \href
  {http://adsabs.harvard.edu/abs/2014MNRAS.445..225C} {445, 225}

\bibitem[\protect\citeauthoryear{{Churchill}, {Steidel}  \& {Vogt}}{{Churchill}
  et~al.}{1996}]{Churchill96}
{Churchill} C.~W.,  {Steidel} C.~C.,   {Vogt} S.~S.,  1996, \mn@doi [\apj]
  {10.1086/177960}, \href {http://adsabs.harvard.edu/abs/1996ApJ...471..164C}
  {471, 164}

\bibitem[\protect\citeauthoryear{{Cresci}, {Mannucci}, {Maiolino}, {Marconi},
  {Gnerucci}  \& {Magrini}}{{Cresci} et~al.}{2010}]{Cresci10}
{Cresci} G.,  {Mannucci} F.,  {Maiolino} R.,  {Marconi} A.,  {Gnerucci} A.,
  {Magrini} L.,  2010, \mn@doi [\nat] {10.1038/nature09451}, \href
  {http://adsabs.harvard.edu/abs/2010Natur.467..811C} {467, 811}

\bibitem[\protect\citeauthoryear{{Crighton} et~al.,}{{Crighton}
  et~al.}{2015}]{Crighton15}
{Crighton} N.~H.~M.,  et~al., 2015, \mn@doi [\mnras] {10.1093/mnras/stv1182},
  \href {http://adsabs.harvard.edu/abs/2015MNRAS.452..217C} {452, 217}

\bibitem[\protect\citeauthoryear{{D'Odorico} et~al.,}{{D'Odorico}
  et~al.}{2013}]{Dodorico13}
{D'Odorico} V.,  et~al., 2013, \mn@doi [\mnras] {10.1093/mnras/stt1365}, \href
  {http://adsabs.harvard.edu/abs/2013MNRAS.435.1198D} {435, 1198}

\bibitem[\protect\citeauthoryear{{Dalla Vecchia} \& {Schaye}}{{Dalla Vecchia}
  \& {Schaye}}{2012}]{Dalla-Vecchia12}
{Dalla Vecchia} C.,  {Schaye} J.,  2012, \mn@doi [\mnras]
  {10.1111/j.1365-2966.2012.21704.x}, \href
  {http://adsabs.harvard.edu/abs/2012MNRAS.426..140D} {426, 140}

\bibitem[\protect\citeauthoryear{{Dessauges-Zavadsky}, {P{\'e}roux}, {Kim},
  {D'Odorico}  \& {McMahon}}{{Dessauges-Zavadsky}
  et~al.}{2003}]{Dessauges-Zavadsky03}
{Dessauges-Zavadsky} M.,  {P{\'e}roux} C.,  {Kim} T.-S.,  {D'Odorico} S.,
  {McMahon} R.~G.,  2003, \mn@doi [\mnras] {10.1046/j.1365-8711.2003.06949.x},
  \href {http://adsabs.harvard.edu/abs/2003MNRAS.345..447D} {345, 447}

\bibitem[\protect\citeauthoryear{{Di Matteo}, {Pipino}, {Lehnert}, {Combes}  \&
  {Semelin}}{{Di Matteo} et~al.}{2009}]{DiMatteo09}
{Di Matteo} P.,  {Pipino} A.,  {Lehnert} M.~D.,  {Combes} F.,   {Semelin} B.,
  2009, \mn@doi [\aap] {10.1051/0004-6361/200911715}, \href
  {http://adsabs.harvard.edu/abs/2009A%26A...499..427D} {499, 427}

\bibitem[\protect\citeauthoryear{{Djorgovski}, {Pahre}, {Bechtold}  \&
  {Elston}}{{Djorgovski} et~al.}{1996}]{Djorgovski96}
{Djorgovski} S.~G.,  {Pahre} M.~A.,  {Bechtold} J.,   {Elston} R.,  1996,
  \mn@doi [\nat] {10.1038/382234a0}, \href
  {http://adsabs.harvard.edu/abs/1996Natur.382..234D} {382, 234}

\bibitem[\protect\citeauthoryear{{Erb}, {Steidel}, {Shapley}, {Pettini},
  {Reddy}  \& {Adelberger}}{{Erb} et~al.}{2006}]{Erb06}
{Erb} D.~K.,  {Steidel} C.~C.,  {Shapley} A.~E.,  {Pettini} M.,  {Reddy} N.~A.,
    {Adelberger} K.~L.,  2006, \mn@doi [\apj] {10.1086/504891}, \href
  {http://adsabs.harvard.edu/abs/2006ApJ...646..107E} {646, 107}

\bibitem[\protect\citeauthoryear{{Francis} \& {Williger}}{{Francis} \&
  {Williger}}{2004}]{Francis04_c}
{Francis} P.~J.,  {Williger} G.~M.,  2004, \mn@doi [\apjl] {10.1086/382590},
  \href {http://adsabs.harvard.edu/abs/2004ApJ...602L..77F} {602, L77}

\bibitem[\protect\citeauthoryear{{Fumagalli}, {O'Meara}, {Prochaska}, {Kanekar}
   \& {Wolfe}}{{Fumagalli} et~al.}{2014}]{Fumagalli14}
{Fumagalli} M.,  {O'Meara} J.~M.,  {Prochaska} J.~X.,  {Kanekar} N.,   {Wolfe}
  A.~M.,  2014, \mn@doi [\mnras] {10.1093/mnras/stu1512}, \href
  {http://adsabs.harvard.edu/abs/2014MNRAS.444.1282F} {444, 1282}

\bibitem[\protect\citeauthoryear{{Fumagalli}, {O'Meara}, {Prochaska},
  {Rafelski}  \& {Kanekar}}{{Fumagalli} et~al.}{2015}]{Fumagalli15}
{Fumagalli} M.,  {O'Meara} J.~M.,  {Prochaska} J.~X.,  {Rafelski} M.,
  {Kanekar} N.,  2015, \mn@doi [\mnras] {10.1093/mnras/stu2325}, \href
  {http://adsabs.harvard.edu/abs/2015MNRAS.446.3178F} {446, 3178}

\bibitem[\protect\citeauthoryear{{Fynbo} et~al.,}{{Fynbo}
  et~al.}{2010}]{Fynbo10}
{Fynbo} J.~P.~U.,  et~al., 2010, \mn@doi [\mnras]
  {10.1111/j.1365-2966.2010.17294.x}, \href
  {http://adsabs.harvard.edu/abs/2010MNRAS.tmp.1315F} {pp 1315--+}

\bibitem[\protect\citeauthoryear{{Fynbo} et~al.,}{{Fynbo}
  et~al.}{2011}]{Fynbo11}
{Fynbo} J.~P.~U.,  et~al., 2011, \mn@doi [\mnras]
  {10.1111/j.1365-2966.2011.18318.x}, \href
  {http://adsabs.harvard.edu/abs/2011MNRAS.413.2481F} {413, 2481}

\bibitem[\protect\citeauthoryear{{Fynbo} et~al.,}{{Fynbo}
  et~al.}{2013}]{Fynbo13}
{Fynbo} J.~P.~U.,  et~al., 2013, \mn@doi [\mnras] {10.1093/mnras/stt1579},
  \href {http://adsabs.harvard.edu/abs/2013MNRAS.436..361F} {436, 361}

\bibitem[\protect\citeauthoryear{{Gauthier}, {Chen}, {Cooksey}, {Simcoe},
  {Seyffert}  \& {O'Meara}}{{Gauthier} et~al.}{2014}]{Gauthier14}
{Gauthier} J.-R.,  {Chen} H.-W.,  {Cooksey} K.~L.,  {Simcoe} R.~A.,  {Seyffert}
  E.~N.,   {O'Meara} J.~M.,  2014, \mn@doi [\mnras] {10.1093/mnras/stt2443},
  \href {http://adsabs.harvard.edu/abs/2014MNRAS.439..342G} {439, 342}

\bibitem[\protect\citeauthoryear{{Gharanfoli}, {Kulkarni}, {Chun}  \&
  {Takamiya}}{{Gharanfoli} et~al.}{2007}]{Gharanfoli07}
{Gharanfoli} S.,  {Kulkarni} V.~P.,  {Chun} M.~R.,   {Takamiya} M.,  2007,
  \mn@doi [\aj] {10.1086/509618}, \href
  {http://adsabs.harvard.edu/abs/2007AJ....133..130G} {133, 130}

\bibitem[\protect\citeauthoryear{{Goldoni}, {Royer}, {Fran{\c c}ois},
  {Horrobin}, {Blanc}, {Vernet}, {Modigliani}  \& {Larsen}}{{Goldoni}
  et~al.}{2006}]{Goldoni06}
{Goldoni} P.,  {Royer} F.,  {Fran{\c c}ois} P.,  {Horrobin} M.,  {Blanc} G.,
  {Vernet} J.,  {Modigliani} A.,   {Larsen} J.,  2006, in Society of
  Photo-Optical Instrumentation Engineers (SPIE) Conference Series. p.~2,
  \mn@doi{10.1117/12.669986}

\bibitem[\protect\citeauthoryear{{Guber} \& {Richter}}{{Guber} \&
  {Richter}}{2016}]{Guber16}
{Guber} C.~R.,  {Richter} P.,  2016, preprint, \href
  {http://adsabs.harvard.edu/abs/2016arXiv160503054G} {} (\mn@eprint {arXiv}
  {1605.03054})

\bibitem[\protect\citeauthoryear{{Haas}, {Schaye}, {Booth}, {Dalla Vecchia},
  {Springel}, {Theuns}  \& {Wiersma}}{{Haas} et~al.}{2013}]{Haas13}
{Haas} M.~R.,  {Schaye} J.,  {Booth} C.~M.,  {Dalla Vecchia} C.,  {Springel}
  V.,  {Theuns} T.,   {Wiersma} R.~P.~C.,  2013, \mn@doi [\mnras]
  {10.1093/mnras/stt1487}, \href
  {http://adsabs.harvard.edu/abs/2013MNRAS.435.2931H} {435, 2931}

\bibitem[\protect\citeauthoryear{{Hartoog}, {Fynbo}, {Kaper}, {De Cia}  \&
  {Bagdonaite}}{{Hartoog} et~al.}{2015}]{Hartoog15}
{Hartoog} O.~E.,  {Fynbo} J.~P.~U.,  {Kaper} L.,  {De Cia} A.,   {Bagdonaite}
  J.,  2015, \mn@doi [\mnras] {10.1093/mnras/stu2578}, \href
  {http://adsabs.harvard.edu/abs/2015MNRAS.447.2738H} {447, 2738}

\bibitem[\protect\citeauthoryear{{Heald} et~al.,}{{Heald}
  et~al.}{2011}]{Heald11}
{Heald} G.,  et~al., 2011, \mn@doi [\aap] {10.1051/0004-6361/201015938}, \href
  {http://adsabs.harvard.edu/abs/2011A%26A...526A.118H} {526, A118}

\bibitem[\protect\citeauthoryear{{Heckman}, {Lehnert}, {Strickland}  \&
  {Armus}}{{Heckman} et~al.}{2000}]{Heckman00}
{Heckman} T.~M.,  {Lehnert} M.~D.,  {Strickland} D.~K.,   {Armus} L.,  2000,
  \mn@doi [\apjs] {10.1086/313421}, \href
  {http://adsabs.harvard.edu/abs/2000ApJS..129..493H} {129, 493}

\bibitem[\protect\citeauthoryear{{Herbert-Fort}, {Prochaska},
  {Dessauges-Zavadsky}, {Ellison}, {Howk}, {Wolfe}  \&
  {Prochter}}{{Herbert-Fort} et~al.}{2006}]{Herbert-Fort06}
{Herbert-Fort} S.,  {Prochaska} J.~X.,  {Dessauges-Zavadsky} M.,  {Ellison}
  S.~L.,  {Howk} J.~C.,  {Wolfe} A.~M.,   {Prochter} G.~E.,  2006, \mn@doi
  [\pasp] {10.1086/507653}, \href
  {http://adsabs.harvard.edu/abs/2006PASP..118.1077H} {118, 1077}

\bibitem[\protect\citeauthoryear{{Ho}, {Kudritzki}, {Kewley}, {Zahid},
  {Dopita}, {Bresolin}  \& {Rupke}}{{Ho} et~al.}{2015}]{Ho15}
{Ho} I.-T.,  {Kudritzki} R.-P.,  {Kewley} L.~J.,  {Zahid} H.~J.,  {Dopita}
  M.~A.,  {Bresolin} F.,   {Rupke} D.~S.~N.,  2015, \mn@doi [\mnras]
  {10.1093/mnras/stv067}, \href
  {http://adsabs.harvard.edu/abs/2015MNRAS.448.2030H} {448, 2030}

\bibitem[\protect\citeauthoryear{{Hopkins}, {Kere{\v s}}, {Murray}, {Quataert}
  \& {Hernquist}}{{Hopkins} et~al.}{2012}]{Hopkins-p12}
{Hopkins} P.~F.,  {Kere{\v s}} D.,  {Murray} N.,  {Quataert} E.,   {Hernquist}
  L.,  2012, \mn@doi [\mnras] {10.1111/j.1365-2966.2012.21981.x}, \href
  {http://adsabs.harvard.edu/abs/2012MNRAS.427..968H} {427, 968}

\bibitem[\protect\citeauthoryear{{Hou}, {Prantzos}  \& {Boissier}}{{Hou}
  et~al.}{2000}]{Hou00}
{Hou} J.~L.,  {Prantzos} N.,   {Boissier} S.,  2000, \aap, \href
  {http://adsabs.harvard.edu/abs/2000A%26A...362..921H} {362, 921}

\bibitem[\protect\citeauthoryear{{Japelj} et~al.,}{{Japelj}
  et~al.}{2015}]{Japelj15}
{Japelj} J.,  et~al., 2015, \mn@doi [\aap] {10.1051/0004-6361/201525665}, \href
  {http://adsabs.harvard.edu/abs/2015A%26A...579A..74J} {579, A74}

\bibitem[\protect\citeauthoryear{{Jenkins}}{{Jenkins}}{2009}]{Jenkins09}
{Jenkins} E.~B.,  2009, \mn@doi [\apj] {10.1088/0004-637X/700/2/1299}, \href
  {http://adsabs.harvard.edu/abs/2009ApJ...700.1299J} {700, 1299}

\bibitem[\protect\citeauthoryear{{Kacprzak}, {Murphy}  \&
  {Churchill}}{{Kacprzak} et~al.}{2010}]{Kacprzak10}
{Kacprzak} G.~G.,  {Murphy} M.~T.,   {Churchill} C.~W.,  2010, \mn@doi [\mnras]
  {10.1111/j.1365-2966.2010.16667.x}, \href
  {http://adsabs.harvard.edu/abs/2010MNRAS.406..445K} {406, 445}

\bibitem[\protect\citeauthoryear{{Kacprzak}, {Muzahid}, {Churchill}, {Nielsen}
  \& {Charlton}}{{Kacprzak} et~al.}{2015}]{Kacprzak15}
{Kacprzak} G.~G.,  {Muzahid} S.,  {Churchill} C.~W.,  {Nielsen} N.~M.,
  {Charlton} J.~C.,  2015, preprint, \href
  {http://adsabs.harvard.edu/abs/2015arXiv151103275K} {} (\mn@eprint {arXiv}
  {1511.03275})

\bibitem[\protect\citeauthoryear{{Kalberla} \& {Kerp}}{{Kalberla} \&
  {Kerp}}{2009}]{Kalberla09}
{Kalberla} P.~M.~W.,  {Kerp} J.,  2009, \mn@doi [\araa]
  {10.1146/annurev-astro-082708-101823}, \href
  {http://adsabs.harvard.edu/abs/2009ARA%26A..47...27K} {47, 27}

\bibitem[\protect\citeauthoryear{{Kashikawa}, {Misawa}, {Minowa}, {Okoshi},
  {Hattori}, {Toshikawa}, {Ishikawa}  \& {Onoue}}{{Kashikawa}
  et~al.}{2014}]{Kashikawa14}
{Kashikawa} N.,  {Misawa} T.,  {Minowa} Y.,  {Okoshi} K.,  {Hattori} T.,
  {Toshikawa} J.,  {Ishikawa} S.,   {Onoue} M.,  2014, \mn@doi [\apj]
  {10.1088/0004-637X/780/2/116}, \href
  {http://adsabs.harvard.edu/abs/2014ApJ...780..116K} {780, 116}

\bibitem[\protect\citeauthoryear{{Kauffmann} et~al.,}{{Kauffmann}
  et~al.}{2003}]{Kauffmann03}
{Kauffmann} G.,  et~al., 2003, \mn@doi [\mnras]
  {10.1111/j.1365-2966.2003.07154.x}, \href
  {http://adsabs.harvard.edu/abs/2003MNRAS.346.1055K} {346, 1055}

\bibitem[\protect\citeauthoryear{{Kausch} et~al.,}{{Kausch}
  et~al.}{2015}]{Kausch15}
{Kausch} W.,  et~al., 2015, \mn@doi [\aap] {10.1051/0004-6361/201423909}, \href
  {http://adsabs.harvard.edu/abs/2015A%26A...576A..78K} {576, A78}

\bibitem[\protect\citeauthoryear{{Kennicutt}}{{Kennicutt}}{1998}]{Kennicutt98}
{Kennicutt} Jr. R.~C.,  1998, \mn@doi [\araa] {10.1146/annurev.astro.36.1.189},
  \href {http://adsabs.harvard.edu/abs/1998ARA%26A..36..189K} {36, 189}

\bibitem[\protect\citeauthoryear{{Kere{\v s}}, {Katz}, {Weinberg}  \&
  {Dav{\'e}}}{{Kere{\v s}} et~al.}{2005}]{Keres05}
{Kere{\v s}} D.,  {Katz} N.,  {Weinberg} D.~H.,   {Dav{\'e}} R.,  2005, \mn@doi
  [\mnras] {10.1111/j.1365-2966.2005.09451.x}, \href
  {http://adsabs.harvard.edu/abs/2005MNRAS.363....2K} {363, 2}

\bibitem[\protect\citeauthoryear{{Kere{\v s}}, {Katz}, {Fardal}, {Dav{\'e}}  \&
  {Weinberg}}{{Kere{\v s}} et~al.}{2009}]{Keres09}
{Kere{\v s}} D.,  {Katz} N.,  {Fardal} M.,  {Dav{\'e}} R.,   {Weinberg} D.~H.,
  2009, \mn@doi [\mnras] {10.1111/j.1365-2966.2009.14541.x}, \href
  {http://adsabs.harvard.edu/abs/2009MNRAS.395..160K} {395, 160}

\bibitem[\protect\citeauthoryear{{Kewley}, {Dopita}, {Sutherland}, {Heisler}
  \& {Trevena}}{{Kewley} et~al.}{2001}]{Kewley01}
{Kewley} L.~J.,  {Dopita} M.~A.,  {Sutherland} R.~S.,  {Heisler} C.~A.,
  {Trevena} J.,  2001, \mn@doi [\apj] {10.1086/321545}, \href
  {http://adsabs.harvard.edu/abs/2001ApJ...556..121K} {556, 121}

\bibitem[\protect\citeauthoryear{{Kewley}, {Rupke}, {Zahid}, {Geller}  \&
  {Barton}}{{Kewley} et~al.}{2010}]{Kewley10}
{Kewley} L.~J.,  {Rupke} D.,  {Zahid} H.~J.,  {Geller} M.~J.,   {Barton} E.~J.,
   2010, \mn@doi [\apjl] {10.1088/2041-8205/721/1/L48}, \href
  {http://adsabs.harvard.edu/abs/2010ApJ...721L..48K} {721, L48}

\bibitem[\protect\citeauthoryear{{Khare}, {Kulkarni}, {Lauroesch}, {York},
  {Crotts}  \& {Nakamura}}{{Khare} et~al.}{2004}]{Khare04}
{Khare} P.,  {Kulkarni} V.~P.,  {Lauroesch} J.~T.,  {York} D.~G.,  {Crotts}
  A.~P.~S.,   {Nakamura} O.,  2004, \mn@doi [\apj] {10.1086/424893}, \href
  {http://adsabs.harvard.edu/abs/2004ApJ...616...86K} {616, 86}

\bibitem[\protect\citeauthoryear{{Kobulnicky}, {Kennicutt}  \&
  {Pizagno}}{{Kobulnicky} et~al.}{1999}]{Kobulnicky99}
{Kobulnicky} H.~A.,  {Kennicutt} Jr. R.~C.,   {Pizagno} J.~L.,  1999, \mn@doi
  [\apj] {10.1086/306987}, \href
  {http://adsabs.harvard.edu/abs/1999ApJ...514..544K} {514, 544}

\bibitem[\protect\citeauthoryear{{Krogager}, {Fynbo}, {M{\o}ller}, {Ledoux},
  {Noterdaeme}, {Christensen}, {Milvang-Jensen}  \& {Sparre}}{{Krogager}
  et~al.}{2012}]{Krogager12}
{Krogager} J.-K.,  {Fynbo} J.~P.~U.,  {M{\o}ller} P.,  {Ledoux} C.,
  {Noterdaeme} P.,  {Christensen} L.,  {Milvang-Jensen} B.,   {Sparre} M.,
  2012, \mn@doi [\mnras] {10.1111/j.1745-3933.2012.01272.x}, \href
  {http://adsabs.harvard.edu/abs/2012MNRAS.424L...1K} {424, L1}

\bibitem[\protect\citeauthoryear{{Krogager} et~al.,}{{Krogager}
  et~al.}{2013}]{Krogager13}
{Krogager} J.-K.,  et~al., 2013, \mn@doi [\mnras] {10.1093/mnras/stt955}, \href
  {http://adsabs.harvard.edu/abs/2013MNRAS.433.3091K} {433, 3091}

\bibitem[\protect\citeauthoryear{{Kulkarni}, {Fall}, {Lauroesch}, {York},
  {Welty}, {Khare}  \& {Truran}}{{Kulkarni} et~al.}{2005}]{Kulkarni05}
{Kulkarni} V.~P.,  {Fall} S.~M.,  {Lauroesch} J.~T.,  {York} D.~G.,  {Welty}
  D.~E.,  {Khare} P.,   {Truran} J.~W.,  2005, \mn@doi [\apj] {10.1086/425956},
  \href {http://adsabs.harvard.edu/abs/2005ApJ...618...68K} {618, 68}

\bibitem[\protect\citeauthoryear{{Kulkarni}, {Woodgate}, {York}, {Thatte},
  {Meiring}, {Palunas}  \& {Wassell}}{{Kulkarni} et~al.}{2006}]{Kulkarni06}
{Kulkarni} V.~P.,  {Woodgate} B.~E.,  {York} D.~G.,  {Thatte} D.~G.,  {Meiring}
  J.,  {Palunas} P.,   {Wassell} E.,  2006, \mn@doi [\apj] {10.1086/497885},
  \href {http://adsabs.harvard.edu/abs/2006ApJ...636...30K} {636, 30}

\bibitem[\protect\citeauthoryear{{Lacy}, {Becker}, {Storrie-Lombardi}, {Gregg},
  {Urrutia}  \& {White}}{{Lacy} et~al.}{2003}]{Lacy03}
{Lacy} M.,  {Becker} R.~H.,  {Storrie-Lombardi} L.~J.,  {Gregg} M.~D.,
  {Urrutia} T.,   {White} R.~L.,  2003, \mn@doi [\aj] {10.1086/378957}, \href
  {http://adsabs.harvard.edu/abs/2003AJ....126.2230L} {126, 2230}

\bibitem[\protect\citeauthoryear{{Lanzetta}, {Wolfe}  \& {Turnshek}}{{Lanzetta}
  et~al.}{1995}]{Lanzetta95}
{Lanzetta} K.~M.,  {Wolfe} A.~M.,   {Turnshek} D.~A.,  1995, \mn@doi [\apj]
  {10.1086/175286}, \href {http://adsabs.harvard.edu/abs/1995ApJ...440..435L}
  {440, 435}

\bibitem[\protect\citeauthoryear{{Lanzetta} et~al.,}{{Lanzetta}
  et~al.}{1997}]{Lanzetta97}
{Lanzetta} K.~M.,  et~al., 1997, \mn@doi [\aj] {10.1086/118567}, \href
  {http://adsabs.harvard.edu/abs/1997AJ....114.1337L} {114, 1337}

\bibitem[\protect\citeauthoryear{{Ledoux}, {Bergeron}  \& {Petitjean}}{{Ledoux}
  et~al.}{2002a}]{Ledoux02a}
{Ledoux} C.,  {Bergeron} J.,   {Petitjean} P.,  2002a, \mn@doi [\aap]
  {10.1051/0004-6361:20020198}, \href
  {http://adsabs.harvard.edu/abs/2002A%26A...385..802L} {385, 802}

\bibitem[\protect\citeauthoryear{{Ledoux}, {Srianand}  \& {Petitjean}}{{Ledoux}
  et~al.}{2002b}]{Ledoux02}
{Ledoux} C.,  {Srianand} R.,   {Petitjean} P.,  2002b, \mn@doi [\aap]
  {10.1051/0004-6361:20021187}, \href
  {http://adsabs.harvard.edu/cgi-bin/nph-bib_query?bibcode=2002A26A...392..781L&db_key=AST}
  {392, 781}

\bibitem[\protect\citeauthoryear{{Ledoux}, {Petitjean}, {Fynbo}, {M{\o}ller}
  \& {Srianand}}{{Ledoux} et~al.}{2006}]{Ledoux06a}
{Ledoux} C.,  {Petitjean} P.,  {Fynbo} J.~P.~U.,  {M{\o}ller} P.,   {Srianand}
  R.,  2006, \mn@doi [\aap] {10.1051/0004-6361:20054242}, \href
  {http://adsabs.harvard.edu/cgi-bin/nph-bib_query?bibcode=2006A26A...457...71L&db_key=AST}
  {457, 71}

\bibitem[\protect\citeauthoryear{{Lehner} et~al.,}{{Lehner}
  et~al.}{2013}]{Lehner13}
{Lehner} N.,  et~al., 2013, \mn@doi [\apj] {10.1088/0004-637X/770/2/138}, \href
  {http://adsabs.harvard.edu/abs/2013ApJ...770..138L} {770, 138}

\bibitem[\protect\citeauthoryear{{Lehnert} \& {Heckman}}{{Lehnert} \&
  {Heckman}}{1996}]{Lehnert96}
{Lehnert} M.~D.,  {Heckman} T.~M.,  1996, \mn@doi [\apj] {10.1086/177180},
  \href {http://adsabs.harvard.edu/abs/1996ApJ...462..651L} {462, 651}

\bibitem[\protect\citeauthoryear{{Lu}, {Sargent}  \& {Barlow}}{{Lu}
  et~al.}{1997}]{Lu97}
{Lu} L.,  {Sargent} W.~L.~W.,   {Barlow} T.~A.,  1997, \apj, \href
  {http://adsabs.harvard.edu/abs/1997ApJ...484..131L} {484, 131}

\bibitem[\protect\citeauthoryear{{Lundgren} et~al.,}{{Lundgren}
  et~al.}{2009}]{Lundgren09}
{Lundgren} B.~F.,  et~al., 2009, \mn@doi [\apj] {10.1088/0004-637X/698/1/819},
  \href {http://adsabs.harvard.edu/abs/2009ApJ...698..819L} {698, 819}

\bibitem[\protect\citeauthoryear{{Maiolino} et~al.,}{{Maiolino}
  et~al.}{2008}]{Maiolino08}
{Maiolino} R.,  et~al., 2008, \mn@doi [\aap] {10.1051/0004-6361:200809678},
  \href {http://adsabs.harvard.edu/abs/2008A%26A...488..463M} {488, 463}

\bibitem[\protect\citeauthoryear{{Martin}}{{Martin}}{2005}]{Martin05}
{Martin} C.~L.,  2005, \mn@doi [\apj] {10.1086/427277}, \href
  {http://adsabs.harvard.edu/abs/2005ApJ...621..227M} {621, 227}

\bibitem[\protect\citeauthoryear{{Martin}, {Shapley}, {Coil}, {Kornei},
  {Bundy}, {Weiner}, {Noeske}  \& {Schiminovich}}{{Martin}
  et~al.}{2012}]{Martin12}
{Martin} C.~L.,  {Shapley} A.~E.,  {Coil} A.~L.,  {Kornei} K.~A.,  {Bundy} K.,
  {Weiner} B.~J.,  {Noeske} K.~G.,   {Schiminovich} D.,  2012, \mn@doi [\apj]
  {10.1088/0004-637X/760/2/127}, \href
  {http://adsabs.harvard.edu/abs/2012ApJ...760..127M} {760, 127}

\bibitem[\protect\citeauthoryear{{Meiring} et~al.,}{{Meiring}
  et~al.}{2006}]{Meiring06}
{Meiring} J.~D.,  et~al., 2006, \mn@doi [\mnras]
  {10.1111/j.1365-2966.2006.10500.x}, \href
  {http://adsabs.harvard.edu/abs/2006MNRAS.370...43M} {370, 43}

\bibitem[\protect\citeauthoryear{{Meiring}, {Lauroesch}, {Kulkarni},
  {P{\'e}roux}, {Khare}, {York}  \& {Crotts}}{{Meiring}
  et~al.}{2007}]{Meiring07}
{Meiring} J.~D.,  {Lauroesch} J.~T.,  {Kulkarni} V.~P.,  {P{\'e}roux} C.,
  {Khare} P.,  {York} D.~G.,   {Crotts} A.~P.~S.,  2007, \mn@doi [\mnras]
  {10.1111/j.1365-2966.2007.11521.x}, \href
  {http://adsabs.harvard.edu/abs/2007MNRAS.376..557M} {376, 557}

\bibitem[\protect\citeauthoryear{{Meiring}, {Lauroesch}, {Kulkarni},
  {P{\'e}roux}, {Khare}  \& {York}}{{Meiring} et~al.}{2009}]{Meiring09}
{Meiring} J.~D.,  {Lauroesch} J.~T.,  {Kulkarni} V.~P.,  {P{\'e}roux} C.,
  {Khare} P.,   {York} D.~G.,  2009, \mn@doi [\mnras]
  {10.1111/j.1365-2966.2009.15064.x}, \href
  {http://adsabs.harvard.edu/abs/2009MNRAS.397.2037M} {397, 2037}

\bibitem[\protect\citeauthoryear{{Moller} \& {Warren}}{{Moller} \&
  {Warren}}{1993}]{Moller93}
{Moller} P.,  {Warren} S.~J.,  1993, \aap, \href
  {http://adsabs.harvard.edu/abs/1993A%26A...270...43M} {270, 43}

\bibitem[\protect\citeauthoryear{{Moller} \& {Warren}}{{Moller} \&
  {Warren}}{1998}]{Moller98a}
{Moller} P.,  {Warren} S.~J.,  1998, \mnras, \href
  {http://adsabs.harvard.edu/abs/1998MNRAS.299..661M} {299, 661}

\bibitem[\protect\citeauthoryear{{M{\o}ller}, {Warren}, {Fall}, {Fynbo}  \&
  {Jakobsen}}{{M{\o}ller} et~al.}{2002}]{Moller02}
{M{\o}ller} P.,  {Warren} S.~J.,  {Fall} S.~M.,  {Fynbo} J.~U.,   {Jakobsen}
  P.,  2002, \mn@doi [\apj] {10.1086/340934}, \href
  {http://adsabs.harvard.edu/abs/2002ApJ...574...51M} {574, 51}

\bibitem[\protect\citeauthoryear{{M{\o}ller}, {Fynbo}, {Ledoux}  \&
  {Nilsson}}{{M{\o}ller} et~al.}{2013}]{Moller13}
{M{\o}ller} P.,  {Fynbo} J.~P.~U.,  {Ledoux} C.,   {Nilsson} K.~K.,  2013,
  \mn@doi [\mnras] {10.1093/mnras/stt067}, \href
  {http://adsabs.harvard.edu/abs/2013MNRAS.430.2680M} {430, 2680}

\bibitem[\protect\citeauthoryear{{Monier}, {Turnshek}  \& {Rao}}{{Monier}
  et~al.}{2009}]{Monier09}
{Monier} E.~M.,  {Turnshek} D.~A.,   {Rao} S.,  2009, \mn@doi [\mnras]
  {10.1111/j.1365-2966.2009.15000.x}, \href
  {http://adsabs.harvard.edu/abs/2009MNRAS.397..943M} {397, 943}

\bibitem[\protect\citeauthoryear{{Murga}, {Zhu}, {M{\'e}nard}  \&
  {Lan}}{{Murga} et~al.}{2015}]{Murga15}
{Murga} M.,  {Zhu} G.,  {M{\'e}nard} B.,   {Lan} T.-W.,  2015, \mn@doi [\mnras]
  {10.1093/mnras/stv1277}, \href
  {http://adsabs.harvard.edu/abs/2015MNRAS.452..511M} {452, 511}

\bibitem[\protect\citeauthoryear{{Neeleman}, {Wolfe}, {Prochaska}  \&
  {Rafelski}}{{Neeleman} et~al.}{2013}]{Neeleman13}
{Neeleman} M.,  {Wolfe} A.~M.,  {Prochaska} J.~X.,   {Rafelski} M.,  2013,
  \mn@doi [\apj] {10.1088/0004-637X/769/1/54}, \href
  {http://adsabs.harvard.edu/abs/2013ApJ...769...54N} {769, 54}

\bibitem[\protect\citeauthoryear{{Nestor}, {Rao}, {Turnshek}  \& {Vanden
  Berk}}{{Nestor} et~al.}{2003}]{Nestor03}
{Nestor} D.~B.,  {Rao} S.~M.,  {Turnshek} D.~A.,   {Vanden Berk} D.,  2003,
  \mn@doi [\apjl] {10.1086/378841}, \href
  {http://adsabs.harvard.edu/abs/2003ApJ...595L...5N} {595, L5}

\bibitem[\protect\citeauthoryear{{Nestor}, {Pettini}, {Hewett}, {Rao}  \&
  {Wild}}{{Nestor} et~al.}{2008}]{Nestor08}
{Nestor} D.~B.,  {Pettini} M.,  {Hewett} P.~C.,  {Rao} S.,   {Wild} V.,  2008,
  \mn@doi [\mnras] {10.1111/j.1365-2966.2008.13857.x}, \href
  {http://adsabs.harvard.edu/abs/2008MNRAS.390.1670N} {390, 1670}

\bibitem[\protect\citeauthoryear{{Nestor}, {Johnson}, {Wild}, {M{\'e}nard},
  {Turnshek}, {Rao}  \& {Pettini}}{{Nestor} et~al.}{2011}]{Nestor11}
{Nestor} D.~B.,  {Johnson} B.~D.,  {Wild} V.,  {M{\'e}nard} B.,  {Turnshek}
  D.~A.,  {Rao} S.,   {Pettini} M.,  2011, \mn@doi [\mnras]
  {10.1111/j.1365-2966.2010.17865.x}, \href
  {http://adsabs.harvard.edu/abs/2011MNRAS.412.1559N} {412, 1559}

\bibitem[\protect\citeauthoryear{{Noterdaeme}, {Petitjean}, {Ledoux}  \&
  {Srianand}}{{Noterdaeme} et~al.}{2009}]{Noterdaeme09dla}
{Noterdaeme} P.,  {Petitjean} P.,  {Ledoux} C.,   {Srianand} R.,  2009, \mn@doi
  [\aap] {10.1051/0004-6361/200912768}, \href
  {http://adsabs.harvard.edu/abs/2009A%26A...505.1087N} {505, 1087}

\bibitem[\protect\citeauthoryear{{Noterdaeme} et~al.,}{{Noterdaeme}
  et~al.}{2012a}]{Noterdaeme12}
{Noterdaeme} P.,  et~al., 2012a, \mn@doi [\aap] {10.1051/0004-6361/201118691},
  \href {http://adsabs.harvard.edu/abs/2012A%26A...540A..63N} {540, A63}

\bibitem[\protect\citeauthoryear{{Noterdaeme} et~al.,}{{Noterdaeme}
  et~al.}{2012b}]{Noterdaeme12dla}
{Noterdaeme} P.,  et~al., 2012b, \mn@doi [\aap] {10.1051/0004-6361/201220259},
  \href {http://adsabs.harvard.edu/abs/2012A%26A...547L...1N} {547, L1}

\bibitem[\protect\citeauthoryear{{Noterdaeme}, {Petitjean}, {P{\^a}ris}, {Cai},
  {Finley}, {Ge}, {Pieri}  \& {York}}{{Noterdaeme} et~al.}{2014}]{Noterdaeme14}
{Noterdaeme} P.,  {Petitjean} P.,  {P{\^a}ris} I.,  {Cai} Z.,  {Finley} H.,
  {Ge} J.,  {Pieri} M.~M.,   {York} D.~G.,  2014, \mn@doi [\aap]
  {10.1051/0004-6361/201322809}, \href
  {http://adsabs.harvard.edu/abs/2014A%26A...566A..24N} {566, A24}

\bibitem[\protect\citeauthoryear{{Oppenheimer} \& {Dav{\'e}}}{{Oppenheimer} \&
  {Dav{\'e}}}{2006}]{Oppenheimer06}
{Oppenheimer} B.~D.,  {Dav{\'e}} R.,  2006, \mn@doi [\mnras]
  {10.1111/j.1365-2966.2006.10989.x}, \href
  {http://adsabs.harvard.edu/abs/2006MNRAS.373.1265O} {373, 1265}

\bibitem[\protect\citeauthoryear{{Oppenheimer}, {Dav{\'e}}, {Kere{\v s}},
  {Fardal}, {Katz}, {Kollmeier}  \& {Weinberg}}{{Oppenheimer}
  et~al.}{2010}]{Oppenheimer10}
{Oppenheimer} B.~D.,  {Dav{\'e}} R.,  {Kere{\v s}} D.,  {Fardal} M.,  {Katz}
  N.,  {Kollmeier} J.~A.,   {Weinberg} D.~H.,  2010, \mn@doi [\mnras]
  {10.1111/j.1365-2966.2010.16872.x}, \href
  {http://adsabs.harvard.edu/abs/2010MNRAS.406.2325O} {406, 2325}

\bibitem[\protect\citeauthoryear{{Osterbrock}}{{Osterbrock}}{1989}]{Osterbrock89book}
{Osterbrock} D.~E.,  1989, {Astrophysics of Gaseous Nebulae and Active Galactic
  Nuclei (Mill Valley: University Science Books)}

\bibitem[\protect\citeauthoryear{{Pagel}, {Edmunds}, {Blackwell}, {Chun}  \&
  {Smith}}{{Pagel} et~al.}{1979}]{Pagel79}
{Pagel} B.~E.~J.,  {Edmunds} M.~G.,  {Blackwell} D.~E.,  {Chun} M.~S.,
  {Smith} G.,  1979, \mn@doi [\mnras] {10.1093/mnras/189.1.95}, \href
  {http://adsabs.harvard.edu/abs/1979MNRAS.189...95P} {189, 95}

\bibitem[\protect\citeauthoryear{{Pei}}{{Pei}}{1992}]{Pei92}
{Pei} Y.~C.,  1992, \mn@doi [\apj] {10.1086/171637}, \href
  {http://adsabs.harvard.edu/abs/1992ApJ...395..130P} {395, 130}

\bibitem[\protect\citeauthoryear{{P{\'e}roux}, {Storrie-Lombardi}, {McMahon},
  {Irwin}  \& {Hook}}{{P{\'e}roux} et~al.}{2001}]{Peroux01}
{P{\'e}roux} C.,  {Storrie-Lombardi} L.~J.,  {McMahon} R.~G.,  {Irwin} M.,
  {Hook} I.~M.,  2001, \mn@doi [\aj] {10.1086/319967}, \href
  {http://adsabs.harvard.edu/abs/2001AJ....121.1799P} {121, 1799}

\bibitem[\protect\citeauthoryear{{P{\'e}roux}, {McMahon}, {Storrie-Lombardi}
  \& {Irwin}}{{P{\'e}roux} et~al.}{2003}]{Peroux03}
{P{\'e}roux} C.,  {McMahon} R.~G.,  {Storrie-Lombardi} L.~J.,   {Irwin} M.~J.,
  2003, \mn@doi [\mnras] {10.1111/j.1365-2966.2003.07129.x}, \href
  {http://adsabs.harvard.edu/cgi-bin/nph-bib_query?bibcode=2003MNRAS.346.1103P&db_key=AST}
  {346, 1103}

\bibitem[\protect\citeauthoryear{{P{\'e}roux}, {Meiring}, {Kulkarni}, {Ferlet},
  {Khare}, {Lauroesch}, {Vladilo}  \& {York}}{{P{\'e}roux}
  et~al.}{2006}]{Peroux06}
{P{\'e}roux} C.,  {Meiring} J.~D.,  {Kulkarni} V.~P.,  {Ferlet} R.,  {Khare}
  P.,  {Lauroesch} J.~T.,  {Vladilo} G.,   {York} D.~G.,  2006, \mn@doi
  [\mnras] {10.1111/j.1365-2966.2006.10865.x}, \href
  {http://adsabs.harvard.edu/abs/2006MNRAS.372..369P} {372, 369}

\bibitem[\protect\citeauthoryear{{P{\'e}roux}, {Meiring}, {Kulkarni}, {Khare},
  {Lauroesch}, {Vladilo}  \& {York}}{{P{\'e}roux} et~al.}{2008}]{Peroux08}
{P{\'e}roux} C.,  {Meiring} J.~D.,  {Kulkarni} V.~P.,  {Khare} P.,  {Lauroesch}
  J.~T.,  {Vladilo} G.,   {York} D.~G.,  2008, \mn@doi [\mnras]
  {10.1111/j.1365-2966.2008.13186.x}, \href
  {http://adsabs.harvard.edu/abs/2008MNRAS.386.2209P} {386, 2209}

\bibitem[\protect\citeauthoryear{{P{\'e}roux}, {Bouch{\'e}}, {Kulkarni}, {York}
   \& {Vladilo}}{{P{\'e}roux} et~al.}{2011a}]{Peroux11a}
{P{\'e}roux} C.,  {Bouch{\'e}} N.,  {Kulkarni} V.~P.,  {York} D.~G.,
  {Vladilo} G.,  2011a, \mn@doi [\mnras] {10.1111/j.1365-2966.2010.17598.x},
  \href {http://adsabs.harvard.edu/abs/2011MNRAS.410.2237P} {410, 2237}

\bibitem[\protect\citeauthoryear{{P{\'e}roux}, {Bouch{\'e}}, {Kulkarni}, {York}
   \& {Vladilo}}{{P{\'e}roux} et~al.}{2011b}]{Peroux11b}
{P{\'e}roux} C.,  {Bouch{\'e}} N.,  {Kulkarni} V.~P.,  {York} D.~G.,
  {Vladilo} G.,  2011b, \mn@doi [\mnras] {10.1111/j.1365-2966.2010.17597.x},
  \href {http://adsabs.harvard.edu/abs/2011MNRAS.410.2251P} {410, 2251}

\bibitem[\protect\citeauthoryear{{P{\'e}roux}, {Bouch{\'e}}, {Kulkarni}, {York}
   \& {Vladilo}}{{P{\'e}roux} et~al.}{2012}]{Peroux12}
{P{\'e}roux} C.,  {Bouch{\'e}} N.,  {Kulkarni} V.~P.,  {York} D.~G.,
  {Vladilo} G.,  2012, \mn@doi [\mnras] {10.1111/j.1365-2966.2011.19947.x},
  \href {http://adsabs.harvard.edu/abs/2012MNRAS.419.3060P} {419, 3060}

\bibitem[\protect\citeauthoryear{{P{\'e}roux}, {Bouch{\'e}}, {Kulkarni}  \&
  {York}}{{P{\'e}roux} et~al.}{2013}]{Peroux13}
{P{\'e}roux} C.,  {Bouch{\'e}} N.,  {Kulkarni} V.~P.,   {York} D.~G.,  2013,
  \mn@doi [\mnras] {10.1093/mnras/stt1760}, \href
  {http://adsabs.harvard.edu/abs/2013MNRAS.436.2650P} {436, 2650}

\bibitem[\protect\citeauthoryear{{P{\'e}roux}, {Kulkarni}  \&
  {York}}{{P{\'e}roux} et~al.}{2014}]{Peroux14}
{P{\'e}roux} C.,  {Kulkarni} V.~P.,   {York} D.~G.,  2014, \mn@doi [\mnras]
  {10.1093/mnras/stt2084}, \href
  {http://adsabs.harvard.edu/abs/2014MNRAS.437.3144P} {437, 3144}

\bibitem[\protect\citeauthoryear{{Peroux} et~al.,}{{Peroux}
  et~al.}{2016}]{Peroux16}
{Peroux} C.,  et~al., 2016, preprint, \href
  {http://adsabs.harvard.edu/abs/2016arXiv160102796P} {} (\mn@eprint {arXiv}
  {1601.02796})

\bibitem[\protect\citeauthoryear{{Petitjean}, {Bergeron}  \&
  {Puget}}{{Petitjean} et~al.}{1992}]{Petitjean92}
{Petitjean} P.,  {Bergeron} J.,   {Puget} J.~L.,  1992, \aap, \href
  {http://adsabs.harvard.edu/abs/1992A%26A...265..375P} {265, 375}

\bibitem[\protect\citeauthoryear{{Pettini} \& {Pagel}}{{Pettini} \&
  {Pagel}}{2004}]{Pettini04}
{Pettini} M.,  {Pagel} B.~E.~J.,  2004, \mn@doi [\mnras]
  {10.1111/j.1365-2966.2004.07591.x}, \href
  {http://adsabs.harvard.edu/abs/2004MNRAS.348L..59P} {348, L59}

\bibitem[\protect\citeauthoryear{{Pettini}, {Ellison}, {Steidel}, {Shapley}  \&
  {Bowen}}{{Pettini} et~al.}{2000}]{Pettini00}
{Pettini} M.,  {Ellison} S.~L.,  {Steidel} C.~C.,  {Shapley} A.~E.,   {Bowen}
  D.~V.,  2000, \mn@doi [\apj] {10.1086/308562}, \href
  {http://adsabs.harvard.edu/abs/2000ApJ...532...65P} {532, 65}

\bibitem[\protect\citeauthoryear{{Pilkington} et~al.,}{{Pilkington}
  et~al.}{2012}]{Pilkington12}
{Pilkington} K.,  et~al., 2012, \mn@doi [\aap] {10.1051/0004-6361/201117466},
  \href {http://adsabs.harvard.edu/abs/2012A%26A...540A..56P} {540, A56}

\bibitem[\protect\citeauthoryear{{Piontek} \& {Steinmetz}}{{Piontek} \&
  {Steinmetz}}{2011}]{Piontek11}
{Piontek} F.,  {Steinmetz} M.,  2011, \mn@doi [\mnras]
  {10.1111/j.1365-2966.2010.17637.x}, \href
  {http://adsabs.harvard.edu/abs/2011MNRAS.410.2625P} {410, 2625}

\bibitem[\protect\citeauthoryear{{Pontzen} et~al.,}{{Pontzen}
  et~al.}{2008}]{Pontzen08}
{Pontzen} A.,  et~al., 2008, \mn@doi [\mnras]
  {10.1111/j.1365-2966.2008.13782.x}, \href
  {http://adsabs.harvard.edu/abs/2008MNRAS.390.1349P} {390, 1349}

\bibitem[\protect\citeauthoryear{{Prochaska}}{{Prochaska}}{1999}]{Prochaska99}
{Prochaska} J.~X.,  1999, \mn@doi [\apjl] {10.1086/311849}, \href
  {http://adsabs.harvard.edu/abs/1999ApJ...511L..71P} {511, L71}

\bibitem[\protect\citeauthoryear{{Prochaska} \& {Wolfe}}{{Prochaska} \&
  {Wolfe}}{1997}]{Prochaska97}
{Prochaska} J.~X.,  {Wolfe} A.~M.,  1997, \mn@doi [\apj] {10.1086/304591},
  \href {http://adsabs.harvard.edu/abs/1997ApJ...487...73P} {487, 73}

\bibitem[\protect\citeauthoryear{{Prochaska}, {Herbert-Fort}  \&
  {Wolfe}}{{Prochaska} et~al.}{2005}]{Prochaska05}
{Prochaska} J.~X.,  {Herbert-Fort} S.,   {Wolfe} A.~M.,  2005, \mn@doi [\apj]
  {10.1086/497287}, \href
  {http://adsabs.harvard.edu/cgi-bin/nph-bib_query?bibcode=2005ApJ...635..123P&db_key=AST}
  {635, 123}

\bibitem[\protect\citeauthoryear{{Puchwein} \& {Springel}}{{Puchwein} \&
  {Springel}}{2013}]{Puchwein13}
{Puchwein} E.,  {Springel} V.,  2013, \mn@doi [\mnras] {10.1093/mnras/sts243},
  \href {http://adsabs.harvard.edu/abs/2013MNRAS.428.2966P} {428, 2966}

\bibitem[\protect\citeauthoryear{{Queyrel} et~al.,}{{Queyrel}
  et~al.}{2012}]{Queyrel12}
{Queyrel} J.,  et~al., 2012, \mn@doi [\aap] {10.1051/0004-6361/201117718},
  \href {http://adsabs.harvard.edu/abs/2012A%26A...539A..93Q} {539, A93}

\bibitem[\protect\citeauthoryear{{Quider}, {Nestor}, {Turnshek}, {Rao},
  {Monier}, {Weyant}  \& {Busche}}{{Quider} et~al.}{2011}]{Quider11}
{Quider} A.~M.,  {Nestor} D.~B.,  {Turnshek} D.~A.,  {Rao} S.~M.,  {Monier}
  E.~M.,  {Weyant} A.~N.,   {Busche} J.~R.,  2011, \mn@doi [\aj]
  {10.1088/0004-6256/141/4/137}, \href
  {http://adsabs.harvard.edu/abs/2011AJ....141..137Q} {141, 137}

\bibitem[\protect\citeauthoryear{{Quiret} et~al.,}{{Quiret}
  et~al.}{2016}]{Quiret16}
{Quiret} S.,  et~al., 2016, preprint, \href
  {http://adsabs.harvard.edu/abs/2016arXiv160202564Q} {} (\mn@eprint {arXiv}
  {1602.02564})

\bibitem[\protect\citeauthoryear{{Rahmani} et~al.,}{{Rahmani}
  et~al.}{2013}]{Rahmani13}
{Rahmani} H.,  et~al., 2013, \mn@doi [\mnras] {10.1093/mnras/stt1356}, \href
  {http://adsabs.harvard.edu/abs/2013MNRAS.435..861R} {435, 861}

\bibitem[\protect\citeauthoryear{{Rahmati}, {Schaye}, {Bower}, {Crain},
  {Furlong}, {Schaller}  \& {Theuns}}{{Rahmati} et~al.}{2015}]{Rahmati15}
{Rahmati} A.,  {Schaye} J.,  {Bower} R.~G.,  {Crain} R.~A.,  {Furlong} M.,
  {Schaller} M.,   {Theuns} T.,  2015, \mn@doi [\mnras]
  {10.1093/mnras/stv1414}, \href
  {http://adsabs.harvard.edu/abs/2015MNRAS.452.2034R} {452, 2034}

\bibitem[\protect\citeauthoryear{{Rao} \& {Turnshek}}{{Rao} \&
  {Turnshek}}{2000}]{Rao00}
{Rao} S.~M.,  {Turnshek} D.~A.,  2000, \mn@doi [\apjs] {10.1086/317344}, \href
  {http://adsabs.harvard.edu/abs/2000ApJS..130....1R} {130, 1}

\bibitem[\protect\citeauthoryear{{Rao}, {Nestor}, {Turnshek}, {Lane}, {Monier}
  \& {Bergeron}}{{Rao} et~al.}{2003}]{Rao03}
{Rao} S.~M.,  {Nestor} D.~B.,  {Turnshek} D.~A.,  {Lane} W.~M.,  {Monier}
  E.~M.,   {Bergeron} J.,  2003, \mn@doi [\apj] {10.1086/377331}, \href
  {http://adsabs.harvard.edu/abs/2003ApJ...595...94R} {595, 94}

\bibitem[\protect\citeauthoryear{{Rao}, {Turnshek}  \& {Nestor}}{{Rao}
  et~al.}{2006}]{Rao06}
{Rao} S.~M.,  {Turnshek} D.~A.,   {Nestor} D.~B.,  2006, \mn@doi [\apj]
  {10.1086/498132}, \href {http://adsabs.harvard.edu/abs/2006ApJ...636..610R}
  {636, 610}

\bibitem[\protect\citeauthoryear{{Rao}, {Belfort-Mihalyi}, {Turnshek},
  {Monier}, {Nestor}  \& {Quider}}{{Rao} et~al.}{2011}]{Rao11}
{Rao} S.~M.,  {Belfort-Mihalyi} M.,  {Turnshek} D.~A.,  {Monier} E.~M.,
  {Nestor} D.~B.,   {Quider} A.,  2011, \mn@doi [\mnras]
  {10.1111/j.1365-2966.2011.19119.x}, \href
  {http://adsabs.harvard.edu/abs/2011MNRAS.416.1215R} {416, 1215}

\bibitem[\protect\citeauthoryear{{Richter}, {Krause}, {Fechner}, {Charlton}  \&
  {Murphy}}{{Richter} et~al.}{2011}]{Richter11}
{Richter} P.,  {Krause} F.,  {Fechner} C.,  {Charlton} J.~C.,   {Murphy} M.~T.,
   2011, \mn@doi [\aap] {10.1051/0004-6361/201015566}, \href
  {http://adsabs.harvard.edu/abs/2011A%26A...528A..12R} {528, A12}

\bibitem[\protect\citeauthoryear{{Rubin}, {Prochaska}, {Koo}  \&
  {Phillips}}{{Rubin} et~al.}{2012}]{Rubin12}
{Rubin} K.~H.~R.,  {Prochaska} J.~X.,  {Koo} D.~C.,   {Phillips} A.~C.,  2012,
  \mn@doi [\apjl] {10.1088/2041-8205/747/2/L26}, \href
  {http://adsabs.harvard.edu/abs/2012ApJ...747L..26R} {747, L26}

\bibitem[\protect\citeauthoryear{{Rupke}, {Kewley}  \& {Barnes}}{{Rupke}
  et~al.}{2010}]{Rupke10}
{Rupke} D.~S.~N.,  {Kewley} L.~J.,   {Barnes} J.~E.,  2010, \mn@doi [\apjl]
  {10.1088/2041-8205/710/2/L156}, \href
  {http://adsabs.harvard.edu/abs/2010ApJ...710L.156R} {710, L156}

\bibitem[\protect\citeauthoryear{{Ryan-Weber}, {Pettini}, {Madau}  \&
  {Zych}}{{Ryan-Weber} et~al.}{2009}]{Ryan-Weber09}
{Ryan-Weber} E.~V.,  {Pettini} M.,  {Madau} P.,   {Zych} B.~J.,  2009, \mn@doi
  [\mnras] {10.1111/j.1365-2966.2009.14618.x}, \href
  {http://adsabs.harvard.edu/abs/2009MNRAS.395.1476R} {395, 1476}

\bibitem[\protect\citeauthoryear{{S{\'a}nchez-Ram{\'{\i}}rez}
  et~al.,}{{S{\'a}nchez-Ram{\'{\i}}rez} et~al.}{2016}]{Sanchez-Ramirez16}
{S{\'a}nchez-Ram{\'{\i}}rez} R.,  et~al., 2016, \mn@doi [\mnras]
  {10.1093/mnras/stv2732}, \href
  {http://adsabs.harvard.edu/abs/2016MNRAS.456.4488S} {456, 4488}

\bibitem[\protect\citeauthoryear{{Sardane}, {Turnshek}  \& {Rao}}{{Sardane}
  et~al.}{2015}]{Sardane15}
{Sardane} G.~M.,  {Turnshek} D.~A.,   {Rao} S.~M.,  2015, \mn@doi [\mnras]
  {10.1093/mnras/stv1506}, \href
  {http://adsabs.harvard.edu/abs/2015MNRAS.452.3192S} {452, 3192}

\bibitem[\protect\citeauthoryear{{Sargent}, {Steidel}  \&
  {Boksenberg}}{{Sargent} et~al.}{1989}]{Sargent89}
{Sargent} W.~L.~W.,  {Steidel} C.~C.,   {Boksenberg} A.,  1989, \mn@doi [\apjs]
  {10.1086/191326}, \href
  {http://adsabs.harvard.edu/cgi-bin/nph-bib_query?bibcode=1989ApJS...69..703S&db_key=AST}
  {69, 703}

\bibitem[\protect\citeauthoryear{{Savaglio} et~al.,}{{Savaglio}
  et~al.}{2005}]{Savaglio05}
{Savaglio} S.,  et~al., 2005, \mn@doi [\apj] {10.1086/497331}, \href
  {http://adsabs.harvard.edu/abs/2005ApJ...635..260S} {635, 260}

\bibitem[\protect\citeauthoryear{{Schaye} et~al.,}{{Schaye}
  et~al.}{2015}]{Schaye15}
{Schaye} J.,  et~al., 2015, \mn@doi [\mnras] {10.1093/mnras/stu2058}, \href
  {http://adsabs.harvard.edu/abs/2015MNRAS.446..521S} {446, 521}

\bibitem[\protect\citeauthoryear{{Sch{\"o}nebeck}, {Puzia}, {Pasquali},
  {Grebel}, {Kissler-Patig}, {Kuntschner}, {Lyubenova}  \&
  {Perina}}{{Sch{\"o}nebeck} et~al.}{2014}]{Schonebeck14}
{Sch{\"o}nebeck} F.,  {Puzia} T.~H.,  {Pasquali} A.,  {Grebel} E.~K.,
  {Kissler-Patig} M.,  {Kuntschner} H.,  {Lyubenova} M.,   {Perina} S.,  2014,
  \mn@doi [\aap] {10.1051/0004-6361/201424196}, \href
  {http://adsabs.harvard.edu/abs/2014A%26A...572A..13S} {572, A13}

\bibitem[\protect\citeauthoryear{{Schroetter}, {Bouch{\'e}}, {P{\'e}roux},
  {Murphy}, {Contini}  \& {Finley}}{{Schroetter} et~al.}{2015}]{Schroetter15}
{Schroetter} I.,  {Bouch{\'e}} N.,  {P{\'e}roux} C.,  {Murphy} M.~T.,
  {Contini} T.,   {Finley} H.,  2015, \mn@doi [\apj]
  {10.1088/0004-637X/804/2/83}, \href
  {http://adsabs.harvard.edu/abs/2015ApJ...804...83S} {804, 83}

\bibitem[\protect\citeauthoryear{{Schulte-Ladbeck}, {Rao}, {Drozdovsky},
  {Turnshek}, {Nestor}  \& {Pettini}}{{Schulte-Ladbeck}
  et~al.}{2004}]{Schulte-Ladbeck04}
{Schulte-Ladbeck} R.~E.,  {Rao} S.~M.,  {Drozdovsky} I.~O.,  {Turnshek} D.~A.,
  {Nestor} D.~B.,   {Pettini} M.,  2004, \mn@doi [\apj] {10.1086/380094}, \href
  {http://adsabs.harvard.edu/abs/2004ApJ...600..613S} {600, 613}

\bibitem[\protect\citeauthoryear{{Schulte-Ladbeck}, {K{\"o}nig}, {Miller},
  {Hopkins}, {Drozdovsky}, {Turnshek}  \& {Hopp}}{{Schulte-Ladbeck}
  et~al.}{2005}]{Schulte-Ladbeck05}
{Schulte-Ladbeck} R.~E.,  {K{\"o}nig} B.,  {Miller} C.~J.,  {Hopkins} A.~M.,
  {Drozdovsky} I.~O.,  {Turnshek} D.~A.,   {Hopp} U.,  2005, \mn@doi [\apjl]
  {10.1086/431324}, \href {http://adsabs.harvard.edu/abs/2005ApJ...625L..79S}
  {625, L79}

\bibitem[\protect\citeauthoryear{{Shull}, {Danforth}  \& {Tilton}}{{Shull}
  et~al.}{2014}]{Shull14}
{Shull} J.~M.,  {Danforth} C.~W.,   {Tilton} E.~M.,  2014, \mn@doi [\apj]
  {10.1088/0004-637X/796/1/49}, \href
  {http://adsabs.harvard.edu/abs/2014ApJ...796...49S} {796, 49}

\bibitem[\protect\citeauthoryear{{Sijacki}, {Springel}, {Di Matteo}  \&
  {Hernquist}}{{Sijacki} et~al.}{2007}]{Sijacki07}
{Sijacki} D.,  {Springel} V.,  {Di Matteo} T.,   {Hernquist} L.,  2007, \mn@doi
  [\mnras] {10.1111/j.1365-2966.2007.12153.x}, \href
  {http://adsabs.harvard.edu/abs/2007MNRAS.380..877S} {380, 877}

\bibitem[\protect\citeauthoryear{{Simcoe}, {Sargent}, {Rauch}  \&
  {Becker}}{{Simcoe} et~al.}{2006}]{Simcoe06}
{Simcoe} R.~A.,  {Sargent} W.~L.~W.,  {Rauch} M.,   {Becker} G.,  2006, \mn@doi
  [\apj] {10.1086/498441}, \href
  {http://adsabs.harvard.edu/abs/2006ApJ...637..648S} {637, 648}

\bibitem[\protect\citeauthoryear{{Simpson}, {Bryan}, {Hummels}  \&
  {Ostriker}}{{Simpson} et~al.}{2015}]{Simpson-c15}
{Simpson} C.~M.,  {Bryan} G.~L.,  {Hummels} C.,   {Ostriker} J.~P.,  2015,
  \mn@doi [\apj] {10.1088/0004-637X/809/1/69}, \href
  {http://adsabs.harvard.edu/abs/2015ApJ...809...69S} {809, 69}

\bibitem[\protect\citeauthoryear{{Smette} et~al.,}{{Smette}
  et~al.}{2015}]{Smette15}
{Smette} A.,  et~al., 2015, \mn@doi [\aap] {10.1051/0004-6361/201423932}, \href
  {http://adsabs.harvard.edu/abs/2015A%26A...576A..77S} {576, A77}

\bibitem[\protect\citeauthoryear{{Som}, {Kulkarni}, {Meiring}, {York},
  {P{\'e}roux}, {Lauroesch}, {Aller}  \& {Khare}}{{Som} et~al.}{2015}]{Som15}
{Som} D.,  {Kulkarni} V.~P.,  {Meiring} J.,  {York} D.~G.,  {P{\'e}roux} C.,
  {Lauroesch} J.~T.,  {Aller} M.~C.,   {Khare} P.,  2015, \mn@doi [\apj]
  {10.1088/0004-637X/806/1/25}, \href
  {http://adsabs.harvard.edu/abs/2015ApJ...806...25S} {806, 25}

\bibitem[\protect\citeauthoryear{{Songaila} \& {Cowie}}{{Songaila} \&
  {Cowie}}{2010}]{Songaila10}
{Songaila} A.,  {Cowie} L.~L.,  2010, \mn@doi [\apj]
  {10.1088/0004-637X/721/2/1448}, \href
  {http://adsabs.harvard.edu/abs/2010ApJ...721.1448S} {721, 1448}

\bibitem[\protect\citeauthoryear{{Springel} et~al.,}{{Springel}
  et~al.}{2005}]{Springel05}
{Springel} V.,  et~al., 2005, \mn@doi [\nat] {10.1038/nature03597}, \href
  {http://adsabs.harvard.edu/abs/2005Natur.435..629S} {435, 629}

\bibitem[\protect\citeauthoryear{{Srianand}, {Hussain}, {Noterdaeme},
  {Petitjean}, {Kr{\"u}hler}, {Japelj}, {P{\^a}ris}  \& {Kashikawa}}{{Srianand}
  et~al.}{2016}]{Srianand16}
{Srianand} R.,  {Hussain} T.,  {Noterdaeme} P.,  {Petitjean} P.,  {Kr{\"u}hler}
  T.,  {Japelj} J.,  {P{\^a}ris} I.,   {Kashikawa} N.,  2016, preprint, \href
  {http://adsabs.harvard.edu/abs/2016MNRAS.tmp..741S} {} (\mn@eprint {arXiv}
  {1604.06475})

\bibitem[\protect\citeauthoryear{{Steidel}, {Dickinson}, {Meyer}, {Adelberger}
  \& {Sembach}}{{Steidel} et~al.}{1997}]{Steidel97}
{Steidel} C.~C.,  {Dickinson} M.,  {Meyer} D.~M.,  {Adelberger} K.~L.,
  {Sembach} K.~R.,  1997, \apj, \href
  {http://adsabs.harvard.edu/abs/1997ApJ...480..568S} {480, 568}

\bibitem[\protect\citeauthoryear{{Steidel}, {Erb}, {Shapley}, {Pettini},
  {Reddy}, {Bogosavljevi{\'c}}, {Rudie}  \& {Rakic}}{{Steidel}
  et~al.}{2010}]{Steidel10}
{Steidel} C.~C.,  {Erb} D.~K.,  {Shapley} A.~E.,  {Pettini} M.,  {Reddy} N.,
  {Bogosavljevi{\'c}} M.,  {Rudie} G.~C.,   {Rakic} O.,  2010, \mn@doi [\apj]
  {10.1088/0004-637X/717/1/289}, \href
  {http://adsabs.harvard.edu/abs/2010ApJ...717..289S} {717, 289}

\bibitem[\protect\citeauthoryear{{Stewart}, {Kaufmann}, {Bullock}, {Barton},
  {Maller}, {Diemand}  \& {Wadsley}}{{Stewart} et~al.}{2011}]{Stewart11}
{Stewart} K.~R.,  {Kaufmann} T.,  {Bullock} J.~S.,  {Barton} E.~J.,  {Maller}
  A.~H.,  {Diemand} J.,   {Wadsley} J.,  2011, \mn@doi [\apjl]
  {10.1088/2041-8205/735/1/L1}, \href
  {http://adsabs.harvard.edu/abs/2011ApJ...735L...1S} {735, L1}

\bibitem[\protect\citeauthoryear{{Straka} et~al.,}{{Straka}
  et~al.}{2015}]{Straka15}
{Straka} L.~A.,  et~al., 2015, \mn@doi [\mnras] {10.1093/mnras/stu2739}, \href
  {http://adsabs.harvard.edu/abs/2015MNRAS.447.3856S} {447, 3856}

\bibitem[\protect\citeauthoryear{{Straka}, {Johnson}, {York}, {Bowen},
  {Florian}, {Kulkarni}, {Lundgren}  \& {P{\'e}roux}}{{Straka}
  et~al.}{2016}]{Straka16}
{Straka} L.~A.,  {Johnson} S.,  {York} D.~G.,  {Bowen} D.~V.,  {Florian} M.,
  {Kulkarni} V.~P.,  {Lundgren} B.,   {P{\'e}roux} C.,  2016, \mn@doi [\mnras]
  {10.1093/mnras/stw508}, \href
  {http://adsabs.harvard.edu/abs/2016MNRAS.458.3760S} {458, 3760}

\bibitem[\protect\citeauthoryear{{Suresh}, {Bird}, {Vogelsberger}, {Genel},
  {Torrey}, {Sijacki}, {Springel}  \& {Hernquist}}{{Suresh}
  et~al.}{2015}]{Suresh15}
{Suresh} J.,  {Bird} S.,  {Vogelsberger} M.,  {Genel} S.,  {Torrey} P.,
  {Sijacki} D.,  {Springel} V.,   {Hernquist} L.,  2015, \mn@doi [\mnras]
  {10.1093/mnras/stu2762}, \href
  {http://adsabs.harvard.edu/abs/2015MNRAS.448..895S} {448, 895}

\bibitem[\protect\citeauthoryear{{Tremonti} et~al.,}{{Tremonti}
  et~al.}{2004}]{Tremonti04}
{Tremonti} C.~A.,  et~al., 2004, \mn@doi [\apj] {10.1086/423264}, \href
  {http://adsabs.harvard.edu/abs/2004ApJ...613..898T} {613, 898}

\bibitem[\protect\citeauthoryear{{Tumlinson} et~al.,}{{Tumlinson}
  et~al.}{2011}]{Tumlinson11}
{Tumlinson} J.,  et~al., 2011, \mn@doi [Science] {10.1126/science.1209840},
  \href {http://adsabs.harvard.edu/abs/2011Sci...334..948T} {334, 948}

\bibitem[\protect\citeauthoryear{{Turnshek}, {Rao}, {Nestor}, {Lane}, {Monier},
  {Bergeron}  \& {Smette}}{{Turnshek} et~al.}{2001}]{Turnshek01}
{Turnshek} D.~A.,  {Rao} S.,  {Nestor} D.,  {Lane} W.,  {Monier} E.,
  {Bergeron} J.,   {Smette} A.,  2001, \mn@doi [\apj] {10.1086/320660}, \href
  {http://adsabs.harvard.edu/abs/2001ApJ...553..288T} {553, 288}

\bibitem[\protect\citeauthoryear{{Turnshek}, {Monier}, {Rao}, {Hamilton},
  {Sardane}  \& {Held}}{{Turnshek} et~al.}{2015}]{Turnshek15}
{Turnshek} D.~A.,  {Monier} E.~M.,  {Rao} S.~M.,  {Hamilton} T.~S.,  {Sardane}
  G.~M.,   {Held} R.,  2015, \mn@doi [\mnras] {10.1093/mnras/stv224}, \href
  {http://adsabs.harvard.edu/abs/2015MNRAS.449.1536T} {449, 1536}

\bibitem[\protect\citeauthoryear{{Tytler}}{{Tytler}}{1982}]{Tytler82}
{Tytler} D.,  1982, \mn@doi [\nat] {10.1038/298427a0}, \href
  {http://adsabs.harvard.edu/abs/1982Natur.298..427T} {298, 427}

\bibitem[\protect\citeauthoryear{{Vernet} et~al.,}{{Vernet}
  et~al.}{2011}]{Vernet11}
{Vernet} J.,  et~al., 2011, \mn@doi [\aap] {10.1051/0004-6361/201117752}, \href
  {http://adsabs.harvard.edu/abs/2011A%26A...536A.105V} {536, A105}

\bibitem[\protect\citeauthoryear{{Vladilo}}{{Vladilo}}{1998}]{Vladilo98}
{Vladilo} G.,  1998, \mn@doi [\apj] {10.1086/305148}, \href
  {http://adsabs.harvard.edu/abs/1998ApJ...493..583V} {493, 583}

\bibitem[\protect\citeauthoryear{{Vogelsberger} et~al.,}{{Vogelsberger}
  et~al.}{2014}]{Vogelsberger14_1}
{Vogelsberger} M.,  et~al., 2014, \mn@doi [\nat] {10.1038/nature13316}, \href
  {http://adsabs.harvard.edu/abs/2014Natur.509..177V} {509, 177}

\bibitem[\protect\citeauthoryear{{Wakker} \& {Mathis}}{{Wakker} \&
  {Mathis}}{2000}]{Wakker00}
{Wakker} B.~P.,  {Mathis} J.~S.,  2000, \mn@doi [\apjl] {10.1086/317316}, \href
  {http://adsabs.harvard.edu/abs/2000ApJ...544L.107W} {544, L107}

\bibitem[\protect\citeauthoryear{{Weatherley}, {Warren}, {M{\o}ller}, {Fall},
  {Fynbo}  \& {Croom}}{{Weatherley} et~al.}{2005}]{Weatherley05}
{Weatherley} S.~J.,  {Warren} S.~J.,  {M{\o}ller} P.,  {Fall} S.~M.,  {Fynbo}
  J.~U.,   {Croom} S.~M.,  2005, \mn@doi [\mnras]
  {10.1111/j.1365-2966.2005.08838.x}, \href
  {http://adsabs.harvard.edu/abs/2005MNRAS.358..985W} {358, 985}

\bibitem[\protect\citeauthoryear{{Weiner} et~al.,}{{Weiner}
  et~al.}{2009}]{Weiner09}
{Weiner} B.~J.,  et~al., 2009, \mn@doi [\apj] {10.1088/0004-637X/692/1/187},
  \href {http://adsabs.harvard.edu/abs/2009ApJ...692..187W} {692, 187}

\bibitem[\protect\citeauthoryear{{Wiersma}, {Schaye}, {Dalla Vecchia}, {Booth},
  {Theuns}  \& {Aguirre}}{{Wiersma} et~al.}{2010}]{Wiersma10}
{Wiersma} R.~P.~C.,  {Schaye} J.,  {Dalla Vecchia} C.,  {Booth} C.~M.,
  {Theuns} T.,   {Aguirre} A.,  2010, \mn@doi [\mnras]
  {10.1111/j.1365-2966.2010.17299.x}, \href
  {http://adsabs.harvard.edu/abs/2010MNRAS.409..132W} {409, 132}

\bibitem[\protect\citeauthoryear{{Wild}, {Hewett}  \& {Pettini}}{{Wild}
  et~al.}{2007}]{Wild07}
{Wild} V.,  {Hewett} P.~C.,   {Pettini} M.,  2007, \mn@doi [\mnras]
  {10.1111/j.1365-2966.2006.11146.x}, \href
  {http://adsabs.harvard.edu/abs/2007MNRAS.374..292W} {374, 292}

\bibitem[\protect\citeauthoryear{{Wolfe}, {Turnshek}, {Smith}  \&
  {Cohen}}{{Wolfe} et~al.}{1986}]{Wolfe86}
{Wolfe} A.~M.,  {Turnshek} D.~A.,  {Smith} H.~E.,   {Cohen} R.~D.,  1986,
  \mn@doi [\apjs] {10.1086/191114}, \href
  {http://adsabs.harvard.edu/cgi-bin/nph-bib_query?bibcode=1986ApJS...61..249W&db_key=AST}
  {61, 249}

\bibitem[\protect\citeauthoryear{{Wolfe}, {Gawiser}  \& {Prochaska}}{{Wolfe}
  et~al.}{2005}]{Wolfe05}
{Wolfe} A.~M.,  {Gawiser} E.,   {Prochaska} J.~X.,  2005, \araa, \href
  {http://adsabs.harvard.edu/cgi-bin/nph-bib_query?bibcode=2005ARA%26A..43..861W&db_key=AST}
  {43, 861}

\bibitem[\protect\citeauthoryear{{Wolfe}, {Prochaska}, {Jorgenson}  \&
  {Rafelski}}{{Wolfe} et~al.}{2008}]{Wolfe08}
{Wolfe} A.~M.,  {Prochaska} J.~X.,  {Jorgenson} R.~A.,   {Rafelski} M.,  2008,
  \mn@doi [\apj] {10.1086/588090}, \href
  {http://ads.iucaa.ernet.in/abs/2008ApJ...681..881W} {681, 881}

\bibitem[\protect\citeauthoryear{{Wolfe}, {Lockman}  \& {Pisano}}{{Wolfe}
  et~al.}{2015}]{Wolfe_s15}
{Wolfe} S.~A.,  {Lockman} F.~J.,   {Pisano} D.~J.,  2015, preprint, \href
  {http://adsabs.harvard.edu/abs/2015arXiv150705237W} {} (\mn@eprint {arXiv}
  {1507.05237})

\bibitem[\protect\citeauthoryear{{Zafar}, {P{\'e}roux}, {Popping}, {Milliard},
  {Deharveng}  \& {Frank}}{{Zafar} et~al.}{2013}]{Zafar13}
{Zafar} T.,  {P{\'e}roux} C.,  {Popping} A.,  {Milliard} B.,  {Deharveng}
  J.-M.,   {Frank} S.,  2013, \mn@doi [\aap] {10.1051/0004-6361/201321154},
  \href {http://adsabs.harvard.edu/abs/2013A%26A...556A.141Z} {556, A141}

\bibitem[\protect\citeauthoryear{{Zych}, {Murphy}, {Pettini}, {Hewett},
  {Ryan-Weber}  \& {Ellison}}{{Zych} et~al.}{2007}]{Zych07}
{Zych} B.~J.,  {Murphy} M.~T.,  {Pettini} M.,  {Hewett} P.~C.,  {Ryan-Weber}
  E.~V.,   {Ellison} S.~L.,  2007, \mn@doi [\mnras]
  {10.1111/j.1365-2966.2007.12015.x}, \href
  {http://adsabs.harvard.edu/abs/2007MNRAS.379.1409Z} {379, 1409}

\bibitem[\protect\citeauthoryear{{Zych}, {Murphy}, {Hewett}  \&
  {Prochaska}}{{Zych} et~al.}{2009}]{Zych09}
{Zych} B.~J.,  {Murphy} M.~T.,  {Hewett} P.~C.,   {Prochaska} J.~X.,  2009,
  \mn@doi [\mnras] {10.1111/j.1365-2966.2008.14157.x}, \href
  {http://adsabs.harvard.edu/abs/2009MNRAS.392.1429Z} {392, 1429}

\bibitem[\protect\citeauthoryear{{van de Voort}, {Schaye}, {Booth}, {Haas}  \&
  {Dalla Vecchia}}{{van de Voort} et~al.}{2011}]{van-de-Voort11}
{van de Voort} F.,  {Schaye} J.,  {Booth} C.~M.,  {Haas} M.~R.,   {Dalla
  Vecchia} C.,  2011, \mn@doi [\mnras] {10.1111/j.1365-2966.2011.18565.x},
  \href {http://adsabs.harvard.edu/abs/2011MNRAS.414.2458V} {414, 2458}

\makeatother
\end{thebibliography}
\appendix
\section{Absorption lines and best obtained Voigt profile fit using X-Shooter data}\label{app_abs_profile}
\begin{figure*}
	\centering	
	\vspace*{-0.6cm}
	\includegraphics[width=0.98\hsize,bb=18 17 594 773,clip=,angle=0]{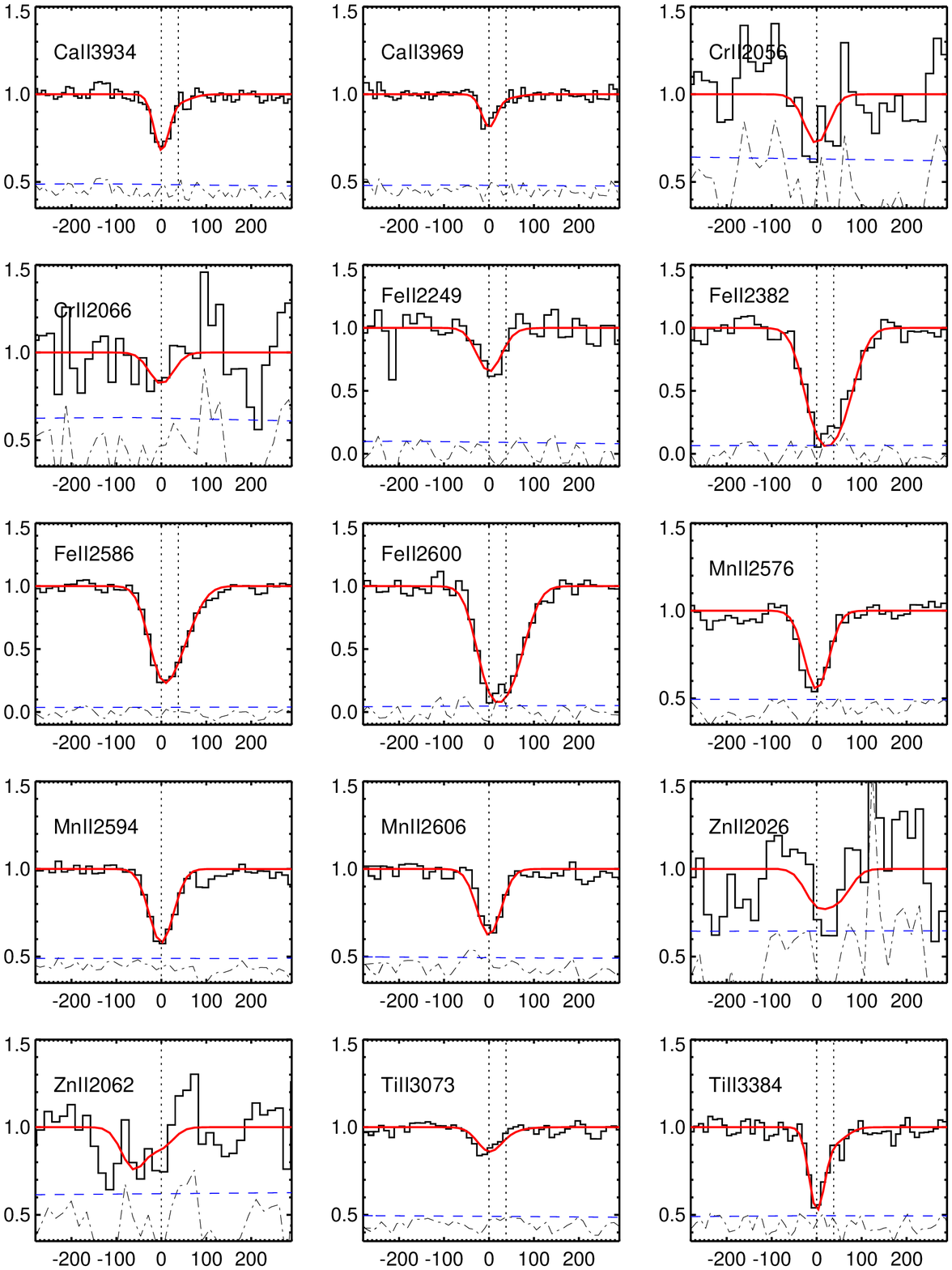}
\vskip 0.5cm
	\caption{ Metal absorption line towards QSO J0957$-$0807.   }
\begin{picture}(400,400)(0,0)
\put( -60,725){\rotatebox{90}{\large Normalized Flux}}
\put( 160,420){\large Velocity (\kms)}
\end{picture}
	\label{fig_0957_abs}
\end{figure*}
\begin{figure*}
	\centering
	\vspace*{-0.6cm}
	\includegraphics[width=0.98\hsize,bb=18 17 594 773,clip=,angle=0]{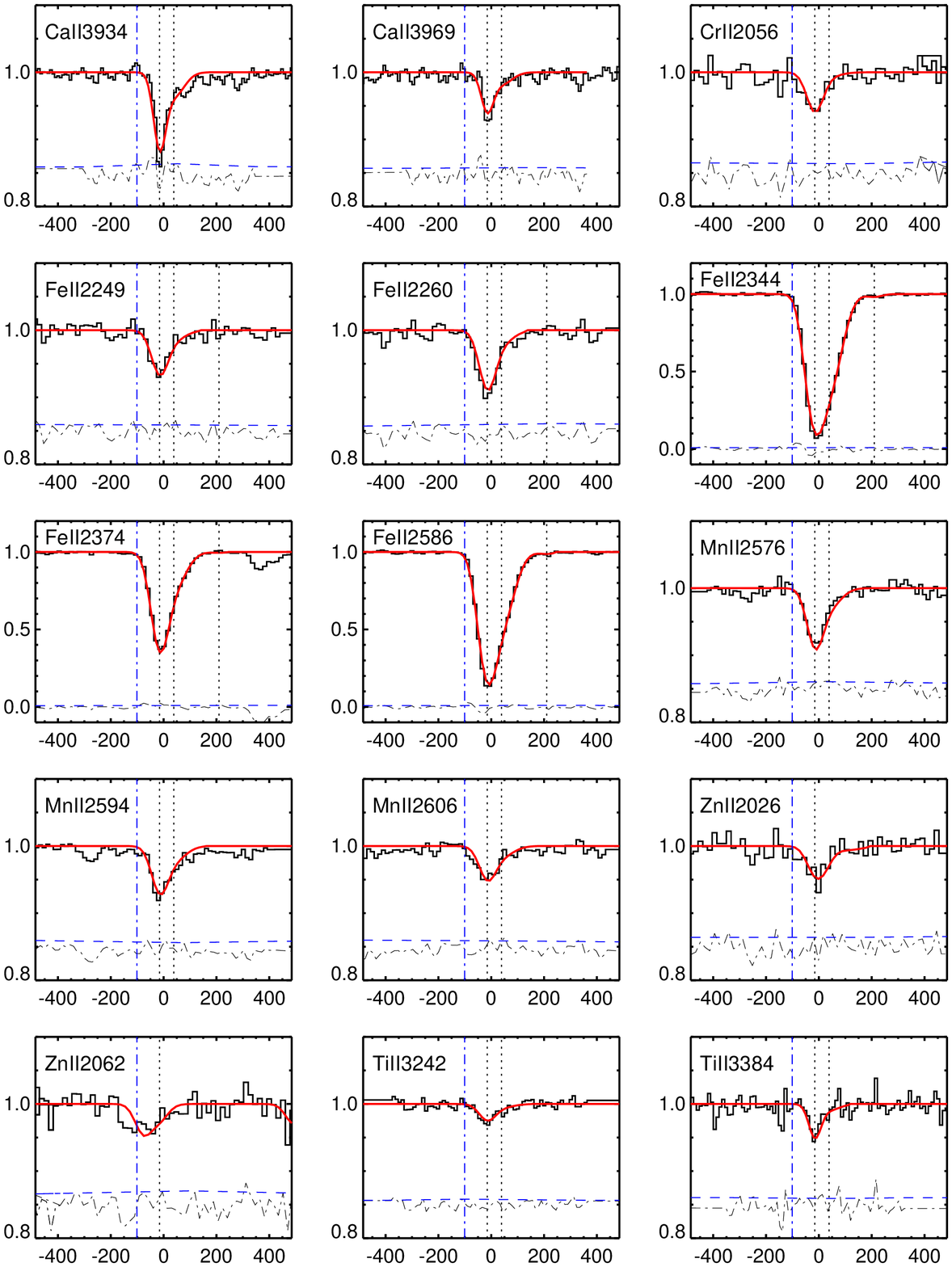}
\vskip 0.5cm
	\caption{Metal absorption lines towards J0958$+$0549. The zero velocity is set at the systemic redshift of the DLA-galaxy at $z$=0.6547. The dashed-dotted line at $\sim -$100 \kms\ marks the velocity of the second galaxy detected in this field (see Fig. \ref{fig_0958_2d}).}
	\begin{picture}(0,0)(0,0)
	\put( -260,350){\rotatebox{90}{\large Normalized Flux}}
	\put( -30,38){\large Velocity (\kms)}
	\end{picture}
	\label{fig_0958_abs}
\end{figure*}
\begin{figure}
	\centering	
	\vspace*{-0.6cm}
	\includegraphics[width=0.8\hsize,bb=164 13 447 778,clip=,angle=0]{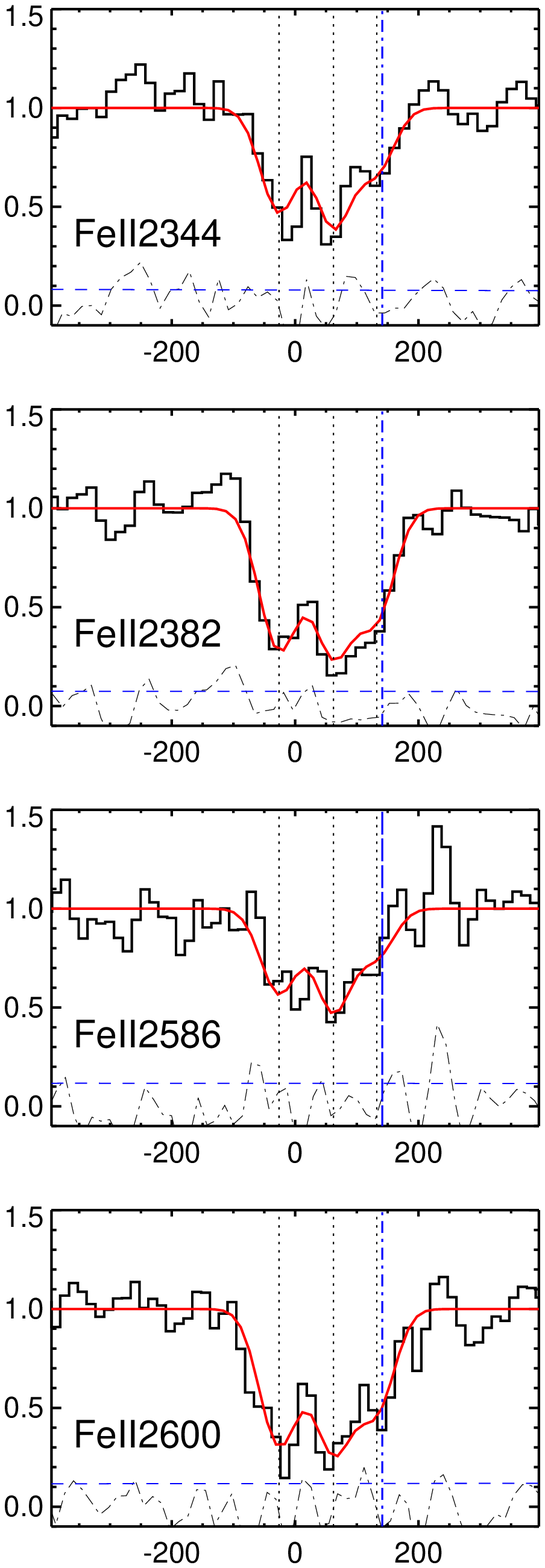}
	\vskip 0.5cm
	\caption{ \FeII\ absorption lines towards QSO J1012$+$0739. The zero velocity is set to the systemic redshift of the emitting galaxy (G2) at $z$ = 0.6154. The dashed-dotted line demonstrate the velocity of the second galaxy detected in this field at an impact parameter of $\sim$ 50 kpc.}
	\begin{picture}(0,0)(0,0)
	\put( -110,300){\rotatebox{90}{\large Normalized Flux}}
	\put( -25,65){\large Velocity (\kms)}
	\end{picture}
	\label{fig_1012_abs}
\end{figure}
\begin{figure*}
	\centering	
	\vspace*{-0.6cm}
	\includegraphics[width=0.98\hsize,bb=18 18 594 774,clip=,angle=0]{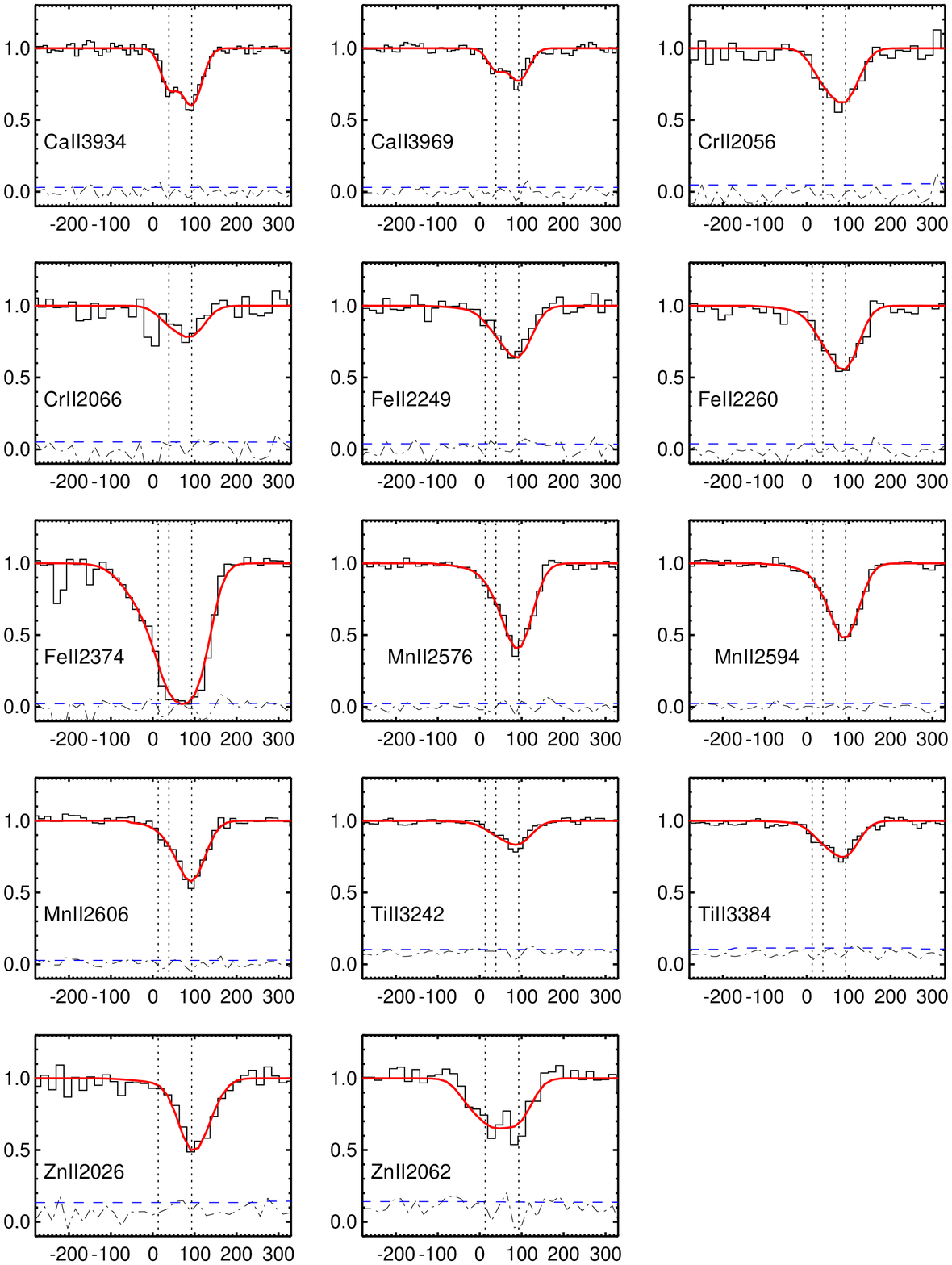}
	\vskip 0.5cm
	\caption{Metal absorption lines towards QSO J1138$+$0139. The reference redshift for the zero velocity is $z$ = 0.6121 which is the systemic redshift of the detected DLA-galaxy.}
	\begin{picture}(0,0)(0,0)
	\put( -260,350){\rotatebox{90}{\large Normalized Flux}}
	\put( -30,38){\large Velocity (\kms)}
	\end{picture}
	\label{fig_1138_abs}
\end{figure*}
\begin{figure*}
	\centering	
	\vspace*{-0.6cm}
	\includegraphics[width=0.98\hsize,bb=8 112 603 678,clip=,angle=0]{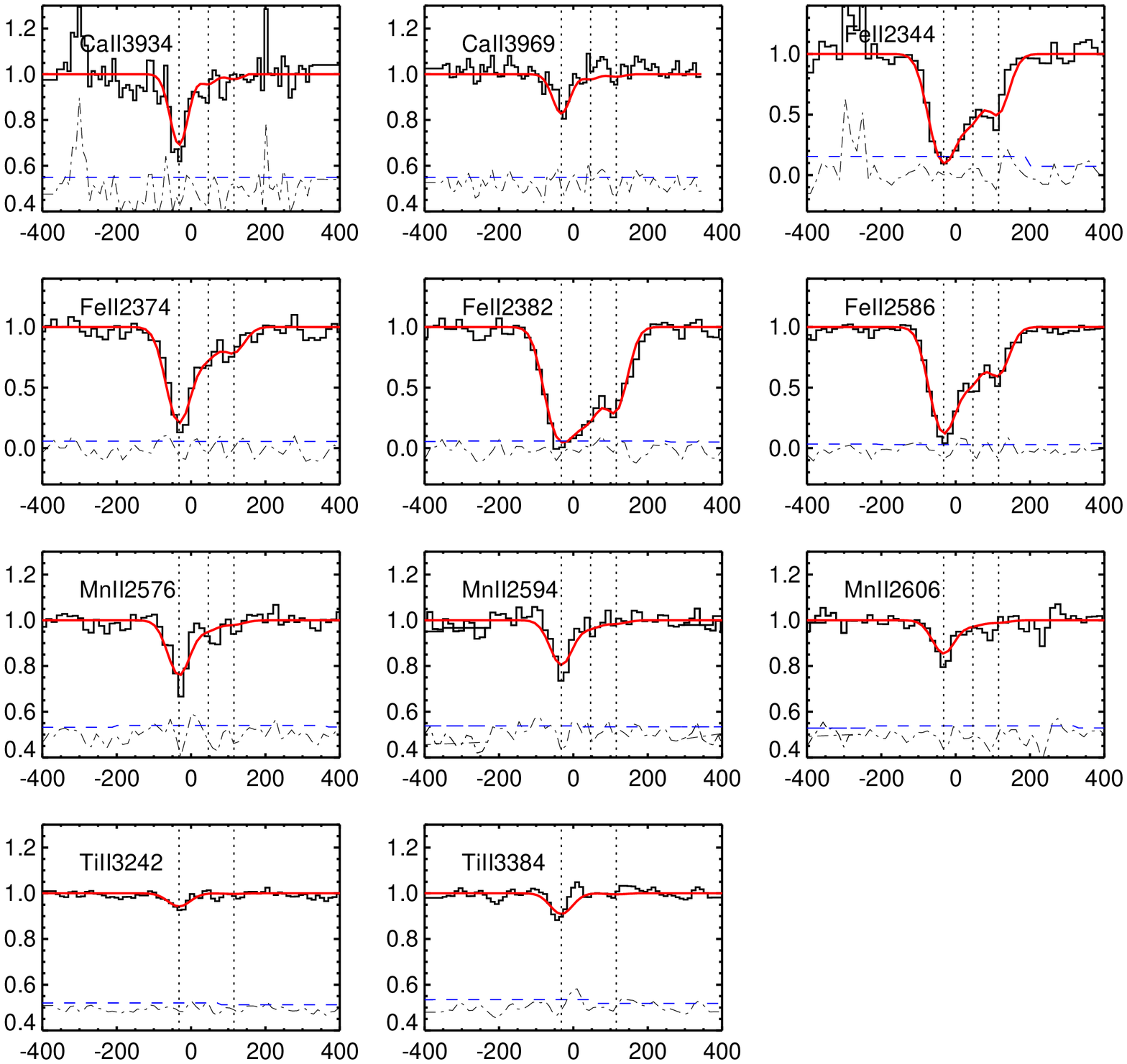}
	\vskip 0.5cm
	\caption{Metal absorption lines towards QSO J1204$+$0953. The reference redshift for the zero velocity is $z$ = 0.6392 which is the systemic redshift of the detected DLA-galaxy.}
	\begin{picture}(0,0)(0,0)
	\put( -260,240){\rotatebox{90}{\large Normalized Flux}}
	\put( -30,45){\large Velocity (\kms)}
	\end{picture}
	\label{fig_1204_abs}
\end{figure*}
\begin{figure*}
	\centering	
	\vspace*{-0.6cm}	
	\includegraphics[width=0.98\hsize,bb=18 17 594 773,clip=,angle=0]{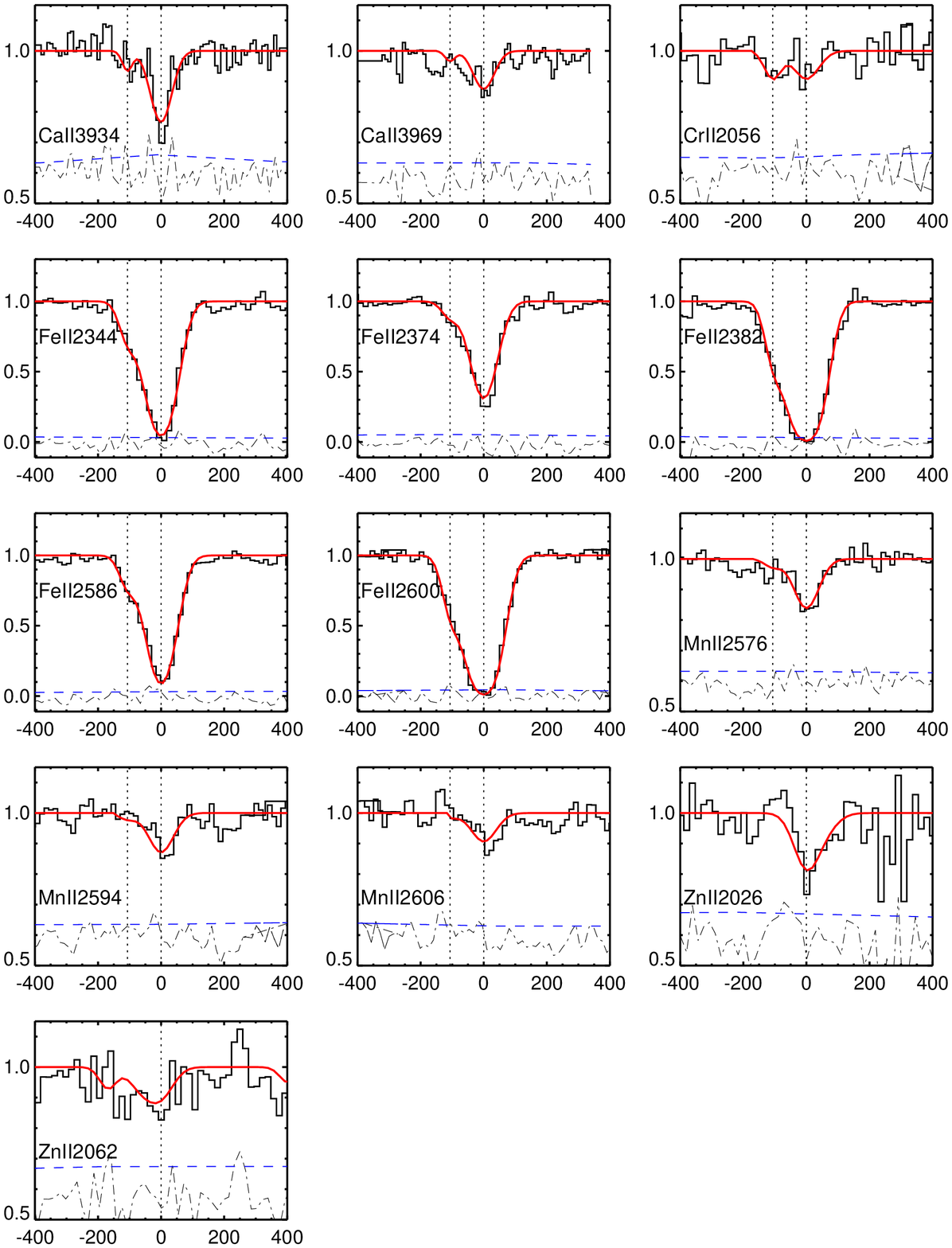}
	\caption{Metal absorption lines towards QSO J1357$+$0525.   }
	\vskip 0.5cm 
	\begin{picture}(0,0)(0,0)
	\put( -260,340){\rotatebox{90}{\large Normalized Flux}}
	\put( -30,45){\large Velocity (\kms)}
	\end{picture}
	\label{fig_1357_abs}
\end{figure*}
\begin{figure}
	\centering	
	\includegraphics[width=0.7\hsize,bb=150 155 461 636,clip=,angle=0]{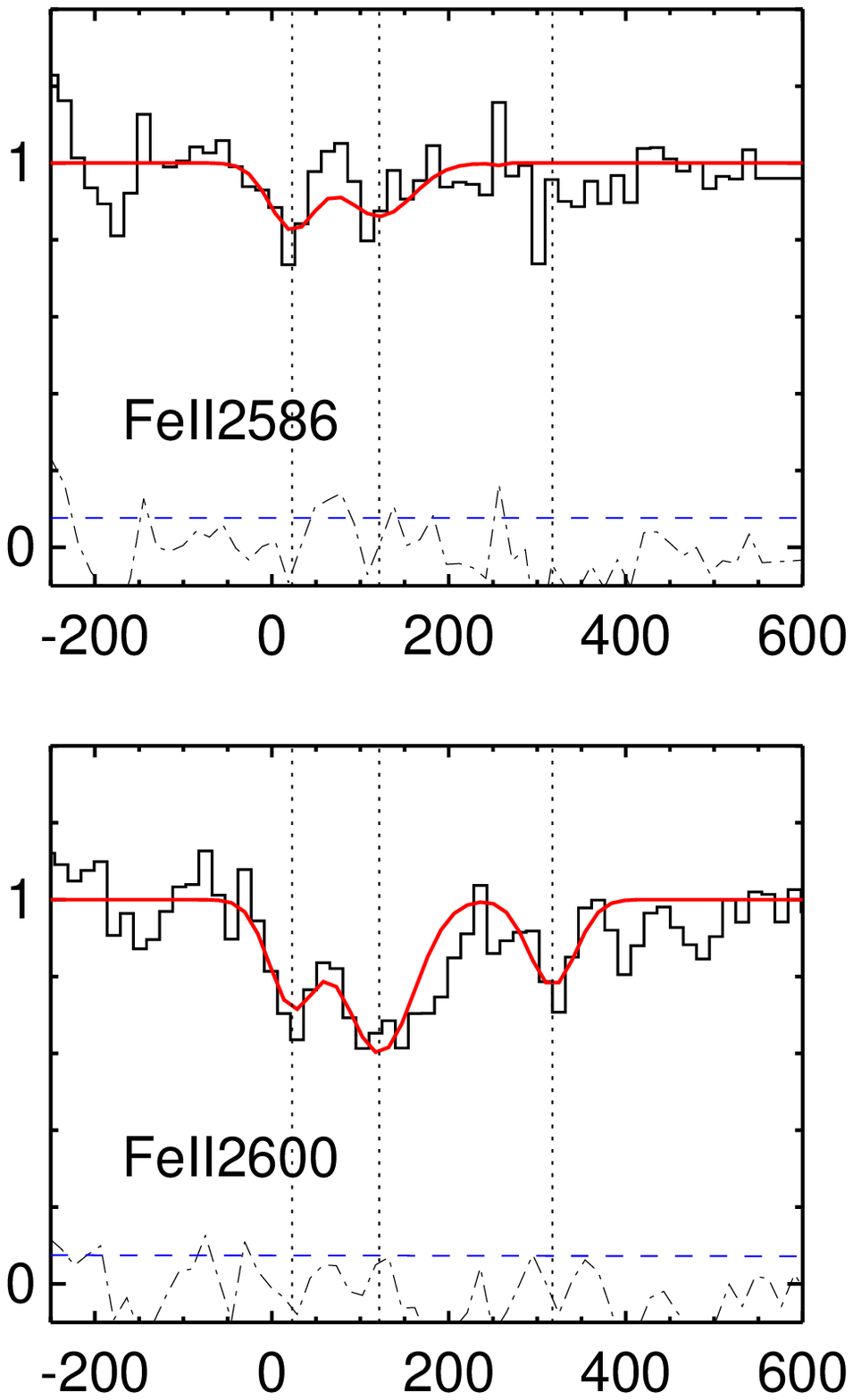}
	\vspace*{0.6cm}	
	\caption{The \FeII\ absorption lines towards QSO J1515$+$0410. The reddest component is seen in the 
		associated \MgII\  lines as well.}
	\vskip 0.5cm 
	\begin{picture}(00,00)(0,0)
	\put( -100,170){\rotatebox{90}{\large Normalized Flux}}
	\put( -30,60){\large Velocity (\kms)}
	\end{picture}
	\label{fig_1515_abs}
\end{figure}
\end{document}